\documentclass[a4paper,12pt]{article}
\usepackage[dvipdfmx]{graphicx}
\usepackage{amsfonts,amsthm,amsmath,amssymb,graphicx,float,mathtools}
\usepackage{xcolor}
\usepackage{cite}
\numberwithin{equation}{section}
\usepackage{subfigure}
\usepackage{color}
\usepackage{appendix}
\usepackage{bbm} 
\usepackage{tensor} 
\usepackage{verbatim} 
\usepackage[
	  pagebackref=false,
	  colorlinks=true,
      linkcolor=blue,
      urlcolor=blue,
      filecolor=black,
      citecolor=red,
      pdfstartview=FitV,
      pdftitle={},
        pdfauthor={},
        pdfsubject={},
        pdfkeywords={},
        pdfpagemode=None,
        bookmarksopen=true
      ]{hyperref}
\usepackage{ascmac}
\usepackage{graphicx} 
\usepackage{ulem}
\usepackage{xcolor}
\DeclareRobustCommand{\erase}{\bgroup\markoverwith{\textcolor{red}{\rule[.5ex]{2pt}{0.4pt}}}\ULon}

\graphicspath{ {Pic/} }

\begin{document}

\newtheorem{definition}{Definition}[section]
\newcommand{\be}{\begin{equation}}
\newcommand{\ee}{\end{equation}}
\newcommand{\bea}{\begin{eqnarray}}
\newcommand{\eea}{\end{eqnarray}}
\newcommand{\LE}{\left[}
\newcommand{\R}{\right]}
\newcommand{\nn}{\nonumber}
\newcommand{\Tr}{\text{Tr}}
\newcommand{\N}{\mathcal{N}}
\newcommand{\G}{\Gamma}
\newcommand{\vf}{\varphi}
\newcommand{\LL}{\mathcal{L}}
\newcommand{\Op}{\mathcal{O}}
\newcommand{\HH}{\mathcal{H}}
\newcommand{\arctanh}{\text{arctanh}}
\newcommand{\up}{\uparrow}
\newcommand{\down}{\downarrow}
\newcommand{\ket}[1]{\left| #1 \right>}
\newcommand{\bra}[1]{\left< #1 \right|}
\newcommand{\ketbra}[1]{\left|#1\right>\left<#1\right|}
\newcommand{\rd}{\partial}
\newcommand{\de}{\partial}
\newcommand{\ba}{\begin{eqnarray}}
\newcommand{\ea}{\end{eqnarray}}
\newcommand{\db}{\bar{\partial}}
\newcommand{\we}{\wedge}
\newcommand{\ca}{\mathcal}
\newcommand{\lr}{\leftrightarrow}
\newcommand{\f}{\frac}
\newcommand{\s}{\sqrt}
\newcommand{\vp}{\varphi}
\newcommand{\hvp}{\hat{\varphi}}
\newcommand{\tvp}{\tilde{\varphi}}
\newcommand{\tp}{\tilde{\phi}}
\newcommand{\ti}{\tilde}
\newcommand{\ap}{\alpha}
\newcommand{\pr}{\approx}
\newcommand{\mb}{\mathbf}
\newcommand{\ddd}{\cdot\cdot\cdot}
\newcommand{\no}{\nonumber \\}
\newcommand{\la}{\langle}
\newcommand{\lb}{\rangle}
\newcommand{\ep}{\epsilon}
 \def\we{\wedge}
 \def\lr{\leftrightarrow}
 \def\f {\frac}
 \def\ti{\tilde}
 \def\ap{\alpha}
 \def\pr{\approx}
 \def\mb{\mathbf}
 \def\ddd{\cdot\cdot\cdot}
 \def\no{\nonumber \\}
 \def\la{\langle}
 \def\lb{\rangle}
 \def\ep{\epsilon}
\newcommand{\mcl}{\mathcal}
 \def\g{\gamma}
\def\Tr{\text{tr}}

\newcommand{\AM}[1]{{\color{orange}{{\bf AM:} #1}}}
\newcommand{\cor}[1]{{\color{green}{#1}}}

\begin{titlepage}
\thispagestyle{empty}

\begin{flushright}

\end{flushright}
\bigskip

\begin{center}
  \noindent{\large \textbf{Entanglement dynamics in 2d HCFTs on the curved background: the case of q-M\"obius Hamiltonian}}\\

\vspace{1cm}
\renewcommand\thefootnote{\mbox{$\fnsymbol{footnote}$}}
Chen Bai\footnote{
baichen22@mails.ucas.ac.cn}${}^{1}$,
Akihiro Miyata\footnote{a.miyata@ucas.ac.cn}${}^{1}$,
and Masahiro Nozaki\footnote{mnozaki@ucas.ac.cn}${}^{1,2}$\\

\vspace{1cm}
${}^{1}${\small \sl Kavli Institute for Theoretical Sciences, University of Chinese Academy of Sciences,
Beijing 100190, China}\\
${}^{2}${\small \sl RIKEN Interdisciplinary Theoretical and Mathematical Sciences (iTHEMS), \\Wako, Saitama 351-0198, Japan}\\

\vskip 4em
\end{center}

\begin{abstract}
     We will explore the dynamical property of non-equilibrium phenomena induced by two-dimensional holographic conformal field theory (2d holographic CFT) Hamiltonian on the curved spacetime by studying the time dependence of the entanglement entropy and mutual information. Here, holographic CFT is the CFT having the gravity dual.
    We will start from the boundary and thermofield double states, evolve the systems in Euclidean time with the Hamiltonian on the curved background, and then evolve them in real-time with the same Hamiltonian. We found that the early- and late-time entanglement structure depends on the curved background, while the entanglement growth does not, and is linear. Furthermore, in the gravity dual for the thermofield double state, this entanglement growth is due to the linear growth of the wormhole, while in the one for the boundary state, it is due to the in-falling of the end of the world brane to the black hole.
    We discussed the low temperature system can be regarded as the dynamical system induced by the multi-joining quenches.
    We also discussed the effective description of the high temperature system, called line tension picture.

\end{abstract}

\end{titlepage} 
\tableofcontents

\section{Introduction and Summary}
Strongly-interacting quantum many-body systems exhibit various interesting dynamical properties.
Information scrambling is one of such properties. Due to the information scrambling, under unitary time evolution of the systems, typically local information is scrambled and spread over a large subsystem of the entire systems after or around some characteristic time scale. Then, local observers who can use only local operations acting on subsystems smaller than the half of the entire systems can not access the original information.
Usually, the ability of the information scrambling is characterized by the Lyapunov exponent of out-of-time-order correlators (OTOC) \cite{Larkin1969QuasiclassicalMI,Roberts:2014ifa,vonKeyserlingk:2017dyr}, and the scrambling time, $t_{\text{scr}}=\lambda_{\text{L}}^{-1}\cdot\log(N)$, is the characteristic time scale of the information scrambling \cite{Maldacena:2015waa,Sekino:2008he,Susskind:2011ap,Roberts:2014ifa}, where $\lambda_{\text{L}},N$ are the Lyapunov exponent and the number of degrees of the freedom in the system, respectively. Recently, the ability is also studied in terms of the Krylov complexity \cite{Parker:2018yvk,Nandy:2024htc,Rabinovici:2020ryf,Chen:2024imd}.

The information scrambling appears in many situations, but the information scrambling on black hole dynamics is prominent and known as maximally chaotic, i.e., the Lyapunov exponent of a semi-classical black hole is proven to saturate an upper bound of the possible Lyapunov exponent within physically reasonable systems \cite{Shenker:2013pqa,Hayden:2007cs}. Such information scrambling on black hole dynamics is closely related to the black hole information paradox \cite{Lashkari:2011yi}. Due to the information scrambling, information on a black hole interior is scrambled and, after some time scale, appears as Hawking radiation, resulting in the apparent loss of the information on the black hole interior, since the Hawking radiation appears to be in a thermal state \cite{Hawking:1975vcx,Hawking:1976ra}. However, recent understanding of Hawking radiation, i.e., the island formula suggests that, after the Page time, the Hawking radiation is not in the thermal state and includes the information on a black hole interior in a very complicated manner \cite{Almheiri:2019hni,Almheiri:2019qdq,Penington:2019kki}. One of the key underlying mechanisms of the phenomena is the information scrambling on black hole dynamics.

From these motivations, previous studies have extensively investigated the information scrambling in quantum many-body systems \cite{vonKeyserlingk:2017dyr,Mascot:2020qep,Hosur:2015ylk,Roberts:2016hpo,Gendiar:2008udd,Hikihara:2011mtb} and holographic settings \cite{Goto:2021gve,Kudler-Flam:2020yml,Mezei:2016wfz,Couch:2019zni,Lin:2024gip,Yu:2022xlh} during time evolution. These studies have provided an understanding of the properties and behavior of information scrambling in typical black holes and the corresponding states of CFTs within holographic settings. However, a general understanding of the conditions under which information scrambling and chaotic dynamics occur in holographic systems has not yet been sufficiently obtained beyond the situations studied so far. Such an understanding is crucial to reveal the essential nature of information scrambling and chaotic dynamics.
To discuss settings beyond the situations studied so far, we need to consider situations that are non-trivial yet controllable to concrete calculations. One such candidate is the dynamics induced by SSD and M\"obius Hamiltonians \cite{Goto:2021sqx,Wen:2018vux,Goto:2023wai,Wen:2020wee,wen2021periodically,Caputa:2020mgb}\footnote{See \cite{Tada:2014kza,Ishibashi:2015jba,Ishibashi:2016bey,Okunishi_2016,2016PhRvB..93w5119W, Gendiar01102009,Gendiar01022010,Hikihara_2011,2011PhRvA..83e2118G,Shibata:2011jup,Shibata_2011,Maruyama_2011,Katsura:2011zyx,Katsura:2011ss,PhysRevB.86.041108,PhysRevB.87.115128} for discussions on the sine-square deformed and M\"obius deformed $2$d CFTs. See also \cite{PhysRevB.97.184309,2019JPhA...52X5401M,2021arXiv211214388G,PhysRevLett.118.260602,2018arXiv180500031W,2020PhRvX..10c1036F,Han_2020,2021PhRvR...3b3044W,2020arXiv201109491F,2021arXiv210910923W,PhysRevB.103.224303,PhysRevResearch.2.023085,PhysRevLett.122.020201,10.21468/SciPostPhys.3.3.019,2016arXiv161104591Z,2020PhRvR...2c3347R,HM,2018PhRvL.120u0604A,2019PhRvB..99j4308M,2022arXiv221100040W,2023arXiv230208009G,2023arXiv231019376N,2023arXiv230501019G,Das:2023xaw,2023arXiv230904665K,Liu:2023tiq,2024arXiv240315851M,2024arXiv240501642L,2024arXiv240216555B,2024arXiv240407884J,Das:2024vqe,Wen:2024bzm} for further developments on their systems.}. These Hamiltonians can be understood as deformations of existing CFT setups and can be interpreted as Hamiltonians of CFTs on curved spacetimes \cite{deBoer:2023lrd,Miyata:2024gvr,Jiang:2024hgt}. Due to the curvature, the spacetime is non-uniform and exhibits inhomogeneity, which can be controlled by the parameters included in the M\"obius Hamiltonian. As a result, through the information scrambling, while information typically spreads uniformly, in this case, due to the inhomogeneity of the SSD and M\"obius Hamiltonians, the information spreads non-uniformly. Therefore, by considering these dynamics, we can concretely calculate and explore whether and how information scrambling and chaotic dynamics occur in such inhomogeneous spacetimes.\footnote{Refer to \cite{Geng:2021mic,Geng:2022dua,Geng:2021iyq} for discussions regarding the entanglement structures of both $d=2$ and $d>2$ holographic CFTs defined on the curved spacetime.} In other words, it allows us to investigate information scrambling and chaotic dynamics on curved spacetimes. Additionally, through the AdS/CFT correspondence \cite{Maldacena:1997re,Witten:1998qj,Gubser:1998bc}, one can construct the bulk spacetimes dual to such inhomogeneous CFTs, and study information scrambling and chaotic dynamics on their spacetimes \cite{Miyata:2024gvr,Goto:2023wai}.
These investigations would reveal understandings that were not visible in uniform CFT setups, and further our understanding that we have obtained in uniform CFT setups.

In \cite{Goto:2023wai,2024arXiv240315851M}, the dynamics of information scrambling related to q-M\"obius Hamiltonian have been studied in the case of $q=1$, corresponding to situations where the vacuum state is the same as that of the uniform case, and the setup is related to the uniform CFT case by a global conformal transformation. In this paper, we focus on the dynamics of information scrambling in more complicated situations, $q=2,3,4,\cdots$, where the vacuum state is not the same as that of the uniform CFT case and the setup can not be obtained by a global conformal transformation from the uniform CFT case. In \cite{Miyata:2024gvr}, thermodynamic and entanglement properties of a thermal state with the q-M\"obius Hamiltonian have been studied, but dynamical aspects related to the Hamiltonian were not discussed. We focus on them in this paper.

An important target for studying the information scrambling associated with the SSD and M\"obius Hamiltonians is the time evolution operator. In this paper, we investigate the properties of these time evolution operators to explore how they exhibit information scrambling. A crucial point in investing these time evolution operators is to focus on their intrinsic properties as much as possible, rather than on the choice of initial states. The quantum quench (see \cite{Mitra:2017fuv} for a review on this topic) and the operator state mapping (the channel-state duality) \cite{Jamiolkowski:1972pzh,Choi:1975nug} are known as such tools, and thus we focus on them in this paper.

In the method using the quantum quench, a boundary state \cite{Cardy:1989ir} is taken as the initial state to introduce a situation with only short-range entanglement, i.e., a situation that excludes the entanglement inherently present in the initial state as much as possible. By considering the time evolution of the boundary state with the SSD and M\"obius Hamiltonians, an inhomogeneous entanglement structure reflecting the structures of these Hamiltonians emerges. Then, by studying this entanglement using entanglement entropy and mutual information, we can investigate the information scrambling associated with the SSD and M\"obius Hamiltonians.

In the method using operator-state mapping (channel-state duality), the time evolution operators for the SSD and M\"obius Hamiltonians are mapped to entangled states in suitable bipartite Hilbert spaces. By investigating the operator entanglement \cite{Mascot:2020qep,Goto:2021gve,Nie:2018dfe} between the bipartite Hilbert spaces using entanglement entropy and mutual information, we can also study the inhomogeneous entanglement dynamics induced by the time evolution operators.

\subsection*{Summary}

In this paper, we Euclidean time evolve the system with the $2$d CFT Hamiltonian on the curved background, so-called q-M\"obius Hamiltonian (For the details of definition, please see Section \ref{sec:preliminary}.), and then Lorentzian time evolve it with the same one.
This inhomogeneous Hamiltonian depends on the two parameters, $\theta$ and $q$.
The initial states considered in this paper are the state with the short-range entanglement, the so-called boundary state, and the state well-approximated as the product of Bell states, the so-called thermofield double state.
We explored the entanglement dynamics during this non-equilibrium process by using the entanglement entropy and the mutual information (See Section \ref{sec:preliminary} for the detail of their definitions.).
We found that the initial and late-time entanglement structure depends on the locations of endpoints, $\theta$, and $q$, while entanglement growth in time does not. 
We proposed the effective description, so-called line tension picture \cite{2017PhRvX...7c1016N,2018arXiv180300089J,2018PhRvX...8c1058R,vonKeyserlingk:2017dyr,2018PhRvD..98j6025M,Kudler-Flam:2019wtv,Kudler-Flam:2020yml,Goto:2021gve,Goto:2022fec,Goto:2023wai}, describing the time dependence of the entanglement entropy in the high temperature regime.
In the low temperature region, we found that the time evolution of the system is approximately the same as in the local and multi-joining quenches \cite{Calabrese:2007mtj,2019JHEP...03..165S,Kudler-Flam:2023ahk}. 
In addition, we explored the gravity dual of the systems considered.
Similar to \cite{HM}, for the thermofield double state, the wormhole, penetrating the black hole, linearly grows, while for the boundary state,  the end of the world brane (EOW) falls to the interior of the black hole.
\section{Preliminary \label{sec:preliminary}}
In this section, we will explain q-M\"obius Hamiltonian, the systems considered, the limits considered, and the quantities considered in this paper.
Define the effective inhomogeneous time operator as
\be
\begin{split}
   U_{\text{eff}}:=e^{-(\epsilon+it)H_{\text{q-M\"obius}}},
\end{split}
\ee
where $\epsilon$ and $t$ are positive real numbers, and $H_{\text{q-M\"obius}}$ is the so-called q-M\"obius Hamiltonian,
\be\label{q-Mobius Hamiltonian}
    H_{\text{q-M\"obius}} = \int^{L}_0 {\frac{dx}{2\pi}} \left[1-\tanh{2\theta}\cos{\left(\f{2q\pi x}{L}\right)}\right] h(x),
\ee
where we consider the spatially-periodic system with the period of $L$, $\theta$ is a {non-negative} real number, $q$ is a positive integer, and $h(x)$ is the Hamiltonian density.
The Hamiltonian considered in this paper is in two-dimensional conformal field theories ($2$d CFTs).
By using the channel-state duality \cite{Jamiolkowski:1972pzh,Choi:1975nug}, define the dual states of $U_{\text{eff}}$ as 
\be \label{eq:dual-state-of-eff-H}
\begin{split}
    &\ket{U_{\text{eff}}}=\f{1}{\sqrt{\Tr e^{-2\epsilon H_{\text{q-M\"obius}}}}}\sum_a e^{-(\epsilon+it)E_a^{\text{q-M\"obius}}}\ket{E_a^{\text{q-M\"obius}}}_1\otimes \ket{E_a^{\text{q-M\"obius} *}}_2,
\end{split}
\ee
where $\ket{E_a^{\text{q-M\"obius}}}_i$ are the eigenstates of $H_{\text{q-M\"obius}}$ on the $i$-th Hilbert space,  and $\ket{\cdot^*}$ is the CPT conjugate of $\ket{\cdot}$. 
The dual states are defined on the double Hilbert space, $\mathcal{H}=\mathcal{H}_1\otimes \mathcal{H}_2$. 

In addition, define the system, evolved from a short-range entangled state, as,
\be\label{eq:Bdy-state}
\ket{\Psi}=\mathcal{N}_2e^{-(\epsilon+it) H_{\text{q-M\"obius}}}\ket{\text{B}},
\ee
where $\epsilon$ is a {positive} real parameter, $\ket{B}$ is a boundary state, and $\mathcal{N}_2$ guarantees that $\left \langle \Psi\big{|}\Psi\right \rangle=1$.
Then, $\mathcal{N}_{2}$ is given by
\be
\mathcal{N}_2=\f{1}{\sqrt{\bra{B}e^{-2\epsilon H_{\text{q-M\"obius}}}\ket{B}}}
\ee

Define the density operators of (\ref{eq:dual-state-of-eff-H}) and (\ref{eq:Bdy-state}) {as}
\be
\begin{split}
    \rho_1 &= \ket{U_{\text{eff}}} \bra{U_{\text{eff}}}=\f{1}{\Tr e^{-2\epsilon H_{\text{q-M\"obius}}}}\sum_{a,b} e^{-\epsilon\left(E_a^{\text{q-M\"obius}}+E_b^{\text{q-M\"obius}}\right)-it\left(E_a^{\text{q-M\"obius}}-E_b^{\text{q-M\"obius}}\right)}\\
    &~~~~~~~~~~~~~~~~~~\times \ket{E_a^{\text{q-M\"obius}}}_1\bra{E_b^{\text{q-M\"obius}}}_1\otimes \ket{E_a^{\text{q-M\"obius} *}}_2\bra{E_b^{\text{q-M\"obius}*}}_2,\\
    \rho_{2}&=\f{1}{\bra{B} e^{-2\epsilon H_{\text{q-M\"obius}}}\ket{B}}e^{-(\epsilon+it) H_{\text{q-M\"obius}}}\ket{\text{B}} \bra{\text{B}}e^{-(\epsilon-it) H_{\text{q-M\"obius}}}.
\end{split}
\ee
Then, divide the Hilbert space into the spatial sub-region, $\mathcal{V}$, and the complement spatial region, $\overline{\mathcal{V}}$.
Subsequently, define the reduced density matrix as 
$\rho_{i=1,2;\mathcal{V}}=\Tr_{\overline{\mathcal{V}}} \rho_{i=1,2}$, and then define the entanglement entropy for $\mathcal{V}$ as the von Neunmann entanglement entropy for $\rho_{i=1,2;\mathcal{V}}$,
\be
S^{i=1,2}_{\mathcal{V}}= \lim_{n\rightarrow 1}S^{i=1,2;(n)}_{\mathcal{V}}= \lim_{n\rightarrow 1}\f{\log{\Tr_{\mathcal{V}}\left(\rho_{i=1,2;\mathcal{V}}\right)^{n}}}{1-n} =-\Tr_{\mathcal{V}}\left(\rho_{i=1,2;\mathcal{V}}\log{\rho_{i=1,2;\mathcal{V}}}\right),
\ee
where we define $n$-th moment of R\'enyi entanglement entropy as $ S^{i=1,2;(n)}_{\mathcal{V}}:= \f{\log{\Tr_{\mathcal{V}}\left(\rho_{i=1,2;\mathcal{V}}\right)^{n}}}{1-n} $.
Define the two spatial intervals as $\mathcal{V}_{j=1,2}$, and then define the mutual information between $\mathcal{V}_{j=1,2}$ by
\be \label{eq:Mutual-information}
I^{i=1,2}_{\mathcal{V}_1,\mathcal{V}_2} = \sum_{j=1,2}S^{i=1,2}_{\mathcal{V}_{j}}-S^{i=1,2}_{\mathcal{V}_{1}\cup \mathcal{V}_{2} }.
\ee

\subsection{Twist operator formalism \label{sec:TOF}}
In this paper, we will employ the twist operator formalism \cite{Calabrese:2009qy,2004JSMTE..06..002C} that suits to analytically compute (R\'enyi) entanglement entropy.
To do so, we begin by defining the Euclidean density matrices that are counterparts of $\rho_{i=1,2}$ as 
\be
\begin{split}
    &\rho^{\text{E}}_{i=1}=\f{1}{\Tr e^{-2\epsilon H_{\text{q-M\"obius}}}}\sum_{a,b}e^{-\f{\left(\epsilon+\tau_1\right)}{2}E^{\text{q-M\"obius}}_{a}}e^{-\f{\left(\epsilon+\tau_2\right)}{2}E^{\text{q-M\"obius}}_{b}}\\
    &~~~~~~~~~~~~~~~~~~{\times \ket{E_a^{\text{q-M\"obius}}}_1\bra{E_b^{\text{q-M\"obius}}}_1\otimes \ket{E_a^{\text{q-M\"obius} *}}_2\bra{E_b^{\text{q-M\"obius}*}}_2,}\\
    & \rho^{\text{E}}_{i=2}=\f{e^{-\tau_a H_{\text{q-M\"obius}}}\ket{B}\bra{B}e^{-\tau_b H_{\text{q-M\"obius}}}}{\bra{B}e^{-2\epsilon H_{\text{q-M\"obius}}}\ket{B}},
\end{split}
\ee
{where $\tau_{1,2},\tau_{a,b}$ are Euclidean times.}
Then, define the Euclidean (R\'enyi) entanglement entropy as 
\be
\begin{split}
&S^{i=1,2;(n)}_{\mathcal{V};\text{E}}:=\f{\log{\Tr_{\mathcal{V}}\left(\rho^{\text{E}}_{i=1,2;\mathcal{V}}\right)^{n}}}{1-n},~S^{i=1,2}_{\mathcal{V};\text{E}}:=\lim_{n\rightarrow 1} S^{i=1,2;(n)}_{\mathcal{V};\text{E}}= -\Tr_{\mathcal{V}}\left(\rho^{\text{E}}_{i=1,2;\mathcal{V}}\log{\rho^{\text{E}}_{i=1,2;\mathcal{V}}}\right).
\end{split}
\ee
In the path-integral formalism, $n$-th moment of Euclidean R\'enyi entanglement entropy results in 
\be
\begin{split}
    S^{i=1,2;(n)}_{\mathcal{V};\text{E}} =\f{1}{1-n}\log{\left[\f{Z_{n}}{Z^n_{1}}\right]},
\end{split}
\ee
where $Z_{n}$ is defined as the partition function on the $n$-sheet geometry, where each sheet is connected along $\mathcal{V}$, while $Z_1$ is the one on the single sheet geometry.
In the twist operator formalism, the $n$-th R\'enyi entanglement entropy is given by evaluating the {$k$-}point function of the primary operators, the so-called twist and anti-twist operators {$\sigma_{n},\overline{\sigma}_{n}$}, instead of $Z_n$ and $Z_1$.
Here, {$k$} is {a positive} integer, and it is determined by the number of subsystem edges. 
Their conformal dimensions are given by $(h_n,h_n)=\left(\f{c(n^2-1)}{24n}, \f{c(n^2-1)}{24n}\right)$.
In the following, we will explain $S^{i=1,2;(n)}_{\mathcal{V},\text{E}}$ for the single interval and double ones in the twist operator formalism.
\subsubsection{Single interval \label{sec:EE-single-interval}}
Now, consider the Euclidean (R\'enyi) entanglement entropy for the single interval.
For $\rho^{i=1}$, we take the subsystem $\mathcal{V}=A_{j=1,2}$ to be a spatial interval on $\mathcal{H}_1$, while for $\rho^{i=2}$, the subsystem is the spatial one.
More concretely, $A_{j=1,2}$ are defined as 
\be \label{eq:subsystems-considered}
A_{1}=\left\{x\big{|}X^f_m<X_2<x<X_1<X^f_{m+1}\right\}, A_{2}=\left\{x\big{|}\hat{X}_2+X^f_m=X_2<x<X_1=\hat{X}_1+X^f_{m+l}\right\},
\ee
where {$0<\hat{X}_1-\hat{X}_2<\f{L}{q}$}, and the fixed points are defined as $X^f_{m}=\f{mL}{q}$, where $m$ are integer numbers and it runs from zero to $q-1$.
The region of $\hat{X}_K$ is ${L/q}>\hat{X}_K>0$.
The integer, $m+l$, is less than or equal to $q-1$.
As in \cite{Nie:2018dfe}, in the twist operator formalism, $S^{i=1,2;(n)}_{\mathcal{V},\text{E}}$ results in 
\be \label{eq:EE-single-interval}
\begin{split}
{S^{i=1;(n)}_{A_{j};E}}&=\f{1}{1-n}\log{\left[\Tr \left(\rho^{\text{E}}_{i=1}\sigma_{n}(X_1)\overline{\sigma}_{n}(X_2)\right)\right]}=\f{1}{1-n} \log{\left[\f{\Tr\left(\sigma_{n}(w_{X_1},\overline{w}_{X_1})\overline{\sigma}_{n}(w_{X_2},\overline{w}_{X_2}) e^{-2\epsilon H_{\text{q-M\"obius}}}\right)}{\Tr e^{-2\epsilon H_{\text{q-M\"obius}}}}\right]}\\
&=\frac{1}{1-n}\log{\left\langle\sigma_n\left(\xi_{X_1},\bar{\xi}_{X_1}\right)\overline{\sigma}_n\left(\xi_{X_2},\bar{\xi}_{X_2}\right)\right\rangle_{\text{q-M\"obius},2\epsilon}}+\frac{h_n}{1-n}\log\left(\prod_{K=1}^2\left|\frac{d\xi_{X_K}}{dw_{X_K}}\right|^2\right),\\
&=\frac{1}{1-n}\log{\left\langle\sigma_n\left(\xi_{X_1},\bar{\xi}_{X_1}\right)\overline{\sigma}_n\left(\xi_{X_2},\bar{\xi}_{X_2}\right)\right\rangle_{\text{q-M\"obius},2\epsilon}}+\frac{c(n+1)}{12n}\log\left(\prod_{K=1}^2f(X_K)\right),\\
{S^{i=2;(n)}_{A_{j};E}}&=\f{1}{1-n}\log{\left[\f{\bra{B}e^{-\tau_b H_{\text{q-M\"obius}}}\sigma_{n}(w_{X_1},\overline{w}_{X_1})\overline{\sigma}_{n}(w_{X_2},\overline{w}_{X_2}) e^{-\tau_a H_{\text{q-M\"obius}}}\ket{B}}{\bra{B}e^{-2\epsilon H_{\text{q-M\"obius}}}\ket{B}}\right]}\\
&=\f{1}{1-n}\log{\left[\f{\bra{B}e^{-(\tau_b+\tau_a) H_{\text{q-M\"obius}}}\sigma_{n,H}(w_{X_1},\overline{w}_{X_1})\overline{\sigma}_{n,H}(w_{X_2},\overline{w}_{X_2}) \ket{B}}{\bra{B}e^{-2\epsilon H_{\text{q-M\"obius}}}\ket{B}}\right]},
\end{split}
\ee
where the expectation value is defined as $\left \langle \cdot \right \rangle_{\text{q-M\"obius},2\epsilon}=\f{\Tr~ \cdot~ e^{-2\epsilon H_{\text{q-M\"obius}}}}{\Tr e^{-2\epsilon H_{\text{q-M\"obius}}}}$ {and evaluated on a torus with the thermal circumstance of $2\epsilon$ and the spatial circumstance of $L_{\text{eff}}=L\cosh(2\theta)$}, and the complex coordinates $(w_{X}, \overline{w}_X)$ are defined as ${(w_{X},\overline{w}_{X})}=(y+iX,y-iX)$, where $y$ and $X$ are real values.
The complex valuables, $(\xi_X, \overline{\xi}_X)$, are defined as 
{\be\label{Mobius-maps}
\begin{split}
     &\xi_X=\frac{L_{\text{eff}}}{2\pi }\log\left[\chi_X\right],~ \overline{\xi}_X=\frac{L_{\text{eff}}}{2\pi }\log\left[\overline{\chi}_X\right],\\
     &\chi_X=\left(\frac{\cosh(\theta)z_X^q-\sinh(\theta)}{\cosh(\theta)-\sinh(\theta)z_X^q}\right)^{\f{1}{q}},~\overline{\chi}_X=\left(\frac{\cosh(\theta)\overline{z}_X^q-\sinh(\theta)}{\cosh(\theta)-\sinh(\theta)\overline{z}_X^q}\right)^{\f{1}{q}},\\
     &z_X=e^{\f{2\pi w}{L}}|_{w=iX}, ~\overline{z}_X=e^{\f{2\pi \overline{w}}{L}}|_{\overline{w}=-iX}.
\end{split}
\ee}
The function, $f(X_K)$, is the envelop function associated with (\ref{q-Mobius Hamiltonian})
\be \label{eq:envelopfunction}
f(X_K) = 1-\tanh{(2\theta)}\cos{\left(\f{2\pi q X_K}{L}\right)}.
\ee
The operator, $\mathcal{O}_{H}$, denotes the primary operator in the Heisenberg picture, and it is given by
\be \label{eq:transformation-lop}
\begin{split}
\mathcal{O}_{H}(w_X,\overline{w}_X)&:=e^{TH_{\text{q-M\"obius}}} \mathcal{O}(w_X,\overline{w}_X)e^{-TH_{\text{q-M\"obius}}}=\left|\f{d(\xi_X+T)}{dw_X}\right|^{2h_{\mathcal{O}}}\mathcal{O}(\xi_X+T,\overline{\xi}_X+T)\\
&=\f{\mathcal{O}(\xi_X+T,\overline{\xi}_X+T)}{f^{2h_{\mathcal{O}}}(X)},
\end{split}
\ee
where we assume that its conformal dimension is given by $h_{\mathcal{O}}=\overline{h}_{\mathcal{O}}$.
This is because in the $(\xi,\overline{\xi})$ coordinates, $H_{\text{q-M\"obius}}$ results in the dilatation operator 
{\be
\begin{split}
    H_{\text{q-M\"obius}}=\f{2\pi}{L_{\text{eff}}}\left[L^{\chi}_0+\overline{L}^{\overline{\chi}}_0-\f{c}{12}\right]+\f{2\pi cq^2}{12L_{\text{eff}}}-\f{2\pi cq^2}{12L},
\end{split}
\ee}
where Virasoro's generators are defined as 
\be
L^{\chi}_{m}=\oint \f{d\chi}{2\pi i}\chi^{1+m}T(\chi), \overline{L}^{\overline{\chi}}_{m}=\oint \f{d\overline{\chi}}{2\pi i}\overline{\chi}^{1+m}\overline{T}(\overline{\chi}),
\ee
where $m$ are integers, and $T(\chi)$ and $\overline{T}(\overline{\chi})$ are the chiral and anti-chiral pieces of the energy density. 
After performing the transformation in (\ref{eq:transformation-lop}) and assuming $\tau_a+\tau_b=2\epsilon$, ${S^{i=2;(n)}_{A_{j};E}}$ reduces to
\be
\begin{aligned}
    {S^{i=2;(n)}_{A_{j};E}}&={\f{c}{24}\cdot\f{n+1}{n} \log{\left[\prod_{K=1,2}f^2(X_K)\right]}}\\
    & \qquad +{\f{1}{1-n}\log{\left\langle\sigma_n(\xi_{X_1}+\tau_a,\overline{\xi}_{X_1}+\tau_a)\overline{\sigma}_n(\xi_{X_2}+\tau_a,\overline{\xi}_{X_2}+\tau_a)\right\rangle_{B}}},
\end{aligned}
\ee
where the expectation value is defined as {$\left \langle \cdot \right \rangle_{B}=\f{\bra{B}~\cdot~e^{-2 \epsilon H_{\text{q-M\"obius}}} \ket{B}}{\bra{B}e^{-2\epsilon H_{\text{q-M\"obius}}}\ket{B}}$}.

\subsubsection{Double intervals \label{sec:EE-double-intervals}}
Now, we will explore the entanglement entropy for the double intervals in the twist operator formalism.
We begin by exploring it for {$\rho_{i=1}$}.
For simplicity, we assume that $A_{j=1,2}$ are the subsystems of $\mathcal{H}_1$.
In addition, we define the subsystem $B$ as the spatial interval of $\mathcal{H}_2$
\be \label{sec:two-intervals}
\begin{split}
    B= \left\{x|0<Y_2<x<Y_1<L\right\}=\begin{cases}
        B_1~&\text{where}~Y_1,Y_2\in(X_n^f,X_{n+1}^f)\\
        B_2~&\text{where}~Y_2\in(X_{n}^f,X_{n+1}^f),~Y_1\in(X_{n+p}^f,X_{n+p+1}^f),
    \end{cases}
\end{split}
\ee
where $n,p$ are integers that run from zero to $q-1$. Then, we consider the entanglement entropy for the union of $A_j$ and $B$, and the Euclidean reduced density matrix associated with {$\mathcal{V}=A_j \cup B$} is given by
{\be
\begin{split}
     \rho^{\text{E}}_{i=1;A_{j=1,2}\cup B}&=\f{1}{\Tr e^{-2\epsilon H_{\text{q-M\"obius}}}}\sum_{a,b}e^{-\f{\left(\epsilon+\tau_1\right)}{2}E^{\text{q-M\"obius}}_{a}}e^{-\f{\left(\epsilon+\tau_2\right)}{2}E^{\text{q-M\"obius}}_{b}}\\
     &\times\Tr_{\overline{A}_{j}}\ket{E^{\text{q-M\"obius}}_{a}}_{1}\bra{E^{\text{q-M\"obius} }_{b}}_1\otimes \Tr_{\overline{B}}\ket{E^{\text{q-M\"obius} *}_{a}}_{2}\bra{E^{\text{q-M\"obius} *}_{b}}_2,
\end{split}
\ee}
where {$\overline{A}_{j}$} ($\overline{B}$) is the complement space of $\mathcal{H}_{1}$ ($\mathcal{H}_{2}$) to {$A_{j}$} ($B$). 
As in \cite{Nie:2018dfe,HM}, in the twist operator formalism, the entanglement entropy associated with $A_{j=1,2}\cup B$ is given by
\be \label{eq:Euclidean-REE-in-operator-formalism}
\begin{split}
    S^{i=1;(n)}_{A_j\cup B;\text{E}}=\f{1}{1-n}\log{\left\langle \sigma_n(X_1)\overline{\sigma}_n(X_2)\overline{\sigma}_n\left(Y_1,-\f{\epsilon+\tau_1}{2}\right)\sigma_n\left(Y_2,-\f{\epsilon+\tau_1}{2}\right)\right \rangle_{\text{q-M\"obius},2\epsilon}}.
\end{split}
\ee
The operators, $\overline{\sigma}_n\left(Y_1,-\f{\epsilon+\tau_1}{2}\right)$ and $\sigma_n\left(Y_2,-\f{\epsilon+\tau_1}{2}\right)$, are defined as 
\be
\begin{split}
    &\sigma_n\left(Y_2,-\f{\epsilon+\tau_1}{2}\right)=e^{-\f{\epsilon+\tau_1}{2}H_{\text{q-M\"obius}}}\sigma_n\left(Y_2\right)e^{\f{\epsilon+\tau_1}{2}H_{\text{q-M\"obius}}},\\
    &\overline{\sigma}_n\left(Y_1,-\f{\epsilon+\tau_1}{2}\right)=e^{-\f{\epsilon+\tau_1}{2}H_{\text{q-M\"obius}}}\overline{\sigma}_n\left(Y_1\right)e^{\f{\epsilon+\tau_1}{2}H_{\text{q-M\"obius}}}.
\end{split}
\ee
In the $(\xi, \overline{\xi})$ coordinates, $S^{i=1;(n)}_{A_j\cup B;\text{E}}$ is given by
\be
\begin{split}
   &S^{i=1;(n)}_{A_j\cup B; \text{E}}\\
   &=\f{1}{1-n}\log{\left\langle \sigma_n(\xi_{X_1},\overline{\xi}_{X_1})\overline{\sigma}_n(\xi_{X_2},\overline{\xi}_{X_2})\overline{\sigma}_n\left(-\tilde{T}+\xi_{Y_1}, -\tilde{T}+\overline{\xi}_{Y_1}\right)\sigma_n\left(-\tilde{T}+\xi_{Y_2}, -\tilde{T}+\overline{\xi}_{Y_2}\right)\right \rangle_{\text{q-M\"obius},2\epsilon}}\\
   &~~~{+\f{c}{12}\cdot \f{n+1}{n}\log{\left(\prod_{K=1,2}f(X_K)f(Y_K)\right)}},
\end{split}
\ee
where the parameter, $\tilde{T}$, is defined as $\tilde{T}=\f{\epsilon+\tau_1}{2}$.

Subsequently, we consider the entanglement entropy of $\rho^{\text{E}}_{i=2}$ for the two intervals.
Define the subsystem $B$ {of $\mathcal{H}_{1}$} as in (\ref{sec:two-intervals}), where we assume that $0<X_2<X_1<Y_2<Y_1<L$.

\subsubsection{Universal piece and theory-dependent piece \label{sec:U.N-T.D}}
Here, we define the universal and theory-dependent pieces, $S^{i;\text{U.}(n)}_{\mathcal{V}=A_j,B,A_j\cup B;\text{E}}$ and $S^{i;\text{T.D.}(n)}_{\mathcal{V}=A_j,B,A_j\cup B;\text{E}}$.
First, we define the universal piece as the piece that does not depend on the details of $2d$ CFT
\be \label{eq:U-Piece}
\begin{split}
    S^{i;\text{U.}(n)}_{\mathcal{V}=A_j,B,A_j\cup B;\text{E}} =\f{c(n+1)}{12n} \times \begin{cases}
        \log{\left[\prod_{K=1}^2 f(X_K)\right]}~&~ \text{for}~\alpha=A_j\\
        \log{\left[\prod_{K=1}^2 f(Y_K)\right]}~&~ \text{for}~\alpha=B\\
         \log{\left[\prod_{K=1}^2 f(X_K)f(Y_K)\right]}~&~ \text{for}~\alpha=A_j\cup B\\
    \end{cases}
\end{split}
\ee
Then, we define the theory-dependent piece, $S^{i;\text{T.D.}(n)}_{\mathcal{V}=A_j,B,A_j\cup B}$, as the piece depending on the details of the $2$d CFTs
\be\label{eq:T-D-Piece}
S^{i;\text{T.D.}(n)}_{\mathcal{V}=A_j,B,A_j\cup B;\text{E}}=S^{i;(n)}_{A_j\cup B; \text{E}}-S^{i;\text{U.}(n)}_{\mathcal{V}=A_j,B,A_j\cup B;\text{E}}.
\ee
By using the definition in (\ref{eq:Mutual-information}), the mutual information is independent of the universal piece,
\be
I^{i}_{A_{j},B;\text{E}}=\lim_{n\rightarrow 1}\left[ {\sum_{\alpha=A_j,B}} S^{i;\text{T.D.}(n)}_{\alpha;\text{E}}-S^{i;\text{T.D.}(n)}_{A_j\cup B;\text{E}}\right]
\ee
Later, we perform the analytic continuation,
\be \label{eq:ac}
\begin{split}
    \tau_{j=1,2}=\epsilon+\begin{cases}
        2it~&~\text{for}~j=1\\
       -2it~&~\text{for}~j=2\\
    \end{cases},~\tau_{\tilde{A}=a,b}=\epsilon+\begin{cases}
        it~&~\text{for}~{\tilde{A}}=a\\
        -it~&~\text{for}~{\tilde{A}}=b\\
    \end{cases},
\end{split}
\ee
and then we will explore the time dependence of the entanglement entropies and mutual information.

\section{Entanglement dynamics of $\ket{U_{\text{eff}}}$\label{Sec:Etanglement-dynamics-Ueff}}
Here, we will report on the time dependence of the entanglement entropy for the single interval and the mutual information for $\ket{U_{\text{eff}}}$.
From here, we take the von Neumann limit, where $n\rightarrow 1$, and then consider the entanglement entropy and mutual information in the two-dimensional holographic conformal field theories ($2$d HCFTs), the CFTs possessing the gravity dual.
We begin by considering the theory-dependent pieces in $2$d holographic CFTs.
As explained in Section \ref{sec:EE-single-interval}, these terms for $\ket{U_{\text{eff}}}$ is 
determined by the multi-point function on the torus.
As in \cite{Witten:1998zw,Miyata:2024gvr}, in 2d HCFT, the gravity dual of the torus exhibits the phase transition concerning $\tau_{\text{eff}}=\f{L_{\text{eff}}}{2\epsilon}$.
In the low effective temperature regime, where $\tau_{\text{eff}}<1$, the gravity dual is given by the thermal $\text{AdS}_{3}$, while in the high effective temperature regime, $\tau_{\text{eff}}>1$, the one is given by the BTZ black hole. In this paper, we consider the high effective temperature region. By employing the celebrated Ryu-Takayanagi formula, \cite{Ryu:2006ef,Ryu:2006bv}, the theory-dependent pieces in this region are given by the geodesic lengths on the BTZ black hole. In the following sections, we explore the theory-dependent pieces in the high effective temperature regime.

\subsection{Entanglement entropy for the single and double intervals\label{Sec:EE-for-singleinterval}}
Now, we will focus on the theory-dependent piece of $S^{i=1}_{A_{j=1,2}}$.
In this case, after the analytic continuation in (\ref{eq:ac}), the reduced density matrix, {$\rho_{i=1;A_j}$}, is independent of time, so that $S^{i=1}_{A_j}$ are independent of time.
Then, the theory-dependent pieces, $S^{i=1;\text{T.D.}(n)}_{\mathcal{V}=A_j,B}$, are determined by
\be
\begin{split}
    S^{i=1;\text{T.D.}}_{\mathcal{V}=A_j} =\f{c}{3}\log{\left(\f{2\epsilon}{\pi}\right)}+\text{Min}\left[ S^{i=1;\text{T.D.}}_{\mathcal{V}=A_j;1}, S^{i=1;\text{T.D.}}_{\mathcal{V}=A_j;2}, S^{i=1;\text{T.D.}}_{\mathcal{V}=A_j;3}\right],
\end{split}
\ee
where the candidates of geodesics, $S^{i=1;\text{T.D.}}_{\mathcal{V}=A_j;I=1,2,3}$, are given by
\be
\begin{split}
    S^{i=1;\text{T.D.}}_{\mathcal{V}=A_j;I=1}&=\frac{c}{6}\log\left(\left|\sin\left[\frac{\pi}{2\epsilon}\left(\xi_{X_1}-\xi_{X_2}\right)\right]\right|^2\right)\\
    &={\frac{c}{6}\log{\left[\sinh\left[\frac{L_{\text{eff}}}{2q\epsilon}\left(\varphi_{X_1}-\varphi_{X_2}\right)\right]\sinh\left[\frac{L_{\text{eff}}}{2q\epsilon}\left(\overline{\varphi}_{X_1}-\overline{\varphi}_{X_2}\right)\right]\right]}},\\
    S^{i=1;\text{T.D.}}_{\mathcal{V}=A_j;I=2}&=\frac{c}{6}\log\left(\left|\sin\left[\frac{\pi}{2\epsilon}\left[\left(\xi_{X_1}-\xi_{X_2}\right)-iL_{\text{eff}}\right]\right]\right|^2\right)+S_{\text{B.H.}}\\
    &={\frac{c}{6}\log{\left[\sinh\left[\frac{L_{\text{eff}}}{2q\epsilon}\left(\varphi_{X_1}-\varphi_{X_2}-q\pi\right)\right]\sinh\left[\frac{L_{\text{eff}}}{2q\epsilon}\left(\overline{\varphi}_{X_1}-\overline{\varphi}_{X_2}+q\pi\right)\right]\right]} } +S_{\text{B.H.}},\\
    S^{i=1;\text{T.D.}}_{\mathcal{V}=A_j;I=3}&=\frac{c}{6}\log\left(\left|\sin\left[\frac{\pi}{2\epsilon}\left[\left(\xi_{X_1}-\xi_{X_2}\right)+iL_{\text{eff}}\right]\right]\right|^2\right)+S_{\text{B.H.}}\\
    &={\frac{c}{6}\log{\left[\sinh\left[\frac{L_{\text{eff}}}{2q\epsilon}\left(\varphi_{X_1}-\varphi_{X_2}+q\pi\right)\right]\sinh\left[\frac{L_{\text{eff}}}{2q\epsilon}\left(\overline{\varphi}_{X_1}-\overline{\varphi}_{X_2}-q\pi\right)\right]\right]} }+S_{\text{B.H.}},\\
\end{split}
\ee
where $S_{\text{B.H.}}$ is defined as the black hole entropy
\be
S_{\text{B.H.}}=\f{c\pi L_{\text{eff}}}{6\epsilon},
\ee
{and $\varphi_{X},\overline{\varphi}_{X}$ are defined by
\be\label{angular-coordinate}
\begin{split}
    &(\chi_X,\overline{\chi}_X)=( e^{\frac{2i}{q} \varphi_X},~e^{-\frac{2i}{q}  \varphi_X}),~ \cos{\varphi_X}=\f{e^{-\theta}\cos{\left(\f{q\pi X}{L}\right)}}{r_X}, ~\sin{\varphi_X}=\f{e^{\theta}\sin{\left(\f{q\pi X}{L}\right)}}{r_X},\\
    &r_X=\sqrt{\left(e^{-\theta}\cos{\left(\f{q\pi X}{L}\right)}\right)^2+\left(e^{\theta}\sin{\left(\f{q\pi X}{L}\right)}\right)^2},\\
    &\left(\xi_X, \overline{\xi}_X\right)=\left(\f{i L_{\text{eff}}\varphi_X}{q \pi },-\f{i L_{\text{eff}}\varphi_X}{q \pi }\right).
\end{split}
\ee
}We will explore $S^{i=1}_{A_{j=1,2}}$ in the high effective temperature regime with large $\theta$, where $\theta \gg 1$.
Furthermore, we will explore the behavior of $S^{i=1}_{A_{j=1,2}}$ in the high and low temperature limits. Here, the high and low temperature limits for $A_j$ are defined as 
\be \label{eq:high-low-temp-limits-single-interval}
\begin{split}
    &\text{High temperature limit:}~ \f{L\sin\left[\frac{q\pi (X_1-X_2)}{L}\right]}{4q\sin\left[\frac{q\pi X_1}{L}\right]\sin\left[\frac{q\pi X_2}{L}\right]}\gg \epsilon,\\
    &\text{Low temperature limit:}~ \f{L\sin\left[\frac{q\pi (X_1-X_2)}{L}\right]}{4q\sin\left[\frac{q\pi X_1}{L}\right]\sin\left[\frac{q\pi X_2}{L}\right]}\ll\epsilon.\\
\end{split}
\ee
As in \cite{2024arXiv240606121M}, after combining the theory-dependent and universal pieces, at the leading order {in the large $\theta$ limit}, $S^{i=1}_{A_{j=1,2}}$ reduce to
\be\label{eq:single-interval-EE}
\begin{split}
    S_{A_1}^{i=1} &\approx \f{c}{6}\log{\left[4\prod_{K=1,2}\sin^2{\left(\f{q\pi X_K}{L}\right)}\right]}+\f{c}{3}\log{\left(\f{2\epsilon}{\pi}\right)}+\f{c}{3}\log{\left[\sinh{\left[\f{L\sin{\left[\f{q\pi (X_1-X_2)}{L}\right]}}{4q\epsilon \sin{\left[\f{q\pi X_1}{L}\right]}\sin{\left[\f{q\pi X_2}{L}\right]}}\right]}\right]}\\
   &\approx \begin{cases}
       \f{c}{3}\log{\left[\f{L \sin{\left[\f{q\pi (X_1-X_2)}{L}\right]}}{q\pi }\right]}~&~\text{for low temperature limit}\\
       \f{c L}{12 q \epsilon}\cdot\left[\f{\sin{\left[\f{q\pi (X_1-X_2)}{L}\right]}}{\sin{\left[\f{q\pi X_1}{L}\right]}\sin{\left[\f{q\pi X_2}{L}\right]}}\right] ~&~\text{for high temperature limit},\\
   \end{cases}\\
    S_{A_2}^{i=1} &\approx   \f{c\pi l\cdot L}{12 q\epsilon}e^{2\theta}~~~~~\text{for both high and low temperature limits}.
\end{split}
\ee

We begin by $S_{A_1}^{i=1}$ in the low temperature limit of the high effective temperature region.
At the leading order of the low temperature limit in the low effective temperature region, the entanglement entropy for the subregion not including {fixed points} $x=\f{mL}{q}$ resembles the vacuum entanglement entropy on the circle with the circumstance of $L/q$. 
We will describe why the entanglement structure in this region resembles the vacuum one in the spatially-periodic system with the size of $L/q$.
As in \cite{2024arXiv240606121M}, except for $x=mL/q$, where $m$ is an integer and runs from $m=1$ to $m=q-1$, at the leading order in the large $\theta$ limit, the $H_{\text{q-M\"obius}}$ is approximately given by
\be \label{eq:approximation-of-H}
\begin{split}
    H_{\text{q-M\"obius}} 
    &\approx \sum^{q-1}_{m=0} { H_{\text{SSD},L/q,m} },
\end{split}
\ee
where $H_{\text{SSD},L/q,m}$ is the SSD Hamiltonian acting on the spatial interval between $x=m L/q$ and $(m+1)L/q$. The length of this interval is $L/q$. 
In the region considered here, the state could be approximated by the vacuum one for $H_{\text{q-M\"obius}}$, and the entanglement structure in the spatial interval between $x=m L/q$ and $(m+1)L/q$ is approximately given by the vacuum one for {$H_{\text{SSD},L/q,m}$}. As in \cite{PhysRevB.97.184309}, the vacuum state for the SSD Hamiltonian on the interval is the same as that for the uniform one with the periodic boundary condition. Therefore, in this region, $S_{A_1}^{i=1}$ is approximately given by the vacuum one for the uniform Hamiltonian on the spatially-periodic system with $L/q$.
Then, we closely look at $S_{A_1}^{i=1}$ in the high temperature limit of the high effective temperature region.
At the leading order in this limit, $S_{A_1}^{i=1}$ is proportional to $1/\epsilon$.
However, contrary to the uniform Hamiltonian \cite{2004JSMTE..06..002C,2009JPhA...42X4005C}, the entanglement entropy is not proportional to the subsystem size. 
The behavior of $S_{A_1}^{i=1}$ depends not only on the subsystem size, but also on the location of the boundaries of the subsystem.
The closer either $X_1$ or $X_2$ or both of them is to $mL/q$, the larger $S_{A_1}^{i=1}$ is.
Finally, we focus on $S_{A_2}^{i=1}$. 
The entanglement entropy is proportional to the number of {fixed} points where the curvature is negatively maximized, and exponentially grows with $\theta$. 


Now, we take the von Neumann limit, $n \rightarrow 1$, perform the analytic continuation in (\ref{eq:ac}), and then explore the time dependence of the analytic-continued entanglement entropy for $A_{j=1,2}\cup B_k$, $S^{i=1}_{A_{j=1,2}\cup B_k}$.
Since only theory-dependent pieces, $S^{i=1}_{A_{j=1,2}\cup B_k}$, depend on the time, we will closely look at the time dependence of $S^{i=1; \text{T.D.}}_{A_{j=1,2}\cup B_k}$.
The time dependence of $S^{i=1; \text{T.D.}}_{A_{j=1,2}\cup B_k}$ should be determined by minimizing all possible candidates satisfying homologous condition, i.e.,
\be
\begin{aligned}
    &S^{i=1; \text{T.D.}}_{A_{j=1,2}\cup B_k}=\f{2c}{3}\log\left(\f{2\epsilon}{\pi}\right)+\text{Min}\{S_{A_j\cup B_k}^{i=1;\text{dis.}},S_{A_j\cup B_k}^{i=1;\text{con.}}\},\\
    &S_{A_j\cup B_k}^{i=1;\text{dis.}}=S_{A_j}^{i=1;\text{T.D.}}+S_{B_k}^{i=1;\text{T.D.}}-\f{2c}{3}\log\left(\f{2\epsilon}{\pi}\right),\\
    &S_{A_j\cup B_k}^{i=1;\text{con.}}=\text{Min}\{S_{A_j\cup B_k}^{i=1;\text{con. (a)}},S_{A_j\cup B_k}^{i=1;\text{con. (b)}},S_{A_j\cup B_k}^{i=1;\text{con. (c)}}\}=\begin{cases}
        S_{A_{j=1,2}\cup B_k}^{i=1;\text{con. (a)}}~&\text{if}~|Y_K-X_K|<\f{L}{2}\\
        S_{A_{j=1,2}\cup B_k}^{i=1;\text{con. (b)}}~&\text{if}~|Y_K-X_K|>\f{L}{2}
    \end{cases},
\end{aligned}
\ee
where we assume that the spatial lengths of subsystems $A_{j=1,2}$, $B_k$ and $A_{j=1,2}\cup B_k$ are smaller than the half of total system size $L$, and $S_{A_j\cup B_k}^{i=1;\text{dis.}}$ is determined by the geodesic ending on the same Hilbert space, while $S_{A_j\cup B_k}^{i=1;\text{con.}}$ is determined by the one connecting the two spatial points of the two Hilbert spaces. 
Their time dependence is given by
\be\label{connected pieces}
\begin{aligned}
    S_{A_{j}\cup B_k}^{i=1;\text{con. (a)}}
    &=\frac{c}{6}\log\left[\f{1}{4}\prod_{K=1}^2\left(\cosh\left[\f{L_{\text{eff}}}{q\epsilon}\left(\varphi_{Y_K}-\varphi_{X_K}\right)\right]+\cosh\left[\f{\pi}{\epsilon}t\right]\right)\right],\\
    S_{A_{j}\cup B_k}^{i=1;\text{con. (b)}}
    &=\frac{c}{6}\sum_{K=1}^2\log\left(\f{1}{2}\left(\cosh\left[\f{L_{\text{eff}}}{q\epsilon}\left(q\pi-\left[\varphi_{Y_K}-\varphi_{X_K}\right] \right)\right]+\cosh\left[\f{\pi}{\epsilon}t\right]\right)\right),\\
    S_{A_{j}\cup B_k}^{i=1;\text{con. (c)}}
    &=\frac{c}{6}\sum_{K=1}^2\log\left(\f{1}{2}\left(\cosh\left[\f{L_{\text{eff}}}{q\epsilon}\left(q\pi+\left[\varphi_{Y_K}-\varphi_{X_K}\right] \right)\right]+\cosh\left[\f{\pi}{\epsilon}t\right]\right)\right),
\end{aligned}
\ee
where we have used $\xi_{X_K}+\bar{\xi}_{X_K}=0$, and all other possibilities are forbidden due to the homologous condition. 
For both $S_{A_{j}\cup B_k}^{i=1;\text{con. (a)}}$ and $S_{A_{j}\cup B_k}^{i=1;\text{con. (b)}}$, they take their minimal values at $t=0$ slice, specifically,
\be\label{minimal-value of connected pieces}
\begin{aligned}
    S_{A_{j}\cup B_k}^{i=1;\text{con. (a)}}(t=0)&\approx \f{c}{6}\sum_{K=1}^{2}\f{L_{\text{eff}}}{q\epsilon}\left(\varphi_{Y_K}-\varphi_{X_K}\right),\\
    S_{A_{j}\cup B_k}^{i=1;\text{con. (b)}}(t=0)&\approx \f{c}{6}\sum_{K=1}^{2}\f{L_{\text{eff}}}{q\epsilon}\left(q\pi-\left[\varphi_{Y_K}-\varphi_{X_K}\right]\right),\\
    S_{A_{j}\cup B_k}^{i=1;\text{con. (c)}}(t=0)&\approx \f{c}{6}\sum_{K=1}^{2}\f{L_{\text{eff}}}{q\epsilon}\left(q\pi+\left[\varphi_{Y_K}-\varphi_{X_K}\right]\right),
\end{aligned}
\ee
where we assume that $L_{\text{eff}}/q\epsilon \gg 1$. Recall that the size of $A_j,~B_k$ and $A_j\cup B_k$ is smaller than the half size of the total system, then we have $\varphi_{Y_K}-\varphi_{X_K}<\f{q\pi}{2}$ such that $S_{A_{j}\cup B_k}^{i=1;\text{con.}}$ is determined by $S_{A_{j}\cup B_k}^{i=1;\text{con. (a)}}$. 

Besides, the generic expression for disconnected term is
\be\label{disconnected-piece}
\begin{aligned}
    S_{A_j\cup B_k}^{i=1;\text{dis.}}&=\f{c}{6}\log\left(\left|\sinh\left[\f{L_{\text{eff}}}{2q\epsilon}\left(\varphi_{X_1}-\varphi_{X_2}\right)\right]\right|^2\cdot\left|\sinh\left[\f{L_{\text{eff}}}{2q\epsilon}\left(\varphi_{Y_1}-\varphi_{Y_2}\right)\right]\right|^2\right)\\
    &=\f{c}{6}\log\left[\f{1}{4}\left(\cosh\left[\f{L_{\text{eff}}}{q\epsilon}\left(\varphi_{X_1}-\varphi_{X_2}\right)\right]-1\right)\cdot\left(\cosh\left[\f{L_{\text{eff}}}{q\epsilon}\left(\varphi_{Y_1}-\varphi_{Y_2}\right)\right]-1\right)\right].
\end{aligned}
\ee
Subsequently, we compare $S_{A_j\cup B_k}^{i=1;\text{dis.}}$ with $S_{A_j\cup B_k}^{i=1;\text{con}}$, and define the exchanging time, $t=t^*$, as the time for them to exchange with the dominance.
In other words,  $t^*$ can be analytically determined by $S_{A_j\cup B_k}^{i=1;\text{dis.}}=S_{A_j\cup B_k}^{i=1;\text{con.}}$,
\be\label{exchanging-time}
\begin{aligned}
    \prod_{K=1}^2&\left(\cosh\left[\f{L_{\text{eff}}}{q\epsilon}\left(\varphi_{Y_K}-\varphi_{X_K}\right)\right]+\cosh\left[\f{\pi}{\epsilon}t^*\right]\right)=\\
    &\left(\cosh\left[\f{L_{\text{eff}}}{q\epsilon}\left(\varphi_{X_1}-\varphi_{X_2}\right)\right]-1\right)\cdot\left(\cosh\left[\f{L_{\text{eff}}}{q\epsilon}\left(\varphi_{Y_1}-\varphi_{Y_2}\right)\right]-1\right),
\end{aligned} 
\ee
where we have assumed $S_{A_j\cup B_k}^{i=1;\text{con.}}=S_{A_j\cup B_k}^{i=1;\text{con. (a)}}$. 
Then, $t^*$ is analytically given by
\be\label{exact-exchanging time}
\begin{aligned}
    &\cosh\left[\f{\pi}{\epsilon}t^*\right]=\f{-b+\sqrt{b^2-4ac}}{2a},~\text{where}\\
    &a=1,~b=\sum_{K=1}^2\cosh\left[\f{L_{\text{eff}}}{q\epsilon}\left(\varphi_{Y_K}-\varphi_{X_K}\right)\right],\\
    &c=\prod_{K=1}^2\cosh\left[\f{L_{\text{eff}}}{q\epsilon}\left(\varphi_{Y_K}-\varphi_{X_K}\right)\right]-\prod_{Z=X,Y}\cosh\left[\f{L_{\text{eff}}}{q\epsilon}\left(\varphi_{Z_1}-\varphi_{Z_2}\right)\right]\\
    &+\sum_{Z=X,Y}\cosh\left[\f{L_{\text{eff}}}{q\epsilon}\left(\varphi_{Z_1}-\varphi_{Z_2}\right)\right]-1,
\end{aligned}
\ee
where the parameters, $a,b,$ and $c$, satisfy $b^2-4ac> 0$ (otherwise there is no such an exchanging time). Then, the time dependence of the entanglement entropy is given by
\be\label{EE-with-exchange of dominance}
\begin{aligned}
    S^{i=1}_{A_j\cup B_k}=S_{A_j\cup B_k}^{i=1;\text{U.}}+\f{2c}{3}\log\left(\f{2\epsilon}{\pi}\right)+\begin{cases}
       S_{A_j\cup B_k}^{i=1;\text{con.}}~&\text{for}~0<t\leq t^*,\\
       S_{A_j\cup B_k}^{i=1;\text{dis.}}~&\text{for}~t> t^*.
    \end{cases}
\end{aligned}
\ee
Consequently, the time dependence of the mutual information results in
\be\label{MI-with-exchange of dominance}
\begin{aligned}
    I_{A_j, B_k}^{i=1}=\begin{cases}
       S_{A_j\cup B_k}^{i=1;\text{dis.}}-S_{A_j\cup B_k}^{i=1;\text{con.}}~&\text{for}~t\leq t^*,\\
       0~&\text{for}~t> t^*.
    \end{cases}
\end{aligned}
\ee
On the other hand, if $S_{A_j\cup B_k}^{i=1;\text{dis.}}\leq S_{A_j\cup B_k}^{i=1;\text{con.}}(t=0)$, then the time dependence of the entanglement entropy is determined by $S_{A_j\cup B_k}^{i=1;\text{dis.}}$, so that the entanglement entropy reduces to the time-{independent} one,
\be\label{EE-without-exchange of dominance}
    S^{i=1}_{A_j\cup B_k}=S_{A_j\cup B_k}^{i=1;\text{U.}}+\f{2c}{3}\log\left(\f{2\epsilon}{\pi}\right)+S_{A_j\cup B_k}^{i=1;\text{dis.}}~~~\text{for}~~~t\ge 0.
\ee
Consequently, the mutual information is time-independent and zero.

\if[0]
\footnote{\textcolor{blue}{CB: this paragraph may be deleted (I forgot to do it after adding line-tension section, sorry). The quasi-particle picture can only capture the exchanging of minimal surfaces but fails to approximate the exact time dependence. Line-tension picture works better for both!}}Note that \eqref{connected pieces} are monotonically increasing functions with respect to time, their minimal values are given by \eqref{minimal-value of connected pieces}, and \eqref{exchanging-time} is hard to solve directly. Since \eqref{connected pieces} are proportional to $\f{c}{3}\log\left(\cosh\left[\f{\pi t}{\epsilon}\right]\right)\approx \f{c\pi}{3\epsilon}t$, we will roughly estimate the connected pieces and exchanging time of dominance by applying quasi-particle picture, i.e., a linear growth function on $t$ with initial value same as $S_{A_j\cup B_k}^{\text{ con.}}(t=0)$,
\be\label{quasiparticle description of connected piece}
\begin{aligned}
    S_{A_j\cup B_k}^{\text{ con.}}&\sim \f{c\pi}{3\epsilon}t+\f{cL_{\text{eff}}}{6q\epsilon}\sum_{K=1}^2\varphi_{Y_K}-\varphi_{X_K},\\
    t^*_{A_j\cup B_k}&\sim S_{A_j\cup B_k}^{\text{ dis.}}-\f{cL_{\text{eff}}}{6q\epsilon}\sum_{K=1}^2\varphi_{Y_K}-\varphi_{X_K}-\f{c\pi}{3\epsilon}t,
\end{aligned}
\ee
here and afterwards we use the symbol ``$\sim$'' to represent estimated results by quasi-particle picture. To better capture the time dependence of entanglement entropy and mutual information, we will use such an approximation except when plotting specific examples. It is easy to see that the precise time dependence of connected pieces in holographic CFT is not described by quasi-particle picture \cite{Nie:2018dfe}. However, due to the monotonicity, we still can use quasi-particle picture to estimate the dominance of different surfaces during the evolution qualitatively. To explain the deviation from quasi-particle picture for holographic CFTs, a toy model was also proposed in \cite{Nie:2018dfe}.
\fi
In the following, we explore the time dependence of the mutual information in the several cases considered in this paper. 
\subsubsection{Symmetric cases}
We begin by considering the time dependence of the mutual information for the symmetric case, where the spacial location of $A$ and $B$ is the same, i.e. $X_K=Y_K$ for $K=1,2$ (and $n=m,p=l$).
In this case, during the early time evolution, $t^{*}>t>0$, the minimal value of connected pieces for $A$ and $B$ is less than the minimal one of disconnected pieces, so that $S^{i=1; \text{T.D.}}_{A_{j=1,2}\cup B_{j=1,2}} ={S^{i=1;\text{con.}}_{A_{j=1,2}\cup B_{j=1,2}}}$.
During the late time evolution, $t>t^{*}$, the time dependence of $S^{i=1; \text{T.D.}}_{A_{j=1,2}\cup B_{j=1,2}}$ is determined by that of $S^{i=1;\text{dis.}}_{A_{j=1,2}\cup B_{j=1,2}}$.
Therefore, the time dependence of the entanglement entropy and the mutual information is determined by \eqref{EE-with-exchange of dominance} and \eqref{MI-with-exchange of dominance}.
Furthermore, the exchanging time $t^*$ is exactly determined by \eqref{exact-exchanging time}, where
\be\label{eq:TFD-two-interval-symmetric-EE-candidates}
    \begin{aligned}
        &S_{A_{j}\cup B_{j}}^{i=1;\text{dis.}}\approx\begin{cases}
            \f{2c}{3}\log\left(\left|\sinh\left[\f{L}{4q\epsilon}\left(\f{\sin\left[\frac{q\pi (X_1-X_2)}{L}\right]}{\sin\left[\frac{q\pi X_1}{L}\right]\sin\left[\frac{q\pi X_2}{L}\right]}\right)\right]\right|\right)~&\text{for}~j=1\\
            \f{c\pi l\cdot L}{6q\epsilon}e^{2\theta}~&\text{for}~j=2,
        \end{cases}\\
        &S_{A_j\cup B_j}^{i=1;\text{con.}}=S_{A_j\cup B_j}^{\text{con. (a)}}=\frac{2c}{3}\log\left(\cosh\left[\f{\pi t}{2\epsilon}\right]\right)\overset{t\gg \epsilon}{\approx}\f{c\pi }{3\epsilon}t.
    \end{aligned}
\ee
Specifically, for $j=1$, at leading-order in the high and low temperature limits, the disconnected piece can be approximated as
\be
\begin{aligned}
    S_{A_1\cup B_1}^{i=1;\text{dis.}}\approx&\f{2c}{3}\log\left(\left|\sinh\left[\f{L}{4q\epsilon}\left(\f{\sin\left[\frac{q\pi (X_1-X_2)}{L}\right]}{\sin\left[\frac{q\pi X_1}{L}\right]\sin\left[\frac{q\pi X_2}{L}\right]}\right)\right]\right|\right)\\
    \approx&\f{2c}{3}\cdot\begin{cases}
        \f{L\sin\left[\frac{q\pi (Z_1-Z_2)}{L}\right]}{4q\epsilon\sin\left[\frac{q\pi Z_1}{L}\right]\sin\left[\frac{q\pi Z_2}{L}\right]}~&\text{for}~\text{high temperature},\\
        \log\left[\f{L}{4q\epsilon}\left(\f{\sin\left[\frac{q\pi (X_1-X_2)}{L}\right]}{\sin\left[\frac{q\pi X_1}{L}\right]\sin\left[\frac{q\pi X_2}{L}\right]}\right)\right]<0~&\text{for}~\text{low temperature},
    \end{cases}
\end{aligned}
\ee
where the low and high temperature limits for state $\ket{U_{\text{eff}}}$ are defined as 
\be \label{eq:high-low-temp-limits}
\begin{split}
    &\text{High temperature limit:}~ \f{L\sin\left[\frac{q\pi (Z_1-Z_2)}{L}\right]}{4q\sin\left[\frac{q\pi Z_1}{L}\right]\sin\left[\frac{q\pi Z_2}{L}\right]}\gg \epsilon,~\text{for }Z=X,Y,\\
    &\text{Low temperature limit:}~ \f{L\sin\left[\frac{q\pi (Z_1-Z_2)}{L}\right]}{4q\sin\left[\frac{q\pi Z_1}{L}\right]\sin\left[\frac{q\pi Z_2}{L}\right]}\ll\epsilon,~\text{for }Z=X,Y.\\
\end{split}
\ee
For $j=2$, it can be approximated by
\be
    S_{A_2\cup B_2}^{i=1;\text{dis.}}\approx\f{c\pi l\cdot L}{6q\epsilon}e^{2\theta},
\ee
which does not depend on different temperature limits within the effective high temperature regime. 
In the symmetric case, if we consider $j=2$ or $j=1$ together with the high temperature limit, during the early-time evolution, $t^*>t>0$, the time dependence of $S^{i=1}_{A_{j=1,2}\cup B_{j=1,2}}$ is determined by $S^{i=1;\text{con.}}_{A_{j=1,2}\cup B_{j=1,2}}$, and it linearly grows with time.
Subsequently, during the late-time evolution, $t>t^*$, the time dependence of $S^{i=1}_{A_{j=1,2}\cup B_{j=1,2}}$ is determined by $S_{A_{j=1,2}\cup B_{j=1,2}}^{i=1;\text{dis.}}$. 
Since $S_{A_{j=1,2}\cup B_{j=1,2}}^{i=1;\text{dis.}}$ is independent of time, $S^{i=1}_{A_{j=1,2}\cup B_{j=1,2}}$ for $t>t^*$ is constant, where in large $\theta$ limit $t^*$ is evaluated by \eqref{exact-exchanging time} as
\be\label{eq:TFD-two-interval-exchanging-time}
    t^*\approx\begin{cases}
        \f{L\sin\left[\frac{q\pi (X_1-X_2)}{L}\right]}{2q\pi\sin\left[\frac{q\pi X_1}{L}\right]\sin\left[\frac{q\pi X_2}{L}\right]}~&\text{for}~j=1\text{ and high temperature limit}\\
        \f{l \cdot L}{2q}e^{2\theta}~&\text{for}~j=2.
    \end{cases}
\ee
Consequently, the mutual information monotonically decreases with $t$ when $0<t<t^*$, and then after $t = t^*$,  it vanishes. On the other hand, when we consider $j=1$ together with the low temperature limit, the disconnected piece is negative-valued and strictly smaller than the connected one. The time dependence disappears and entanglement entropy remains unchanged during the evolution. As a result, the mutual information between $A_j$ and $B_j$ is zero.

\subsubsection{Asymmetric cases \label{sec:asymmetric-cases}}
Now, we consider the time dependence of the entanglement entropy and mutual information in the following two cases. For simplicity, we assume that $n=m$ and $p\geq l$. In the first case, and the spatial locations of endpoints satisfy
\be \label{eq:first-case}
    0<Y_2\leq X_2<X_1<Y_1<\f{L}{2},~X_2,Y_2\in(X_m^f,X_{m+1}^f),
\ee
where we assume that the sizes of all subsystems are strictly smaller than half of the whole system, for simplicity. In the second case, we assume that the endpoint locations of subsystems satisfy
\be \label{eq:second-case}
    0<X_2<Y_2<X_1<Y_1<\f{L}{2},~X_2,Y_2\in(X_m^f,X_{m+1}^f).
\ee
We call them the asymmetric cases.


We closely look at the theory-dependent pieces of $S^{i=1}_{A_{j=1,2}\cup B_{k=1,2}}$.
We begin by considering the disconnected pieces.
According to \eqref{disconnected-piece}, in the first case (\ref{eq:first-case}), their disconnected pieces are given by
\be\label{Asymmetric-disconnected-pieces}
\begin{aligned}
    S_{A_j\cup B_k}^{i=1;\text{dis.}}\approx\begin{cases}
        \f{c}{3}\sum_{Z=X,Y}\log\left(\left|\sinh\left[\f{L}{4q\epsilon}\left(\f{\sin\left[\frac{q\pi (Z_1-Z_2)}{L}\right]}{\sin\left[\frac{q\pi Z_1}{L}\right]\sin\left[\frac{q\pi Z_2}{L}\right]}\right)\right]\right|\right),~&\text{for}~j=k=1\\
        \f{c\pi (p+l)\cdot L}{12q\epsilon} e^{2\theta},~&\text{for}~j=k=2\\
        \f{c}{3}\log\left(\left|\sinh\left[\f{L}{4q\epsilon}\left(\f{\sin\left[\frac{q\pi (X_1-X_2)}{L}\right]}{\sin\left[\frac{q\pi X_1}{L}\right]\sin\left[\frac{q\pi X_2}{L}\right]}\right)\right]\right|\right)+\f{c\pi p\cdot L}{12q\epsilon} e^{2\theta},~&\text{for}~j=1,~k=2
    \end{cases}.
\end{aligned}
\ee
 Furthermore, one can approximate \eqref{Asymmetric-disconnected-pieces} in the various temperature limits as
\be
    \begin{aligned}
        &S_{A_1\cup B_1}^{i=1;\text{dis.}}\approx\begin{cases}
            \f{cL}{12q\epsilon}\sum_{Z=X,Y}\f{\sin\left[\frac{q\pi (Z_1-Z_2)}{L}\right]}{\sin\left[\frac{q\pi Z_1}{L}\right]\sin\left[\frac{q\pi Z_2}{L}\right]}~&\text{for high temperature limit}\\
            \f{c}{3}\sum_{Z=X,Y}\log\left(\f{L}{4q\epsilon}\left[\f{\sin\left[\frac{q\pi (Z_1-Z_2)}{L}\right]}{\sin\left[\frac{q\pi Z_1}{L}\right]\sin\left[\frac{q\pi Z_2}{L}\right]}\right]\right)<0~&\text{for low temperature limit},
        \end{cases}\\
        &S_{A_2\cup B_2}^{i=1;\text{dis.}}\approx\f{c\pi (p+l)\cdot L}{12q\epsilon} e^{2\theta} ~~~~~\text{for both low and high temperature limits,}\\
        &S_{A_1\cup B_2}^{i=1;\text{dis.}}\approx\begin{cases}
            \f{cL}{12q\epsilon}\f{\sin\left[\frac{q\pi (X_1-X_2)}{L}\right]}{\sin\left[\frac{q\pi X_1}{L}\right]\sin\left[\frac{q\pi X_2}{L}\right]}+\f{c\pi p\cdot L}{12q\epsilon} e^{2\theta}~&\text{for high temperature limit}\\
            \f{c}{3}\log\left(\f{L}{4q\epsilon}\left[\f{\sin\left[\frac{q\pi (X_1-X_2)}{L}\right]}{\sin\left[\frac{q\pi X_1}{L}\right]\sin\left[\frac{q\pi X_2}{L}\right]}\right]\right)+\f{c\pi p\cdot L}{12q\epsilon} e^{2\theta}~&\text{for low temperature limit}.
        \end{cases}
    \end{aligned}
\ee

Then, we move to the time dependence of the connected pieces, which depends on the cases considered here.
First, we consider the first case in (\ref{eq:first-case}).
In this case, the connected pieces result in
    \be
        \begin{aligned}
            S_{A_j\cup B_k}^{i=1;\text{con.}}=\begin{cases}
                \frac{c}{6}\sum_{K=1}^2\log\left(\f{1}{2}\left(\cosh\left[\f{L}{2q\epsilon}\f{\sin\left[\frac{q\pi (Y_K-X_K)}{L}\right]}{\sin\left[\frac{q\pi X_K}{L}\right]\sin\left[\frac{q\pi Y_K}{L}\right]}\right]+\cosh\left[\f{\pi}{\epsilon}t\right]\right)\right)~&\text{for}~j=k=1\\
                S_{A_2\cup B_2}^{i=1;\text{con.}}~&\text{for}~j=k=2\\
                S_{A_1\cup B_2}^{i=1;\text{con.}}~&\text{for}~j=1,~k=2,
            \end{cases}
        \end{aligned}
    \ee
    where
    \be
        \begin{aligned}
            &S_{A_2\cup B_2}^{i=1;\text{con.}}\approx\begin{cases}
                &\frac{c}{6}\sum_{K=1}^2\log \left( {\f{1}{2}} \left(\cosh\left[\frac{L}{2q\epsilon}\left(\f{\sin\left[\frac{q\pi (Y_K-X_K)}{L}\right]}{\sin\left[\frac{q\pi Y_K}{L}\right]\sin\left[\frac{q\pi X_K}{L}\right]}\right)\right]+\cosh\left[\f{\pi}{\epsilon}t\right]\right) \right)~~~~~\text{for}~p=l,~n=m\\
                &\frac{c}{6}\log\left( {\f{1}{2}}\left(\cosh\left[\f{\pi(p-l)\cdot L}{2q\epsilon}e^{2\theta}\right]+\cosh\left[\f{\pi}{\epsilon}t\right]\right)\right)+\\
                &~~~\f{c}{6}\log \left( {\f{1}{2}} \left(\cosh\left[\frac{L}{2q\epsilon}\left(\f{\sin\left[\frac{q\pi (Y_2-X_2)}{L}\right]}{\sin\left[\frac{q\pi Y_2}{L}\right]\sin\left[\frac{q\pi X_2}{L}\right]}\right)\right]+\cosh\left[\f{\pi}{\epsilon}t\right]\right)\right)~~~~~\text{for}~p>l,~n=m,
            \end{cases}\\
            &S_{A_1\cup B_2}^{i=1;\text{con.}}\approx\f{c}{6}\log\left( {\f{1}{2}}\left(\cosh\left[\f{\pi p\cdot L}{2q\epsilon}e^{2\theta}\right]+\cosh\left[\f{\pi}{\epsilon}t\right]\right)\right)+\\
            &~~~~~~~\f{c}{6}\log\left( {\f{1}{2}}\left(\cosh\left[\f{L}{2q\epsilon}\left(\f{\sin\left[\frac{q\pi (Y_2-X_2)}{L}\right]}{\sin\left[\frac{q\pi Y_2}{L}\right]\sin\left[\frac{q\pi X_2}{L}\right]}\right)\right]+\cosh\left[\f{\pi}{\epsilon}t\right]\right)\right)~\text{for}~n=m,
        \end{aligned}
    \ee
   which are identical for both of the two asymmetric cases.
   
   Furthermore, in the high temperature limit defined in (\ref{eq:high-low-temp-limits}), the initial (and also minimal due to their monotonically increasing properties over time) values of connected pieces are approximately given by
    \be
        \begin{aligned}
            &S_{A_1\cup B_1,\text{ min.}}^{i=1;\text{con.}}=S_{A_1\cup B_1}^{i=1;\text{con.}}(t=0)\approx \f{cL}{12q\epsilon}\left(\f{\left|\sin\left[\frac{q\pi (Y_1-X_1)}{L}\right]\right|}{\sin\left[\frac{q\pi Y_1}{L}\right]\sin\left[\frac{q\pi X_1}{L}\right]}+\f{\left|\sin\left[\frac{q\pi (X_2-Y_2)}{L}\right]\right|}{\sin\left[\frac{q\pi Y_2}{L}\right]\sin\left[\frac{q\pi X_2}{L}\right]}\right),\\
            &S_{A_2\cup B_2,\text{ min.}}^{i=1;\text{con.}}=S_{A_2\cup B_2}^{i=1;\text{con.}}(t=0)\approx\begin{cases}
                \f{cL}{12q\epsilon}\sum_{K=1}^2\f{\left|\sin\left[\frac{q\pi (Y_K-X_K)}{L}\right]\right|}{\sin\left[\frac{q\pi Y_K}{L}\right]\sin\left[\frac{q\pi X_K}{L}\right]}~&\text{for}~p=l,~n=m\\
                \f{cL}{12q\epsilon}\f{\left|\sin\left[\frac{q\pi (Y_2-X_2)}{L}\right]\right|}{\sin\left[\frac{q\pi Y_2}{L}\right]\sin\left[\frac{q\pi X_2}{L}\right]}+\f{c\pi (p-l)\cdot L}{12q\epsilon} e^{2\theta}~&\text{for}~p>l,~n=m,            
            \end{cases}\\
            &S_{A_1\cup B_2,\text{ min.}}^{i=1;\text{con.}}=S_{A_1\cup B_2}^{i=1;\text{con.}}(t=0)\approx\f{cL}{12q\epsilon}\f{\left|\sin\left[\frac{q\pi (Y_2-X_2)}{L}\right]\right|}{\sin\left[\frac{q\pi Y_2}{L}\right]\sin\left[\frac{q\pi X_2}{L}\right]}+\f{c\pi p\cdot L}{12q\epsilon} e^{2\theta}~~~\text{for}~n=m.
        \end{aligned}
    \ee
    Upon the case-by-case comparison with \eqref{Asymmetric-disconnected-pieces}, we discover that the initial values of connected pieces are smaller than the values of disconnected pieces, i.e. $S_{A_{j=1,2}\cup B_{k=1,2}}^{i=1;\text{con.}}(t=0)<S_{A_{j=1,2}\cup B_{k=1,2}}^{i=1;\text{dis.}}$, and hence the connected pieces contribute first. Then, due to the monotonically increasing nature, the connected pieces start to grow with time $t$ until $t=t^*$ when they are equal to the values of disconnected pieces. After that, the entanglement entropy saturates and the disconnected pieces start to dominate. Therefore, for all cases with asymmetric settings, their entanglement entropy and mutual information are given by \eqref{EE-with-exchange of dominance} and \eqref{MI-with-exchange of dominance},
    \be\label{eq:EE-MI-asymmetric-cases}
    \begin{aligned}
    S^{i=1}_{A_j\cup B_k}&=S_{A_j\cup B_k}^{i=1;\text{U.}}+\f{2c}{3}\log\left(\f{2\epsilon}{\pi}\right)+\begin{cases}
       S_{A_j\cup B_k}^{i=1;\text{con.}}~&\text{for}~0<t\leq t^*\\
       S_{A_j\cup B_k}^{i=1;\text{dis.}}~&\text{for}~t> t^*,
    \end{cases}~~\text{and}\\
    I_{A_j, B_k}^{i=1}&=\begin{cases}
       S_{A_j\cup B_k}^{i=1;\text{dis.}}-S_{A_j\cup B_k}^{i=1;\text{con.}}~&\text{for}~0<t\leq t^*\\
       0~&\text{for}~t> t^*.
    \end{cases}
    \end{aligned}
    \ee
    The exchanging time $t^*$ is solved by using \eqref{exact-exchanging time}, that is
    \be\label{eq:exchanging-time-asymmetric-cases}
        \begin{split}
            t^*=& \f{\epsilon}{\pi}\cosh^{-1}\left(\f{-b+\sqrt{b^2-4c}}{2}\right),~~b=\sum_{K=1}^2\cosh\left[\f{L_{\text{eff}}}{q\epsilon}(\varphi_{Y_K}-\varphi_{X_K})\right],\\
            c=&\prod_{K=1}^2\cosh\left[\f{L_{\text{eff}}}{q\epsilon}\left(\varphi_{Y_K}-\varphi_{X_K}\right)\right]-\prod_{Z=X,Y}\cosh\left[\f{L_{\text{eff}}}{q\epsilon}\left(\varphi_{Z_1}-\varphi_{Z_2}\right)\right]\\
            &+\sum_{Z=X,Y}\cosh\left[\f{L_{\text{eff}}}{q\epsilon}\left(\varphi_{Z_1}-\varphi_{Z_2}\right)\right]-1,
        \end{split}
    \ee
    where we have
    \be
    \begin{split}
        &\varphi_{Y_1}-\varphi_{X_1}\approx\begin{cases}
            \f{e^{-2\theta}\sin\left[\f{q\pi(Y_1-X_1)}{L}\right]}{\sin\left[\f{q\pi Y_1}{L}\right]\sin\left[\f{q\pi X_1}{L}\right]}~&\text{for}~A_1\cup B_1\\
            (p-l)\cdot\pi~&\text{for}~A_2\cup B_2\\
            p\cdot \pi~&\text{for}~A_1\cup B_2,
        \end{cases}~\varphi_{Y_2}-\varphi_{X_2}\approx\begin{cases}
            \f{e^{-2\theta}\sin\left[\f{q\pi(Y_2-X_2)}{L}\right]}{\sin\left[\f{q\pi Y_2}{L}\right]\sin\left[\f{q\pi X_2}{L}\right]}~&\text{for}~A_1\cup B_1\\
            \f{e^{-2\theta}\sin\left[\f{q\pi(Y_2-X_2)}{L}\right]}{\sin\left[\f{q\pi Y_2}{L}\right]\sin\left[\f{q\pi X_2}{L}\right]}~&\text{for}~A_2\cup B_2\\
            \f{e^{-2\theta}\sin\left[\f{q\pi(Y_2-X_2)}{L}\right]}{\sin\left[\f{q\pi Y_2}{L}\right]\sin\left[\f{q\pi X_2}{L}\right]}~&\text{for}~A_1\cup B_2,
        \end{cases}\\
        &\varphi_{X_1}-\varphi_{X_2}\approx\begin{cases}
            \f{e^{-2\theta}\sin\left[\f{q\pi(X_1-X_2)}{L}\right]}{\sin\left[\f{q\pi X_1}{L}\right]\sin\left[\f{q\pi X_2}{L}\right]}~&\text{for}~A_1\cup B_1\\
            l\cdot\pi~&\text{for}~A_2\cup B_2\\
            \f{e^{-2\theta}\sin\left[\f{q\pi(X_1-X_2)}{L}\right]}{\sin\left[\f{q\pi X_1}{L}\right]\sin\left[\f{q\pi X_2}{L}\right]}~&\text{for}~A_1\cup B_2,
        \end{cases}~\varphi_{Y_1}-\varphi_{Y_2}\approx\begin{cases}
            \f{e^{-2\theta}\sin\left[\f{q\pi(Y_1-Y_2)}{L}\right]}{\sin\left[\f{q\pi Y_1}{L}\right]\sin\left[\f{q\pi Y_2}{L}\right]}~&\text{for}~A_1\cup B_1\\
            p\cdot\pi~&\text{for}~A_2\cup B_2\\
            p\cdot\pi~&\text{for}~A_1\cup B_2,
        \end{cases}
    \end{split}
    \ee
    and note that $n= m$, $p>l$ have been considered.

\subsubsection{Disjoint cases \label{sec:disjoint-case}}
We will close this section by considering the disjoint cases, where $A$ and $B$ have no spatial overlaps, i.e., $X_2<X_1<Y_2<Y_1$.
In this case, the disconnected pieces of $S^{i=1}_{A_j\cup B_k}$ are same as \eqref{Asymmetric-disconnected-pieces}. 
To determine the time dependence of the entanglement entropy and mutual information, we need to consider the connected pieces as well. The time dependence of the connected pieces is given by
\be
    \begin{aligned}
        S_{A_j\cup B_k}^{i=1;\text{con.}}=\begin{cases}
            \frac{c}{6}\sum_{K=1}^2\log\left(\cosh\left[\f{L}{2q\epsilon}\left(\f{\sin\left[\frac{q\pi (Y_K-X_K)}{L}\right]}{\sin\left[\frac{q\pi X_K}{L}\right]\sin\left[\frac{q\pi Y_K}{L}\right]}\right)\right]+\cosh\left[\f{\pi}{\epsilon}t\right]\right)~&\text{for}~j=k=1\\
            S_{A_2\cup B_2}^{i=1;\text{con.}}~&\text{for}~j=k=2\\
            S_{A_1\cup B_2}^{i=1;\text{con.}}~&\text{for}~j=1,~k=2,
        \end{cases}
    \end{aligned}
\ee
where
\be
        \begin{aligned}
            &S_{A_2\cup B_2}^{i=1;\text{con.}}\approx\frac{c}{6}\log\left(\cosh\left[\f{\pi\left(n-m+p-l\right)\cdot L}{2q\epsilon}e^{2\theta}\right]+\cosh\left[\f{\pi}{\epsilon}t\right]\right)+\\
            &~~~\frac{c}{6}\log\left(\cosh\left[\f{\pi(n-m)\cdot L}{2q\epsilon}e^{2\theta}\right]+\cosh\left[\f{\pi}{\epsilon}t\right]\right)~~~\text{with}~n>m+l+1,\quad n+p+1<q,\\
            &S_{A_1\cup B_2}^{i=1;\text{con.}}\approx\f{c}{6}\times\begin{cases}
            &\log\left(\cosh\left[\f{\pi p\cdot L}{2q\epsilon}e^{2\theta}\right]+\cosh\left[\f{\pi}{\epsilon}t\right]\right)+\\
            &~~~\log\left(\cosh\left[\f{L}{2q\epsilon}\left(\f{\sin\left[\frac{q\pi (Y_2-X_2)}{L}\right]}{\sin\left[\frac{q\pi Y_2}{L}\right]\sin\left[\frac{q\pi X_2}{L}\right]}\right)\right]+\cosh\left[\f{\pi}{\epsilon}t\right]\right)~~~~~\text{for}~m=n,\\
            &\log\left(\cosh\left[\f{\pi(n-m+p)\cdot L}{2q\epsilon}e^{2\theta}\right]+\cosh\left[\f{\pi}{\epsilon}t\right]\right)+\\
            &~~~\log\left(\cosh\left[\f{\pi(n-m)\cdot L}{2q\epsilon}e^{2\theta}\right]+\cosh\left[\f{\pi}{\epsilon}t\right]\right)~~~~~\text{for}~m+1<n.
    \end{cases}
    \end{aligned}
    \ee
In this case, the connected piece monotonically grows with $t$. At $t=0$, the connected piece is larger than the disconnected one. In other words, during the time evolution, the behavior of the theory-dependent piece of $S^{i=1}_{A_j\cup B_k}$ is determined by the disconnected one. This disconnected piece is the same as the sum of the theory-dependent pieces of $S^{i=1}_{A_j}$ and $S^{i=1}_{B_k}$. Consequently, the mutual information is independent of time, and zero.

\section{Entanglement production on curved spacetimes \label{sec:entropy-production}}
In this section, we will consider the inhomogeneous version of the global quench considered in \cite{2005JSMTE..04..010C}.
In other words, the system is in (\ref{eq:Bdy-state}).

\subsection{Entanglement entropy of $\ket{\Psi(t)}$ for the single interval\label{Sec:EE-BS-for-singleinterval}}

Now, we closely look at {$S^{i=2;(n)}_{A_j; \text{E}}$} in the path integral formalism.
In this formalism, {$S^{i=2;(n)}_{A_j; \text{E}}$} is given by (\ref{eq:EE-single-interval}).
By performing the transformation in (\ref{eq:transformation-lop}) and then taking the von Neumann limit, where $n\rightarrow 1$, {$S^{i=2;(n)}_{A_j; \text{E}}$} results in
\be \label{eq:Renyi-BS-single}
\begin{split}
&S^{i=2}_{A_j; \text{E}}={\lim_{n\to 1}}\f{1}{1-n}\log{\left[\f{\bra{B}e^{-\tau_b H_{\text{q-M\"obius}}}\sigma_n(w_{X_1},\overline{w}_{X_1})\overline{\sigma}_n(w_{X_2},\overline{w}_{X_2})e^{-\tau_a H_{\text{q-M\"obius}}}\ket{B}}{\bra{B}e^{-2\epsilon H_{\text{q-M\"obius}}}\ket{B}}\right]}\\
&=\f{c}{12}\log{\left[\prod_{K=1,2}f^2(X_K)\right]}\\
&\quad +{\lim_{n\to 1}}\f{1}{1-n}\log{\left[\f{\bra{\text{B}} {e^{-2\epsilon H_{\text{q-M\"obius}}}}  \sigma_n(\xi_{X_1}+\tau_a,\overline{\xi}_{X_1}+\tau_a)\overline{\sigma}_n(\xi_{X_2}+\tau_a,\overline{\xi}_{X_2}+\tau_a)\ket{\text{B}}}{\bra{B}e^{-2\epsilon H_{\text{q-M\"obius}}}\ket{B}}\right]},
\end{split}
\ee
where we call the first and second terms in the last line  $S^{i=2;\text{U.}}_{A_j; \text{E}}$ and  $S^{i=2;\text{T.D.}}_{A_j; \text{E}}$, respectively.
The universal piece, $S^{i=2;\text{U.}}_{A_j; \text{E}}$, does not depend on the details of CFTs, while the theory-dependent piece, $S^{i=2;\text{T.D.}}_{A_j; \text{E}}$, does, and it is determined by the two-point function on the cylinder with the width of $2\epsilon$ and the circumstance of $L_{\text{eff}}$. 
\subsubsection{The universal piece\label{Sec:UP}}
We begin by considering the behavior of the universal piece in the large $\theta$ limit.
At the leading order in this limit, $S^{i=2;\text{U.}}_{A_j}=S^{i=2;\text{U.}}_{A_j;\text{E}}$ reduces to 
\be
\begin{split}
    S^{i=2;\text{U.}}_{A_j}&=\f{c}{6}\log{\left[\prod_{K=1,2}\left(1-\cos{\left(\f{2q\pi X_K}{L}\right)}\tanh{(2\theta)}\right)\right]}\\
    &\underset{\theta \gg 1}{\approx}\f{c}{6}\log{\left[4\prod_{K=1,2}\sin^2{\left(\f{q\pi X_K}{L}\right)}\right]}.
\end{split}
\ee

\subsubsection{2d holographic CFTs\label{Sec:EE-for-SI-BS-in-2d-hCFTs}}
Here, we will report the time dependence of $S^{i=2}_{A_j}$ in $2$d holographic CFTs after performing analytic continuation (\ref{eq:ac}). 
We can compute the theory-dependent piece in AdS/BCFT context \cite{2011PhRvL.107j1602T,2011JHEP...11..043F,2012JHEP...06..066N}. 
However, in this paper, we employ the method of image that guarantees the conformal invariance of the partition function, ${\bra{B}e^{-2\epsilon H_{\text{q-M\"obius}}}\ket{B}}$. 
In the method of image, the two-point function on the finite cylinder is equivalent to the four-point function on the torus with the thermal circumstance of $4\epsilon$ and the spatial circumstance of $L_{\text{eff}}$. 
The four-point function is constructed of the twist and anti-twist operators and their copies.
The conformal symmetry determines the location of copies.
If the location of the original operator is $(\xi,\overline{\xi})$, the location of its copy is given by $(4\epsilon-\overline{\xi},4\epsilon-\xi)$.
Consequently, the theory-dependent piece is determined by
\be
\begin{split}
  &S^{i=2;\text{T.D.}}_{A_j; \text{E}}\\=  &\lim_{n\rightarrow 1}\f{1}{2(1-n)}\log{\left[\left\langle\sigma_n(\xi^{\tau_a}_{X_1},\overline{\xi}^{\tau_a}_{X_1})\overline{\sigma}_n(\xi^{\tau_a}_{X_2},\overline{\xi}^{\tau_a}_{X_2})\overline{\sigma}_n(4\epsilon-\overline{\xi}^{\tau_a}_{X_1},4\epsilon-\xi^{\tau_a}_{X_1})\sigma_n(4\epsilon-\overline{\xi}^{\tau_a}_{X_2},4\epsilon-\xi^{\tau_a}_{X_2})\right \rangle_{\text{torus}}\right]},
\end{split}
\ee
where the complex coordinates are defined as  $(\xi^{\tau_a}_{X_K},\overline{\xi}^{\tau_a}_{X_K})=(\tau_a+\xi_{X_K},\tau_a+\overline{\xi}_{X_K})$.
We assume that $L_{\text{eff}}> 4\epsilon$.
In this parameter region, $S^{i=2;\text{T.D.}}_{A_j; \text{E}}$, is determined by the geodesic length on the BTZ black hole geometry.
After performing the analytic continuation (\ref{eq:ac}), the geodesics are divided into two-folds: the time dependent ones; the time-independent ones. 
The former is given by the geodesic length connecting the original operators with their copies,
\be \label{eq:t-d-p-single-bs}
\begin{split}
 S^{i=2;\text{T.D.}}_{A_j; \text{Dep.}}&= \f{c}{3}\log{\left(\f{4\epsilon}{\pi}\right)}+\text{Min}[S^{\text{Dep.}}_1,S^{\text{Dep.}}_2],\\
    S^{\text{Dep.}}_1&=\f{c}{12}\cdot\sum_{K=1,2}\log{\left|\sin{\left[\f{\pi(-4\epsilon+2\tau_a+\xi_{X_K}+\overline{\xi}_{X_K})}{4\epsilon}\right]}\right|^2}=\f{c}{3}\log{\left[\cosh{\left(\f{\pi t}{2\epsilon}\right)}\right]},\\
    S^{\text{Dep.}}_2&=\f{c}{12}\cdot\sum_{K=1,2}\log{\left|\sin{\left[\f{\pi(-4\epsilon+2\tau_a+\xi_{X_K}+\overline{\xi}_{X_K}\pm i L_{\text{eff}})}{4\epsilon}\right]}\right|^2}\\
    &=\f{c}{6}\log{\left[\f{1}{2}\left(\cosh{\left(\f{\pi t}{\epsilon}\right)}+\cosh{\left(\f{\pi L_{\text{eff}}}{2\epsilon}\right)}\right)\right]}>S^{\text{Dep.}}_1
\end{split}
\ee
where we have used \eqref{eq:ac} and the fact that $\xi_{X_K}+\overline{\xi}_{X_K}=0$. 
Thus, $S^{i=2;\text{T.D.}}_{A_j}$ is independent of the location of the subsystems. 
Let us focus on the time dependence of $S^{i=2;\text{T.D.}}_{A_j}$  and then consider the one in the time regime, $t\gg \epsilon \gg 1$.
In this regime, $S^{i=2;\text{T.D.}}_{A_j}$ reduces to
\be\label{eq:t-d-p-single-bs-linearized-approx}
\begin{split}
   S^{i=2;\text{T.D.}}_{A_j; \text{Dep.}}\approx  \f{c}{3}\log{\left(\f{4\epsilon}{\pi}\right)}+ \f{c \pi t}{6 \epsilon}.
\end{split}
\ee
The latter is determined by one of the geodesic connecting the original operators,
\be \label{eq:t-d-single-i2}
\begin{split}
 &S^{i=2;\text{T.D.}}_{A_j; \text{Stat.}}= \f{c}{3}\log{\left(\f{4\epsilon}{\pi}\right)}+\text{Min}[S^{\text{Stat.}}_{A_j; 1},S^{\text{Stat.}}_{A_j; 2}],\\
    &S^{\text{Stat.}}_{A_j; 1}=\f{c}{6}\cdot\log{\left|\sin{\left[\f{\pi(\xi_{X_1}-\xi_{X_2})}{4\epsilon}\right]}\right|^2},\\
    &S^{\text{Stat.}}_{A_j; 2}=\f{c}{6}\cdot\log{\left|\sin{\left[\f{\pi(\xi_{X_1}-\xi_{X_2}\pm iL_{\text{eff}})}{4\epsilon}\right]}\right|^2}.
\end{split}
\ee
We closely look at the static non-universal piece. 
By rewriting $S^{\text{Stat.}}_{A_j;1,2}$ in terms of $\varphi_X$ {given by \eqref{angular-coordinate}}, $S^{\text{Stat.}}_{A_j;1,2}$ reduce to
\be \label{eq:stat-single-bs}
\begin{split}
 &S^{\text{Stat.}}_{A_j; 1}=\f{c}{3}\cdot\log{\left[\sinh{\left[\f{L_{\text{eff}}(\varphi_{X_1}-\varphi_{X_2})}{4q\epsilon}\right]}\right]},~S^{\text{Stat.}}_{A_j; 2}=\f{c}{3}\cdot\log{\left[\sinh{\left[\f{L_{\text{eff}}(\varphi_{X_1}-\varphi_{X_2}\pm q\pi)}{4q\epsilon}\right]}\right]},
\end{split}
\ee
where $\varphi_{X_1}-\varphi_{X_2}$ are given by
\be
\begin{split}
    \sin{\left(\varphi_{X_1}-\varphi_{X_2}\right)}&=\f{\sin{\left[\f{q\pi (X_1-X_2)}{L}\right]}}{\prod_{K=1,2}r_{X_K}}, \\
    \cos{\left(\varphi_{X_1}-\varphi_{X_2}\right)}&=\f{e^{-2\theta}\prod_{K=1,2}\cos{\left(\f{q\pi X_K}{L}\right)}+e^{2\theta}\prod_{K=1,2}\sin{\left(\f{q\pi X_K}{L}\right)}}{\prod_{K=1,2}r_{X_K}}.
\end{split}
\ee
First, we consider $S^{i=2;\text{T.D.}}_{A_1; \text{Stat.}}$, and then take the large $\theta$ limit, $\theta\gg 1$.
At the leading order of this limit, $\varphi_{X_1}-\varphi_{X_2}$ is approximately given by
\be
\varphi_{X_1}-\varphi_{X_2} \approx\f{e^{-2\theta}\sin{\left[\f{q\pi (X_1-X_2)}{L}\right]}}{\prod_{K=1,2}\left|\sin{\left(\f{q\pi X_K}{L}\right)}\right|}.
\ee
Then, at the leading order in this limit, $S^{i=2;\text{T.D.}}_{A_1; \text{Stat.}}$ results in
\be
\begin{split}
    S^{i=2;\text{T.D.}}_{A_1; \text{Stat.}} &\underset{\theta \gg1}{ \approx}\f{c}{3}\log{\left(\f{4\epsilon}{\pi}\right)}+ \f{c}{3}\cdot \log{\left[\sinh{\left[\f{L \sin{\left[\f{q\pi (X_1-X_2)}{L}\right]}}{8q\epsilon  \prod_{K=1,2}\sin{\left[\f{q\pi (\hat{X}_K)}{L}\right]}}\right]}\right]}\\
&\approx \f{c}{3}\log{\left(\f{4\epsilon}{\pi}\right)}+\f{c}{3}\cdot\begin{cases}
\log{\left[\f{L \sin{\left[\f{q\pi (X_1-X_2)}{L}\right]}}{8q\epsilon  \prod_{K=1,2}\sin{\left[\f{q\pi (\hat{X}_K)}{L}\right]}}\right]}~&~\text{for}~~\f{L \sin{\left[\f{q\pi (X_1-X_2)}{L}\right]}}{8q\epsilon  \prod_{K=1,2}\sin{\left[\f{q\pi (\hat{X}_K)}{L}\right]}} \ll 1\\
\f{L \sin{\left[\f{q\pi (X_1-X_2)}{L}\right]}}{8q\epsilon  \prod_{K=1,2}\sin{\left[\f{q\pi (\hat{X}_K)}{L}\right]}} ~&~\text{for}~~\f{L \sin{\left[\f{q\pi (X_1-X_2)}{L}\right]}}{8q\epsilon  \prod_{K=1,2}\sin{\left[\f{q\pi (\hat{X}_K)}{L}\right]}} \gg 1\\
\end{cases},
\end{split}
\ee
where we assume that $0<\hat{X}_K <\f{L}{q}$.
For $A_1$, in the low temperature limit, $\f{L \sin{\left[\f{q\pi (X_1-X_2)}{L}\right]}}{8q\epsilon  \prod_{K=1,2}\sin{\left[\f{q\pi (\hat{X}_K)}{L}\right]}} \ll 1$, the theory-dependent piece is independent of time, and it is given by
\be
S^{i=2;\text{T.D.}}_{A_1} \underset{\theta \gg 1}  {\approx}\f{c}{3}\log{\left(\f{4\epsilon}{\pi}\right)}+\f{c}{3}\cdot\log{\left[\f{L \sin{\left[\f{q\pi (X_1-X_2)}{L}\right]}}{8q\epsilon  \prod_{K=1,2}\sin{\left[\f{q\pi (\hat{X}_K)}{L}\right]}}\right]}
\ee
Consequently, in this low temperature limit, the entanglement entropy reduces to
\be \label{eq:EE-lagth-lowte}  
S_{A_1}^{i=2} {\approx} \f{c}{3}\log{\left(\f{L}{ q\pi}\sin{\left[\f{q\pi (X_1-X_2)}{L}\right]}\right)}
\ee 
As in \cite{Calabrese:2009qy,2004JSMTE..06..002C}, (\ref{eq:EE-lagth-lowte}) is equivalent to the entanglement entropy in the finite circle system with the size of $L/q$.
The mechanism for the emergence of the vacuum entanglement structure was explained in Section \ref{Sec:EE-for-singleinterval}. 
Then, we closely look at the time dependence of $S_{A_1}$ in the high-temperature regime, $\f{L \sin{\left[\f{q\pi (X_1-X_2)}{L}\right]}}{8q\epsilon  \prod_{K=1,2}\sin{\left[\f{q\pi (\hat{X}_K)}{L}\right]}} \gg 1$.
The time dependence of $S_{A_1}^{i=1}$ is approximately given by
\be\label{eq:EE-lagth-highte} 
\begin{split}
    S^{i=2}_{A_1} \approx& \f{c}{3}\log{\left(\f{4\epsilon}{\pi}\right)}+\f{c}{6}\log{\left[4\prod_{K=1,2}\sin^2{\left(\f{q\pi X_K}{L}\right)}\right]}\\
    &+\f{c}{3}\begin{cases}
        \f{\pi t}{2\epsilon}  ~&~\text{for}~\f{L\sin{\left[\f{q\pi (X_1-X_2)}{L}\right]}}{4q\pi \prod_{K=1,2}\sin{\left[\f{q\pi \hat{X}_K}{L}\right]}}> t\gg \epsilon\\
        \left[\f{L \sin{\left[\f{q\pi (X_1-X_2)}{L}\right]}}{8q\epsilon  \prod_{K=1,2}\sin{\left[\f{q\pi (\hat{X}_K)}{L}\right]}}\right]~&~\text{for}~t> \f{L\sin{\left[\f{q\pi (X_1-X_2)}{L}\right]}}{4q\pi \prod_{K=1,2}\sin{\left[\f{q\pi \hat{X}_K}{L}\right]}}
    \end{cases}.
\end{split}
\ee

Subsequently, we consider the time dependence of $S^{i=2}_{A_2}$ in the large $\theta$ limit.
First, we closely look at the behavior of $S^{i=2;\text{T.D.}}_{A_2; \text{Stat.}}$ in the large $\theta$ limit.
At the leading in the large $\theta$ limit, $\varphi_{X_1}-\varphi_{X_2}$ is approximated by 
\be
\varphi_{X_1}-\varphi_{X_2}\approx l\cdot \pi
\ee
The leading behavior of $S^{i=2;\text{T.D.}}_{A_2; \text{Stat.}}$ is given by
\be
\begin{split}
S^{i=2;\text{T.D.}}_{A_2; \text{Stat.}} \approx \f{c}{3}\log{\left(\f{4\epsilon}{\pi}\right)}+\begin{cases}
    \f{c\pi l \cdot L }{24 q \epsilon}e^{2\theta} ~&~\text{for}~\f{q}{2}>l\\
     \f{c\pi (q-l) \cdot L }{24 q \epsilon}e^{2\theta} ~&~\text{for}~l>\f{q}{2}\\
\end{cases},
\end{split}
\ee
where we assume that $q$ is an even integer.
Then, the time dependence of $S^{i=2}_{A_2}$ is determined by
\be \label{eq:EE-A2-BS} 
\begin{split}
     S^{i=2}_{A_2} \approx& \f{c}{3}\log{\left(\f{4\epsilon}{\pi}\right)}+\f{c}{6}\log{\left[4\prod_{K=1,2}\sin^2{\left(\f{q\pi X_K}{L}\right)}\right]}\\
       &+\f{c}{3}\begin{cases}
        \f{\pi t}{2\epsilon}  ~&~\text{for}~\f{q}{2}>l~\text{and}~\f{l Le^{2\theta}}{4q}> t\gg \epsilon\\
        \f{c\pi l\cdot L }{8 q \epsilon}e^{2\theta}~&~\text{for}~\f{q}{2}>l~\text{and}~t>\f{l Le^{2\theta}}{4q}\\
        \f{\pi t}{2\epsilon}  ~&~\text{for}~l>\f{q}{2}~\text{and}~\f{(q-l) Le^{2\theta}}{4q}> t\gg \epsilon\\
        \f{c\pi(q- l)\cdot L }{8 q \epsilon}e^{2\theta}~&~\text{for}~\f{q}{2}>l~\text{and}~ t>\f{(q-l) Le^{2\theta}}{4q}\\
    \end{cases}.\\
\end{split}
\ee
At $t=0$, $S^{i=2}_{A_2}$ can be expressed by 
\be
\begin{split}
    S_{A_{2}}^{i=2}
    &= \f{c}{3} \log{\left[\f{L}{q\pi}\sin{\left(\f{q\pi \left(\hat{X}_1-X^f_{m+l}\right)}{L}\right)}\right]}+\f{c}{3} \log{\left[\f{L}{q\pi}\sin{\left(\f{q\pi \left(\hat{X}_2-X^f_{m}\right)}{L}\right)}\right]}\\
    &+\f{c}{3}\log{\left(\f{16\epsilon}{\pi}\right)}+\f{2c}{3}\log{\left(\f{q\pi}{L}\right)}.
\end{split}
\ee
Here, the first two terms are the same as the vacuum entanglement entropies for the SSD Hamiltonian on the circle system with the size of $L/q$. 
The first term is the vacuum entanglement entropy for the spatial interval region between $X^f_{m+l}$ and $\hat{X}^f_{1}$ of the system with $L/q$, while the second term is that for the interval between $X^{f}_m$ and $\hat{X}^f_2$.

\subsubsection{Interpretation \label{sec:interpretation}}
In this section, we will give interpretations on the findings in the low or high temperature temperatures and large $\theta$ limit.
Note that the definition of high or low temperature limits is (\ref{eq:high-low-temp-limits}).
In the high temperature with large $\theta$, the entanglement dynamics can be described by the propagation of quasiparticles with the speed of light.
In the low temperature with large $\theta$, the non-equilibrium process can be considered as one of the multi-joining quenches \cite{2007JSMTE..10....4C,2013JHEP...05..080N,2019JHEP...03..165S,2019JHEP...09..018C,2023arXiv230904665K}.

We begin by considering one possible interpretation on the entanglement entropy in the high effective temperature and high temperature region, $L\cosh{2\theta}/\beta>1 $ and $L/\beta>1$.
Here, we also assume that $\theta \gg 1$. This interpretation describes the time dependence of entanglement entropy at the $\mathcal{O}(1/\epsilon)$ in the high temperature limit.
Here, we consider the gravity dual of the system considered.
The detail of the gravity dual will be explained in Section \ref{sec:gravity-dual}.
The geometry is given by 
\be \label{eq:bulk-metric-in-xi}
ds^2=\f{dr^{2}}{r^{2}-r_{0}^{2}}  +r^{2} dx^2_{\xi} - (r^{2}-r_{0}^{2} ) dt^2,
\ee
where $r$ is the radial coordinate, $dx_{\xi}$ is defined as $\f{ d\xi -d\bar{\xi} }{2i} $, $\tau$ is the real time associated with $H_{\text{q-M\"obius}}$, and $r_0$ is the black hole radius. 
The boundary metric, around $r\rightarrow \infty$, of the bulk geometry in (\ref{eq:bulk-metric-in-xi}) is given by
\be
ds^2_{\text{Boundary}}=dx^2_{\xi}-dt^2. 
\ee
Note that as in \cite{2024arXiv240606121M}, the constant slice, where the system considered here is, should be taken along $\tilde{r}=\f{r}{f(x)}$. However, the difference between geodesic length ending at the constant $\tilde{r}$ slice and that ending at the $r$ slice is at $\mathcal{O}(1)$. Therefore, we can neglect the contribution from the endpoint of the geodesic.

Let us define the effective subsystem size, $l^{\text{effective}}_{A_j}$ in this coordinate as
\be\label{eq:effective-subsystem-size}
l^{\text{effective}}_{A_j}=\int^{x_{\xi}(X_1)}_{x_{\xi}(X_  2)}dx_{\xi}=\f{L\cosh{2\theta}}{q\pi}\left[\arctan{\left(e^{2\theta}\tan{\left(\f{q\pi X_1}{L}\right)} \right)}-\arctan{\left(e^{2\theta}\tan{\left(\f{q\pi X_2}{L}\right)} \right)}\right],
\ee
and the effective length between two arbitrary points $X,Y\in[0,L]$ as
\be\label{eq:effective-length}
l^{\text{effective}}(X,Y):=\left|\int^{x_{\xi}(X)}_{x_{\xi}(Y)}dx_{\xi}\right|=\left|x_{\xi}(X)-x_{\xi}(Y)\right|.
\ee
Then, take the large $\theta$ limit, such that $l_{A_j}^{\text{effective}}$ reduces to
\be
\begin{split}
    l^{\text{effective}}_{A_j} \approx \begin{cases}
        \f{L}{2q\pi }\f{\sin{\left[\f{q\pi (X_1-X_2)}{L}\right]}}{\prod_{K=1,2}\sin{\left[\f{q\pi (X_K)}{L}\right]}}~&~\text{for}~j=1\\
        \f{l \cdot L}{2q}e^{2\theta}~&~\text{for}~j=2
    \end{cases}.
\end{split}
\ee
Up to the constant proportional to the central charge, in the $(\xi,\overline{\xi})$ coordinates, the q-M\"obius Hamiltonian is the same as the uniform Hamiltonian, or the CFT Hamiltonian on the flat space. Therefore, we assume that for $\ket{U_{\text{eff}}}$, the quasiparticles homogeneously distribute at $t=0$, while for $\ket{\Psi(t)}$, the entangled pairs homogeneously emerge. Furthermore, we assume that they propagate at the speed of light, and at the leading order in the high temperature limit, the entanglement entropy for $A_j$ is given by the quasiparticle density times the size of the effective subsystem defined as the interval between $x_{\xi}(x_1)$ and $x_{\xi}(x_2)$.
Here, we assume that for $\ket{U_{\text{eff}}}$ the quasiparticle density is $\f{c \pi}{6\epsilon}$, while for $\ket{\Psi(t)}$ the one is $\f{c \pi}{12\epsilon}$ as in the case of the uniform Hamiltonian. This quasiparticle picture can describe the time evolution of the entanglement entropies and mutual information at $\mathcal{O}\left(\f{1}{\epsilon}\right)$ in the high temperature limit. 
To explain the contributions at $\mathcal{O}(1)$, we need an effective description beyond the quasiparticle picture. 
To find the effective description beyond the quasiparticle picture would be interesting. 
We will leave it as a future problem.

Then, we discuss the relation between the system considered here and the multi-joining quench. We consider the low temperature region with the large $\theta$ in the high effective temperature. In the large $\theta$ limit, q-M\"obius Hamiltonian reduces to the sum of the SSD Hamiltonians as in (\ref{eq:approximation-of-H}). Note that at the leading order, the Hamiltonian densities around fixed points $x=X^f_m$ can not be reduced to SSD ones. In the low temperature limit, the system at $t=0$ would be approximately given by the vacuum state for the q-M\"obius Hamiltonian, 
\be
\ket{\Psi}\underset{ \epsilon \gg 1} {\approx}\ket{0},
\ee
where $\ket{0}$ denotes the vacuum state for the q-M\"obius Hamiltonian.
Then, at $t=0$, the system approximately results in the product of the SSD vacuum states
\be
\ket{\Psi} \approx \prod_{m=0}^{q-1} \ket{0}_m,
\ee
where $\ket{0}_m$ is the vacuum state for the spatial interval from $x=X^f_{m}$ to $x=X^f_{m+1}$.
In other words, at $t=0$, the system is constructed of $q$ independent systems.
If we closely look at the entanglement structure around $x=X^f_{m}$, it is a short-entangled one.
If this intuitive picture is true, during the time evolution induced by $H_{\text{q-M\"obius}}$, the entanglement structure associated with the spatial region far from $x=X^f_m$ should not change.
The behavior of $S_{A_j}^{i=2}$ in (\ref{eq:EE-lagth-lowte}) is consistent with this intuitive picture.
Contrary to the entanglement structure associated with $A_1$, the time evolution operator of $H_{\text{q-M\"obius}}$ can change the entanglement structure associated with $A_2$ because the initial entanglement structure around $x=X^{f}_X$ is approximately short-range entangled. 
This may be the reason for $S_{A_2}^{i=2}$ to linearly grow in time.
One possible interpretation of this non-equilibrium process is the multi-joining quenches, inducing the highly entangled structure around the points connecting the independent systems during the time evolution.

\subsection{Entanglement entropy of $\ket{\Psi(t)}$ for the double interval\label{Sec:EE-BS-for-DI}}
Then, we consider the entanglement entropy of $\rho_{i=2}$ for the union of two spatial intervals, $A_j\cup B$ in the Euclidean path-integral.
By employing the twist-operator formalism, $n$-th moment of Euclidean R\'enyi entanglement entropy for the union of the double intervals, $S^{i=2;(n)}_{A_j\cup B;\text{E}}$, is given by
\be \label{eq:Renyi-BS-double}
\begin{split}
&S^{i=2;(n)}_{A_j\cup B;\text{E}}\\
&=\f{1}{1-n}\log{\left[\f{\bra{B}e^{-\tau_b H_{\text{q-M\"obius}}}\sigma_n(w_{X_1},\overline{w}_{X_1})\overline{\sigma}_n(w_{X_2},\overline{w}_{X_2})\sigma_n(w_{Y_1},\overline{w}_{Y_1})\overline{\sigma}_n(w_{Y_2},\overline{w}_{Y_2})e^{-\tau_a H_{\text{q-M\"obius}}}\ket{B}}{\bra{B}e^{-2\epsilon H_{\text{q-M\"obius}}}\ket{B}}\right]}\\
&=-\f{c}{24}\cdot \f{n+1}{n}\log{\left(\prod_{K=1,2}f^{-2}(X_K)\cdot f^{-2}(Y_K) \right)}\\
&+\f{1}{1-n}\log{\left[\f{\bra{\text{B}}e^{-2\epsilon H_{\text{q-M\"obius}}} \sigma_n(\xi_{X_1}^{\tau_a},\overline{\xi}_{X_1}^{\tau_a})\overline{\sigma}_n(\xi_{X_2}^{\tau_a},\overline{\xi}_{X_2}^{\tau_a})\sigma_n(\xi_{Y_1}^{\tau_a},\overline{\xi}_{Y_1}^{\tau_a})\overline{\sigma}_n(\xi_{Y_2}^{\tau_a},\overline{\xi}_{Y_2}^{\tau_a})\ket{\text{B}}}{\bra{B}e^{-2\epsilon H_{\text{q-M\"obius}}}\ket{B}}\right]},
\end{split}
\ee
where we assume that $L>Y_1>Y_2>X_1>X_2>0$.
In the last line, first and second terms are universal and theory-dependent terms, respectively.
As in the case of the single interval, at the leading order in the large $\theta$ limit, the universal piece  simply reduces to
\be
\begin{split}
    S^{i=2;\text{U.}(n)}_{A_{j}\cup B;\text{ E}}&=\f{c(n+1)}{12n}\log{\left[\prod_{K=1,2}\left(1-\cos{\left(\f{2q\pi X_K}{L}\right)}\tanh{2\theta}\right)\left(1-\cos{\left(\f{2q\pi Y_K}{L}\right)}\tanh{2\theta}\right)\right]}\\
    &\underset{\theta \gg 1}{\approx}\f{c(n+1)}{12n}\log{\left[2^4\prod_{K=1,2}\sin^2{\left(\f{q\pi X_K}{L}\right)}\sin^2{\left(\f{q\pi Y_K}{L}\right)}\right]}\\
    &\underset{n\rightarrow 1}{=}\f{c}{6}\log{\left[2^4\prod_{K=1,2}\sin^2{\left(\f{q\pi X_K}{L}\right)}\sin^2{\left(\f{q\pi Y_K}{L}\right)}\right]}.
\end{split}
\ee
\subsubsection{The theory-dependent pieces in $2$d holographic CFT\label{Sec:EE-for-DI-BS-in-2d-hCFTs}}
Here, we will present the time dependence of the theory-dependence piece, $S^{i=2; \text{T.D.}}_{A_j\cup B;\text{E}}$, in 2d holographic CFTs. 
To do so, first, we employ the method of image for the computation of the non-universal piece of Euclidean R\'enyi entanglement entropy, and then take the von Neumann limit, $n\rightarrow 1$.
Then, the behavior of $S^{i=2; \text{T.D.}}_{A_j\cup B}$ is determined by the eight-point function of twist and anti-twist operators on the thermal torus,
\be
\begin{split}
    &S^{i=2; \text{T.D.}}_{A_j\cup B;\text{E}}\\=  &\lim_{n \rightarrow 1}\f{1}{2(1-n)}\log\bigg{[}\bigg{\langle}\sigma_n(\xi^{\tau_a}_{X_1},\overline{\xi}^{\tau_a}_{X_1})\overline{\sigma}_n(\xi^{\tau_a}_{X_2},\overline{\xi}^{\tau_a}_{X_2})\sigma_n(\xi^{\tau_a}_{Y_1},\overline{\xi}^{\tau_a}_{Y_1})\overline{\sigma}_n(\xi^{\tau_a}_{Y_2},\overline{\xi}^{\tau_a}_{Y_2})\\
    &\times\overline{\sigma}_n(4\epsilon-\overline{\xi}^{\tau_a}_{X_1},4\epsilon-\xi^{\tau_a}_{X_1})\sigma_n(4\epsilon-\overline{\xi}^{\tau_a}_{X_2},4\epsilon-\xi^{\tau_a}_{X_2})\overline{\sigma}_n(4\epsilon-\overline{\xi}^{\tau_a}_{Y_1},4\epsilon-\xi^{\tau_a}_{Y_1})\sigma_n(4\epsilon-\overline{\xi}^{\tau_a}_{Y_2},4\epsilon-\xi^{\tau_a}_{Y_2})\bigg{ \rangle}_{\text{torus}}\bigg{]}.
\end{split}
\ee

In $2$d holographic CFT, the theory-dependent piece is determined by the length of minimal geodesics.
We perform the analytic continuation in (\ref{eq:ac}).
Then, as in the case of the single interval, the geodesics ending on the boundaries of $A_j$ and $B$ can be divided into time-dependent and time-independent ones. 
The minimal time-dependent geodesics are determined by
\be\label{eq:time dependent-geodesics-BDS-EE}
\begin{split}
    S^{i=2; \text{T.D.}}_{A_j\cup B\text{;Dep.}} &= \f{2c}{3}\log{\left(\f{4\epsilon}{\pi}\right)}+\text{Min}\left[2S^{\text{Dep.}}_1,2S^{\text{Dep.}}_2\right],
\end{split}
\ee
where $S^{\text{Dep.}}_{\tilde{i}=1,2}$ are defined in (\ref{eq:t-d-p-single-bs}).
Let us closely look at the static geodesics.
Addition to the subsystems, $A_j$ and $B$, we define $C$ and $D$ as 
\be
C=\left\{x|X_1<x<Y_2\right\}, D=\overline{A_j\cup B\cup C}.
\ee
The minimal surface of the static geodesics is given by
\be
\begin{split}
    S^{i=2; \text{T.D.}}_{A_j\cup B\text{;Stat.}}=\text{Min}\left[S^{i=2;\text{T.D.}}_{A_j; \text{Stat.}}+S^{i=2;\text{T.D.}}_{B; \text{Stat.}},S^{i=2;\text{T.D.}}_{C; \text{Stat.}}+S^{i=2;\text{T.D.}}_{D; \text{Stat.}}\right],
\end{split}
\ee
where $S^{i=2;\text{T.D.}}_{A_j; \text{Stat.}}$ is defined in (\ref{eq:t-d-single-i2}), and $S^{i=2;\text{T.D.}}_{B; \text{Stat.}}$ is given by replacing $X_K$ of $S^{i=2;\text{T.D.}}_{A_j; \text{Stat.}}$ with $Y_K$.
In addition,  $S^{i=2;\text{T.D.}}_{C; \text{Stat.}}$ and $S^{i=2;\text{T.D.}}_{D; \text{Stat.}}$ are defined as 
\be
\begin{split}
    &S^{i=2;\text{T.D.}}_{C; \text{Stat.}}=\text{Min}\left[S^{\text{Stat.}}_{C;1},S^{\text{Stat.}}_{C;2}\right],~ S^{i=2;\text{T.D.}}_{D; \text{Stat.}}=\text{Min}\left[S^{\text{Stat.}}_{D;1},S^{\text{Stat.}}_{D;2}\right],\\
     &S^{\text{Stat.}}_{C; 1}=\f{c}{6}\cdot\log{\left|\sin{\left[\f{\pi(\xi_{Y_2}-\xi_{X_1})}{4\epsilon}\right]}\right|^2}, S^{\text{Stat.}}_{C; 2}=\f{c}{6}\cdot\log{\left|\sin{\left[\f{\pi(\xi_{Y_2}-\xi_{X_1}\pm iL_{\text{eff}})}{4\epsilon}\right]}\right|^2},\\
     &S^{\text{Stat.}}_{D; 1}=\f{c}{6}\cdot\log{\left|\sin{\left[\f{\pi(\xi_{Y_1}-\xi_{X_2})}{4\epsilon}\right]}\right|^2}, S^{\text{Stat.}}_{D; 2}=\f{c}{6}\cdot\log{\left|\sin{\left[\f{\pi(\xi_{Y_1}-\xi_{X_2}\pm iL_{\text{eff}})}{4\epsilon}\right]}\right|^2}.\\
\end{split}
\ee
Let us define $R_m$ as the spatial region where $X^f_{m}>x>X^f_{m-1}$, where $m$ runs from $1$ to $q$.

\subsubsection{Case studies on the time dependence of mutual information\label{Sec:Case-study}}
Now, we present the time dependence of the mutual information for (\ref{eq:Bdy-state}) for various cases:
\begin{itemize}
    \item [{\bf Case.\ 1:}] In Case.\ 1, the subsystems, $A$ and $B$, are contained in $R_m$. Here, we assume that 
    \be
X^f_{m-1}<X_2<X_1<Y_2<Y_1<X^f_m.
    \ee
      \item [{\bf Case.\ 2:}] In Case.\ 2, the subsystems, $A$ and $B$, are contained in $R_m$ and $R_{m+l}$, where we assume that $l$ is a positive integer. In other words, the subsystems are determined by 
    \be
X^f_{m-1}<X_2<X_1<X^f_m<X^f_{m+l-1}<Y_2<Y_1<X^f_{m+l}.
    \ee
     \item [{\bf Case.\ 3:}] In Case.\ 3, the subsystem $A$ is contained in $R_m$, while $B$ contains $n\ge 1$ fixed points where the curvature is negatively maximized. 
     More concretely, the subsystems are determined by 
    \be
X^f_{m-1}<X_2<X_1<X^f_m<X^f_{m+l-1}<Y_2<X^f_{m+l}<X^f_{m+l+n-1}<Y_1<X^f_{m+l+n}.
    \ee
    \item [{\bf Case.\ 4:}] In Case.\ 4, $A$ and $B$ are spatially-disjoint intervals. We assume that $A$ contains $n_1\ge 1$ fixed points where the curvature is negatively maximized, while $B$ does $n_2\ge 1$ fixed points. 
    The subsystems, $A$ and $B$, are determined by
    \be
    \begin{split}
&X^f_{m-1}<X_2<X^f_{m}<X^f_{m+n_1-1}<X_1<X^f_{m+n_1},\\
&X^f_{m+l+n_1-1}<Y_2<X^f_{m+l+n_1}<X^f_{m+l+\sum_{i=1,2}n_i-1}<Y_1<X^f_{m+l+\sum_{i=1,2}n_i}.
    \end{split}
    \ee
\end{itemize}

\subsubsection*{Case.\ 1\label{Sec:TI-in-Rm}}
Here, we explore the time dependence of $S_{A_1\cup B}^{i=2}$ for Case.\ 1. In the low temperature region of the large $\theta$ limit, where $\f{L \sin{\left[\f{q\pi (X_1-X_2)}{L}\right]}}{8q\epsilon  \prod_{K=1,2}\sin{\left[\f{q\pi (\hat{X}_K)}{L}\right]}} \ll 1$ and $\f{L \sin{\left[\f{q\pi (Y_1-Y_2)}{L}\right]}}{8q\epsilon  \prod_{K=1,2}\sin{\left[\f{q\pi (\hat{Y}_K)}{L}\right]}} \ll 1$, $S^{i=2;\text{T.D.}}_{A_1\cup B}$ is determined by one of the static geodesics, so that $S_{A_1\cup B}^{i=2}$ is given by
\be
\begin{split}
    S_{A_1\cup B}^{i=2}\approx  \f{c}{3}\cdot\text{Min}\bigg{[}&\log{\left(\f{L}{q\pi}\sin{\left[\f{q\pi (X_1-X_2)}{L}\right]}\right)}+ \log{\left(\f{L}{q\pi}\sin{\left[\f{q\pi (Y_1-Y_2)}{L}\right]}\right)},\\
&\log{\left(\f{L}{q\pi}\sin{\left[\f{q\pi (Y_1-X_2)}{L}\right]}\right)}+ \log{\left(\f{L}{q\pi}\sin{\left[\f{q\pi (Y_2-X_1)}{L}\right]}\right)}\bigg{]}.
\end{split}
\ee
In this case, the mutual information between $A_1$ and $B$ is approximated by the vacuum mutual information in the finite system with the size of $L/q$. 
In the high temperature region of the large $\theta$ limit, $S^{i=2;\text{T.D.}}_{A_1\cup B\text{;Stat.}}$ is determined by
\be \label{eq:EE-BS-Rm-high}
\begin{split}
    S^{i=2;\text{T.D.}}_{A_1\cup B\text{;Stat.}}\approx  \f{cL}{24q\epsilon}\cdot\text{Min}\bigg{[}&\left(\f{\sin{\left[\f{q\pi (X_1-X_2)}{L}\right]}}{  \prod_{K=1,2}\sin{\left[\f{q\pi (\hat{X}_K)}{L}\right]}}+\f{\sin{\left[\f{q\pi (Y_1-Y_2)}{L}\right]}}{ \prod_{K=1,2}\sin{\left[\f{q\pi (\hat{Y}_K)}{L}\right]}}\right), \\
    &\left(\f{ \sin{\left[\f{q\pi (Y_1-X_2)}{L}\right]}}{\sin{\left[\f{q\pi (\hat{Y}_1)}{L}\right]} \sin{\left[\f{q\pi (\hat{X}_2)}{L}\right]}}+\f{\sin{\left[\f{q\pi (Y_2-X_1)}{L}\right]}}{\sin{\left[\f{q\pi (\hat{Y}_2)}{L}\right]} \sin{\left[\f{q\pi (\hat{X}_1)}{L}\right]}}\right)\bigg{]} 
\end{split}
\ee
Let $S^{i=2;\text{T.D.}}_{A_1\cup B,1\text{;Stat.}}$ and $S^{i=2;\text{T.D.}}_{A_1\cup B,2\text{;Stat.}}$ denote the first and second terms of (\ref{eq:EE-BS-Rm-high}), respectively.
The difference between them is given by
\be
\begin{split}
   S^{i=2;\text{T.D.}}_{A_1\cup B,1\text{;Stat.}}-S^{i=2;\text{T.D.}}_{A_1\cup B,2\text{;Stat.}}=\f{cL}{12q\epsilon}\left[\f{\sin{\left(\f{q\pi (\hat{X}_1-\hat{Y}_2)}{L}\right)}}{\sin{\left(\f{q\pi \hat{Y}_2}{L}\right)}\sin{\left(\f{q\pi \hat{X}_1}{L}\right)}}\right]<0.
\end{split}
\ee
Therefore, $S^{i=2;\text{T.D.}}_{A_1\cup B\text{;Stat.}}$ is given by
\be
S^{i=2;\text{T.D.}}_{A_1\cup B\text{;Stat.}}\approx \f{cL}{24q\epsilon}\cdot\left(\f{\sin{\left[\f{q\pi (X_1-X_2)}{L}\right]}}{  \prod_{K=1,2}\sin{\left[\f{q\pi (\hat{X}_K)}{L}\right]}}+\f{\sin{\left[\f{q\pi (Y_1-Y_2)}{L}\right]}}{ \prod_{K=1,2}\sin{\left[\f{q\pi (\hat{Y}_K)}{L}\right]}}\right)=S^{i=2;\text{T.D.}}_{A_1\text{;Stat.}}+S^{i=2;\text{T.D.}}_{B\text{;Stat.}}
\ee
Consequently, the time dependence of $S_{A_1\cup B}^{i=2}$ is approximated by
\be
\begin{split}
    S_{A_1\cup B}^{i=2}&\approx\f{c}{6}\log{\left[2^4\prod_{K=1,2}\sin^2{\left(\f{q\pi X_K}{L}\right)}\sin^2{\left(\f{q\pi Y_K}{L}\right)}\right]}+ \f{2c}{3}\log{\left(\f{4\epsilon}{\pi}\right)}\\
    &+\f{c}{6\epsilon}\cdot\begin{cases}
    2\pi t ~&~\text{for}~t_*>t\gg \epsilon\\
    \f{L}{4q}\cdot\left(\f{\sin{\left[\f{q\pi (X_1-X_2)}{L}\right]}}{  \prod_{K=1,2}\sin{\left[\f{q\pi (\hat{X}_K)}{L}\right]}}+\f{\sin{\left[\f{q\pi (Y_1-Y_2)}{L}\right]}}{ \prod_{K=1,2}\sin{\left[\f{q\pi (\hat{Y}_K)}{L}\right]}}\right) ~&~\text{for}~t>t_*
    \end{cases}
\end{split},
\ee
where we assume that $t_*$ is smaller than $L/2$, and
$t_*$ is defined as 
\be
t_*\approx\f{L}{8\pi q}\cdot\left(\f{\sin{\left[\f{q\pi (X_1-X_2)}{L}\right]}}{  \prod_{K=1,2}\sin{\left[\f{q\pi (\hat{X}_K)}{L}\right]}}+\f{\sin{\left[\f{q\pi (Y_1-Y_2)}{L}\right]}}{ \prod_{K=1,2}\sin{\left[\f{q\pi (\hat{Y}_K)}{L}\right]}}\right).
\ee
In this case, the mutual information is independent of time, and its value is zero.

In the above discussion, we assumed the case $\f{L \sin{\left[\f{q\pi (Y_2-X_1)}{L}\right]}}{8q\epsilon  \sin{\left[\f{q\pi (\hat{Y}_2)}{L}\right]} \sin{\left[\f{q\pi (\hat{X}_1)}{L}\right]} } \gg 1$.
Then, we consider the different limit, where $\f{L \sin{\left[\f{q\pi (Y_2-X_1)}{L}\right]}}{8q\epsilon  \sin{\left[\f{q\pi (\hat{Y}_2)}{L}\right]} \sin{\left[\f{q\pi (\hat{X}_1)}{L}\right]} } \ll 1$ with keeping $\f{L \sin{\left[\f{q\pi (X_1-X_2)}{L}\right]}}{8q\epsilon  \prod_{K=1,2}\sin{\left[\f{q\pi (\hat{X}_K)}{L}\right]}} \gg 1$ and $\f{L \sin{\left[\f{q\pi (Y_1-Y_2)}{L}\right]}}{8q\epsilon  \prod_{K=1,2}\sin{\left[\f{q\pi (\hat{Y}_K)}{L}\right]}} \gg 1$. 
In this case, $S^{i=2;\text{T.D.}}_{A_1\cup B\text{;Stat.}}$ is given by
\be
\begin{split}
    &S^{i=2;\text{T.D.}}_{A_1\cup B\text{;Stat.}}\\
    &\approx  \text{Min}\bigg{[}\f{cL}{24q\epsilon}\cdot\left(\f{\sin{\left[\f{q\pi (X_1-X_2)}{L}\right]}}{  \prod_{K=1,2}\sin{\left[\f{q\pi (\hat{X}_K)}{L}\right]}}+\f{\sin{\left[\f{q\pi (Y_1-Y_2)}{L}\right]}}{ \prod_{K=1,2}\sin{\left[\f{q\pi (\hat{Y}_K)}{L}\right]}}\right), \\
    & \qquad \qquad \f{cL}{24q\epsilon}\cdot\f{ \sin{\left[\f{q\pi (Y_1-X_2)}{L}\right]}}{\sin{\left[\f{q\pi (\hat{Y}_1)}{L}\right]} \sin{\left[\f{q\pi (\hat{X}_2)}{L}\right]}}+\f{c}{3}\log \left( \sinh \left[\f{L\sin{\left[\f{q\pi (Y_2-X_1)}{L}\right]}}{8q\epsilon \sin{\left[\f{q\pi (\hat{Y}_2)}{L}\right]} \sin{\left[\f{q\pi (\hat{X}_1)}{L}\right]}} \right]\right)\bigg{]}.
\end{split}
\ee
If $\f{L\sin{\left[\f{q\pi (Y_2-X_1)}{L}\right]}}{8q\epsilon \sin{\left[\f{q\pi (\hat{Y}_2)}{L}\right]} \sin{\left[\f{q\pi (\hat{X}_1)}{L}\right]}} < \log \sqrt{3} $, then the second factor gives a smaller contribution. Thus, $S^{i=2;\text{T.D.}}_{A_1\cup B\text{;Stat.}}$ becomes 
\be
    S^{i=2;\text{T.D.}}_{A_1\cup B\text{;Stat.}} \approx  \f{cL}{24q\epsilon}\cdot\f{ \sin{\left[\f{q\pi (Y_1-X_2)}{L}\right]}}{\sin{\left[\f{q\pi (\hat{Y}_1)}{L}\right]} \sin{\left[\f{q\pi (\hat{X}_2)}{L}\right]}}+\f{c}{3}\log \left( \sinh \left[\f{L\sin{\left[\f{q\pi (Y_2-X_1)}{L}\right]}}{8q\epsilon \sin{\left[\f{q\pi (\hat{Y}_2)}{L}\right]} \sin{\left[\f{q\pi (\hat{X}_1)}{L}\right]}} \right]\right).
\ee
Therefore, the time dependence of $S_{A_1\cup B}^{i=2}$ for the present case is approximated by
\be
\begin{split}
    S_{A_1\cup B}^{i=2}&\approx\f{c}{6}\log{\left[2^4\prod_{K=1,2}\sin^2{\left(\f{q\pi X_K}{L}\right)}\sin^2{\left(\f{q\pi Y_K}{L}\right)}\right]}+ \f{2c}{3}\log{\left(\f{4\epsilon}{\pi}\right)}\\
    &+\f{c}{6\epsilon}\cdot\begin{cases}
   2\pi t ~&~\text{for}~t_*'>t\gg \epsilon\\
    \f{L}{4q}\cdot\f{ \sin{\left[\f{q\pi (Y_1-X_2)}{L}\right]}}{\sin{\left[\f{q\pi (\hat{Y}_1)}{L}\right]} \sin{\left[\f{q\pi (\hat{X}_2)}{L}\right]}}+2\epsilon \log \left( \sinh \left[\f{L\sin{\left[\f{q\pi (Y_2-X_1)}{L}\right]}}{8q\epsilon \sin{\left[\f{q\pi (\hat{Y}_2)}{L}\right]} \sin{\left[\f{q\pi (\hat{X}_1)}{L}\right]}} \right]\right) ~&~\text{for}~t>t_*'
    \end{cases}
\end{split},
\ee
where we assume that $t_*'$ is smaller than $L/2$, and
$t_*'$ is defined as 
\be
t_*'\approx\f{L}{8\pi q}\cdot\left(\f{ \sin{\left[\f{q\pi (Y_1-X_2)}{L}\right]}}{\sin{\left[\f{q\pi (\hat{Y}_1)}{L}\right]} \sin{\left[\f{q\pi (\hat{X}_2)}{L}\right]}}+\f{8q\epsilon}{L} \log \left( \sinh \left[\f{L\sin{\left[\f{q\pi (Y_2-X_1)}{L}\right]}}{8q\epsilon \sin{\left[\f{q\pi (\hat{Y}_2)}{L}\right]} \sin{\left[\f{q\pi (\hat{X}_1)}{L}\right]}} \right]\right)\right).
\ee
For $t>t_*'$, the mutual information is given by
\be
I_{A_{1},B}^{i=2} \approx -\f{c}{3}\log \left( \sinh \left[\f{L\sin{\left[\f{q\pi (Y_2-X_1)}{L}\right]}}{8q\epsilon \sin{\left[\f{q\pi (\hat{Y}_2)}{L}\right]} \sin{\left[\f{q\pi (\hat{X}_1)}{L}\right]}} \right]\right)-\f{cL}{24q\epsilon}\cdot\f{ \sin{\left[\f{q\pi (Y_2-X_1)}{L}\right]}}{\sin{\left[\f{q\pi (\hat{Y}_2)}{L}\right]} \sin{\left[\f{q\pi (\hat{X}_1)}{L}\right]}}\geq 0.
\ee

\subsubsection*{Case.\ 2}
Here, we explore the time dependence of $S_{A_1\cup B}^{i=2}$ for Case.\ 2. In the low temperature region of the large $\theta$ limit, where $\f{L \sin{\left[\f{q\pi (X_1-X_2)}{L}\right]}}{8q\epsilon  \prod_{K=1,2}\sin{\left[\f{q\pi (\hat{X}_K)}{L}\right]}} \ll 1$ and $\f{L \sin{\left[\f{q\pi (Y_1-Y_2)}{L}\right]}}{8q\epsilon  \prod_{K=1,2}\sin{\left[\f{q\pi (\hat{Y}_K)}{L}\right]}} \ll 1$. After minimization, the entanglement entropy $S_{A_1\cup B}$ is given by
\be
    S_{A_1\cup B}^{i=2}\approx  \f{c}{3}\cdot\left[\log{\left(\f{L}{q\pi}\sin{\left[\f{q\pi (X_1-X_2)}{L}\right]}\right)}+ \log{\left(\f{L}{q\pi}\sin{\left[\f{q\pi (Y_1-Y_2)}{L}\right]}\right)}\right].
\ee
In this case, the mutual information between $A_1$ and $B$ is simply zero and $A_1$ decouples with $B$. In the high temperature region of the large $\theta$ limit, $S^{i=2;\text{T.D.}}_{A_1\cup B\text{;Stat.}}$ is determined by
\be \label{eq:EE-BS-Rm-Rm+l-high}
\begin{split}
    S^{i=2;\text{T.D.}}_{A_1\cup B\text{;Stat.}}\approx&  \f{cL}{24q\epsilon}\cdot\text{Min}\left[\left(\f{\sin{\left[\f{q\pi (X_1-X_2)}{L}\right]}}{  \prod_{K=1,2}\sin{\left[\f{q\pi (\hat{X}_K)}{L}\right]}}+\f{\sin{\left[\f{q\pi (Y_1-Y_2)}{L}\right]}}{ \prod_{K=1,2}\sin{\left[\f{q\pi (\hat{Y}_K)}{L}\right]}}\right), 2\pi le^{2\theta}\right]\\
    =&\f{cL}{24q\epsilon}\left(\f{\sin{\left[\f{q\pi (X_1-X_2)}{L}\right]}}{  \prod_{K=1,2}\sin{\left[\f{q\pi (\hat{X}_K)}{L}\right]}}+\f{\sin{\left[\f{q\pi (Y_1-Y_2)}{L}\right]}}{ \prod_{K=1,2}\sin{\left[\f{q\pi (\hat{Y}_K)}{L}\right]}}\right)=S^{i=2;\text{T.D.}}_{A_1\text{;Stat.}}+S^{i=2;\text{T.D.}}_{B\text{;Stat.}}.
\end{split}
\ee
Consequently, the time dependence of $S_{A_1\cup B}^{i=2}$ is approximated by
\be
\begin{split}
    S_{A_1\cup B}^{i=2}&\approx\f{c}{6}\log{\left[2^4\prod_{K=1,2}\sin^2{\left(\f{q\pi X_K}{L}\right)}\sin^2{\left(\f{q\pi Y_K}{L}\right)}\right]}+ \f{2c}{3}\log{\left(\f{4\epsilon}{\pi}\right)}\\
    &+\f{c}{6\epsilon}\cdot\begin{cases}
    2\pi t ~&~\text{for}~t_*'>t\gg \epsilon\\
    \f{L}{4q}\cdot\left(\f{\sin{\left[\f{q\pi (X_1-X_2)}{L}\right]}}{  \prod_{K=1,2}\sin{\left[\f{q\pi (\hat{X}_K)}{L}\right]}}+\f{\sin{\left[\f{q\pi (Y_1-Y_2)}{L}\right]}}{ \prod_{K=1,2}\sin{\left[\f{q\pi (\hat{Y}_K)}{L}\right]}}\right) ~&~\text{for}~t>t_*'
    \end{cases}
\end{split},
\ee
where we assume that $t_*'$ is smaller than $L/2$, and
$t_*'$ is defined as 
\be
t_*'\approx\f{L}{8\pi q}\cdot\left(\f{\sin{\left[\f{q\pi (X_1-X_2)}{L}\right]}}{  \prod_{K=1,2}\sin{\left[\f{q\pi (\hat{X}_K)}{L}\right]}}+\f{\sin{\left[\f{q\pi (Y_1-Y_2)}{L}\right]}}{ \prod_{K=1,2}\sin{\left[\f{q\pi (\hat{Y}_K)}{L}\right]}}\right).
\ee
In this case, the mutual information first shows a linear decrease until $t=t^*$ and then becomes zero.

\subsubsection*{Case.\ 3}
Here, we explore the time dependence of $S_{A_1\cup B}^{i=2}$ for Case.\ 3, and assume that $n+l<\f{q}{2}$ for simplicity. In the low temperature region of the large $\theta$ limit, where $\f{L \sin{\left[\f{q\pi (X_1-X_2)}{L}\right]}}{8q\epsilon  \prod_{K=1,2}\sin{\left[\f{q\pi (\hat{X}_K)}{L}\right]}} \ll 1$. The minimal surface of static geodesic is given by
\be
\begin{split}
    S_{A_1\cup B\text{;Stat.}}^{i=2;\text{T.D.}}\approx \text{Min}\Bigg[& \f{c}{3}\log\left(\f{L \sin{\left[\f{q\pi (X_1-X_2)}{L}\right]}}{8q\epsilon  \prod_{K=1,2}\sin{\left[\f{q\pi (\hat{X}_K)}{L}\right]}}\right)+\f{c}{3}\log\left(\sinh\left[\f{\pi l\cdot L }{8q\epsilon}e^{2\theta}\right]\right),\\
    &\f{c}{3}\log\left(\sinh\left[\f{\pi l\cdot L }{8q\epsilon}e^{2\theta}\right]\right)+\f{c}{3}\log\left(\sinh\left[\f{\pi (l+n)\cdot L }{8q\epsilon}e^{2\theta} \right]\right)         \Bigg],
\end{split}
\ee
and the first term is picked up after minimization. Note that $S_{A_1\cup B\text{;Stat.}}^{i=2;\text{T.D.}}$ can be both negative or positive. If it takes a negative value, then
\be
  S_{A_1\cup B}^{i=2}= S_{A_1\cup B}^{i=2;\text{U.}} +S_{A_1\cup B\text{;Stat.}}^{i=2;\text{T.D.}},
\ee
and the mutual information vanishes. In contrast, if $S_{A_1\cup B\text{;Stat.}}^{i=2;\text{T.D.}}>0$, the entanglement entropy is obtained as
\be
\begin{split}
    S_{A_1\cup B}^{i=2}&\approx\f{c}{6}\log{\left[2^4\prod_{K=1,2}\sin^2{\left(\f{q\pi X_K}{L}\right)}\sin^2{\left(\f{q\pi Y_K}{L}\right)}\right]}+ \f{2c}{3}\log{\left(\f{4\epsilon}{\pi}\right)}\\
    &+\f{c}{3\epsilon}\cdot\begin{cases}
    2\epsilon\log{\left[\cosh{\left(\f{\pi t}{2\epsilon}\right)}\right]} ~&~\text{for}~t_*'>t\geq 0\\
    \epsilon\cdot\left[\log\left(\f{L \sin{\left[\f{q\pi (X_1-X_2)}{L}\right]}}{8q\epsilon  \prod_{K=1,2}\sin{\left[\f{q\pi (\hat{X}_K)}{L}\right]}}\right)+\f{\pi l\cdot L }{8q\epsilon}e^{2\theta}\right] ~&~\text{for}~t>t_*',
    \end{cases}
\end{split}
\ee
where $t_*'$ can be determined by solving 
\be
    2\log{\left[\cosh{\left(\f{\pi t_*'}{2\epsilon}\right)}\right]}\approx\log\left(\f{L \sin{\left[\f{q\pi (X_1-X_2)}{L}\right]}}{8q\epsilon  \prod_{K=1,2}\sin{\left[\f{q\pi (\hat{X}_K)}{L}\right]}}\right)+\f{\pi l\cdot L }{8q\epsilon}e^{2\theta}.
\ee
Correspondingly, the mutual information initially decreases monotonically until $t\approx t_*'$, and then it becomes zero. 

Moreover, if we consider high temperature region of the large $\theta$ limit, i.e., $\epsilon\ll 1$ and $\f{L \sin{\left[\f{q\pi (X_1-X_2)}{L}\right]}}{8q\epsilon  \prod_{K=1,2}\sin{\left[\f{q\pi (\hat{X}_K)}{L}\right]}} \gg 1$,
The minimal surface of static geodesic is given by
\be
\begin{split}
    S_{A_1\cup B\text{;Stat.}}^{i=2;\text{T.D.}}\approx \text{Min}\left[ \f{cL \sin{\left[\f{q\pi (X_1-X_2)}{L}\right]}}{24q\epsilon  \prod_{K=1,2}\sin{\left[\f{q\pi (\hat{X}_K)}{L}\right]}}+\f{c\pi l\cdot L }{24q\epsilon}e^{2\theta},~\f{c\pi  (2l+n)\cdot L}{24q\epsilon}e^{2\theta} \right],
\end{split}
\ee
such that the time dependence of entanglement entropy reads
\be
\begin{split}
    S_{A_1\cup B}^{i=2}&\approx\f{c}{6}\log{\left[2^4\prod_{K=1,2}\sin^2{\left(\f{q\pi X_K}{L}\right)}\sin^2{\left(\f{q\pi Y_K}{L}\right)}\right]}+ \f{2c}{3}\log{\left(\f{4\epsilon}{\pi}\right)}\\
    &+\f{c}{6\epsilon}\cdot\begin{cases}
    2\pi t ~&~\text{for}~t_*'>t\gg \epsilon\\
    \f{L}{4q}\cdot\left(\f{ \sin{\left[\f{q\pi (X_1-X_2)}{L}\right]}}{\prod_{K=1,2}\sin{\left[\f{q\pi (\hat{X}_K)}{L}\right]}}+l\pi e^{2\theta}\right) ~&~\text{for}~t>t_*',
    \end{cases}
\end{split}
\ee
where the exchanging time is determined as
\be
    t_*'\approx\f{L}{8\pi q}\cdot\left(\f{ \sin{\left[\f{q\pi (X_1-X_2)}{L}\right]}}{\prod_{K=1,2}\sin{\left[\f{q\pi (\hat{X}_K)}{L}\right]}}+l\pi e^{2\theta}\right).
\ee
In this case, the mutual information evolves downward initially and vanishes when $t\geq t_*'$.

\subsubsection*{Case.\ 4}
Here, we explore the time dependence of $S_{A_2\cup B}^{i=2}$ for Case.\ 4. Within the effective high temperature regime, the minimal surface with respect to static geodesics is given by
\be
\begin{split}
    S_{A_2\cup B\text{;Stat.}}^{i=2;\text{T.D.}}\approx\f{c\pi L e^{2\theta}}{24q\epsilon}\cdot&\text{Min}\left[  \sum_{i=1,2}n_i,~q-\sum_{i=1,2}n_i,~2l+\sum_{i=1,2}n_i,~q-|n_1-n_2|\right]\\
    =&\f{c\pi L e^{2\theta}}{24q\epsilon}\cdot \begin{cases}
        \sum_{i=1,2}n_i~&\text{if}~\sum_{i=1,2}n_i\leq\f{q}{2}\\
        q-\sum_{i=1,2}n_i~&\text{if}~\sum_{i=1,2}n_i>\f{q}{2}.
    \end{cases}
\end{split}
\ee
Therefore, the entanglement entropy is given by
\be
    \begin{split}
    S_{A_2\cup B}^{i=2}&\approx\f{c}{6}\log{\left[2^4\prod_{K=1,2}\sin^2{\left(\f{q\pi X_K}{L}\right)}\sin^2{\left(\f{q\pi Y_K}{L}\right)}\right]}+ \f{2c}{3}\log{\left(\f{4\epsilon}{\pi}\right)}\\
    &+\f{c}{6\epsilon}\cdot\begin{cases}
    2\pi t ~&~\text{for}~t_*'>t\gg \epsilon\\
    \f{\pi L e^{2\theta}}{4q}\cdot (n_1+n_2) ~&~\text{for}~t>t_*'\text{ and }\sum_{i=1,2}n_i\leq\f{q}{2}\\
    \f{\pi L e^{2\theta}}{4q}\cdot (q-n_1-n_2) ~&~\text{for}~t>t_*'\text{ and }\sum_{i=1,2}n_i>\f{q}{2},
    \end{cases}
\end{split}
\ee
where the exchanging time can be obtained as
\be
    t_*'\approx\f{ L e^{2\theta}}{8q}\cdot\begin{cases}
        \sum_{i=1,2}n_i~&\text{if}~\sum_{i=1,2}n_i\leq\f{q}{2}\\
        q-\sum_{i=1,2}n_i~&\text{if}~\sum_{i=1,2}n_i>\f{q}{2}.
    \end{cases}
\ee
The mutual information decreases during $0<t< t_*'$ and becomes zero at $t=t_*'$.

\section{Gravity dual \label{sec:gravity-dual}}

In this section, we will describe the gravity dual of the system considered in this paper, and then holographically calculate the time dependence of entanglement entropy and mutual information.
Let $r_0$ and $L_{\text{AdS}}$ denote the black hole radius and AdS radius, respectively.
Then, we assume that  $L_{\text{AdS}}=1$.
We focus on the system associated with $\mathcal{H}_i$ of $\ket{U_{\text{eff}}}\bra{U_{\text{eff}}}$, and then consider the reduce density matrix associated with $\mathcal{H}_i$, $\rho_{\mathcal{H}_i=1,2}=\Tr_{\mathcal{H}_{j\neq i}} \ket{U_{\text{eff}}}\bra{U_{\text{eff}}}$.
Then, it results in the thermal state on the curved geometry as in \cite{2024arXiv240606121M},
\be \label{eq:rm-oneside}
\rho_{\mathcal{H}_i=1,2} =\f{ e^{-2\epsilon H_{\text{q-M\"obius}}}}{\Tr e^{-2\epsilon H_{\text{q-M\"obius}}}}
\ee
The gravity dual of (\ref{eq:rm-oneside}) is given by the BTZ black hole geometry
\be\label{eq:Euclidean-BTZ-BH-metric}
ds^2=\f{dr^2}{r^2-r_0^2}+\left(r^2-r_0^2\right)d\tau_{\xi_{\tau}}^2+r^2dx_{\xi_{\tau}}^2=\f{dr^2}{r^2-r_0^2}+\left(r^2-r_0^2\right)d\tau^2+r^2dx_{\xi}^2
\ee
where $x_{\xi_{\tau}}=\f{\xi_{\tau}-\bar{\xi}_{\tau}}{2i}=x_{\xi}$, $\tau_{\xi_{\tau}}=\f{\xi_{\tau}+\bar{\xi}_{\tau}}{2}$, $(\xi_{{\tau}},\overline{\xi}_{{\tau}})=(\xi+{{\tau}},\overline{\xi}+{{\tau}})$ and $x_{\xi}\sim x_{\xi}+L_{\text{eff}}$.
In the last equation, we used $\xi+\overline{\xi}=0$.
 After the analytic continuation in (\ref{eq:ac}), the geometry results in the geometry with Lorentzian signature
\be\label{Lorentzian-BTZ-black hole-metric}
    ds^2=\f{dr^2}{r^2-r_0^2}-\left(r^2-r_0^2\right)dt^2+r^2dx_{\xi}^2,
\ee
where  $t\in\mathbb{R}$ is the time associated with $H_{\text{q-M\"obius}}$. 
If we set $r=r_0\cosh\rho$ and re-scale $t\to\tilde{t}=r_0t$, $x_{\xi}\to\tilde{x}_{\xi}=r_0x_{\xi}$, one can obtain the metric for the exterior region of a black hole in the global coordinates, i.e.
\be \label{eq:exterior-geometry-1}
    ds^2_{\text{Exterior}}=d\rho^2-\sinh^2\rho d\tilde{t}^2+\cosh^2\rho d\tilde{x}_{\xi},
\ee
where $\rho\in[0,\infty)$, $\tilde{x}_{\xi}\sim\tilde{x}_{\xi}+r_0L_{\text{eff}}$ and $\tilde{t}\in\mathbb{R}$. 

The location of horizon is at $\rho=0$. We also recall that the Poincar\'e patch reads
\be\label{Poincare-patch}
    ds^2=\f{dz^2-dx_0^2+dx_1^2}{z^2},
\ee
which is related to the exterior metric by the following relations
\be\label{First-exterior to Poincare Patch}
\begin{aligned}
    &z=e^{\tilde{x}_{\xi}}\f{r_0}{r}=e^{\tilde{x}_{\xi}}\text{sech}\rho\in\mathbb{R}^+,\\
    &x_0=e^{\tilde{x}_{\xi}}\sinh\tilde{t}\tanh\rho\in\mathbb{R},\\
    &x_1=e^{\tilde{x}_{\xi}}\cosh\tilde{t}\tanh\rho\in[0,\infty).
\end{aligned}
\ee
In Poincar\'e patch, $z(\rho\to\infty)=0$ and $z(\rho=0)=e^{\tilde{x}_{\xi}}$ represent the conformal boundary and horizon, respectively. Near the conformal boundary , $\rho\to\infty$
, one can find
\be
    x_1\pm x_0\approx e^{\tilde{x}_{\xi}\pm \tilde{t}},~\f{1}{z}\approx \f{e^{\rho-\tilde{x}_{\xi}}}{2},
\ee
which also lead to
\be\label{BTZ-black hole-Exterior-patch}
    \text{Exterior:}~~x_0^2-x_1^2\leq 0.
\ee
Note that (\ref{eq:exterior-geometry-1}) describes one of exterior regions, i.e., $x_1\geq 0$. The second exterior region, $x_1\leq 0$, is mapped to \eqref{Poincare-patch} by applying the analytic continuation $\tilde{t}\to-\tilde{t}+i\pi$, and we have
\be\label{Second-exterior to Poincare Patch}
\begin{aligned}
    &z=e^{\tilde{x}_{\xi}}\f{r_0}{r}=e^{\tilde{x}_{\xi}}\text{sech}\rho\in\mathbb{R}^+,\\
    &x_0=e^{\tilde{x}_{\xi}}\sinh\tilde{t}\tanh\rho\in\mathbb{R},\\
    &x_1=-e^{\tilde{x}_{\xi}}\cosh\tilde{t}\tanh\rho\in(-\infty,0].
\end{aligned}
\ee

Then, let us consider the patch covering the interior of the black hole\footnote{Here, by ``interior'' we mean the interior living in the future direction.}. By assuming that the metric is continuous when it goes through the horizon, and thus applying the continuation $\rho=ip$, $\tilde{t}=T-\f{i\pi}{2}$, and $p,T\in\mathbb{R}$, one obtains the interior metric
\be
    ds^2_{\text{Interior}}=-dp^2+\sin^2p dT^2+\cos^2p d\tilde{x}_{\xi}^2,
\ee
which is related to the Poincar\'e patch by
\be\label{Interior to Poincare Patch}
\begin{aligned}
    &z=e^{\tilde{x}_{\xi}}\sec p\in\mathbb{R},\\
    &x_0=e^{\tilde{x}_{\xi}}\cosh T\tan p\in\mathbb{R},\\
    &x_1=e^{\tilde{x}_{\xi}}\sinh T\tan p\in\mathbb{R}.
\end{aligned}
\ee
The interior metric covers the patch
\be\label{BTZ-black hole-Interior-patch}
    \text{Interior:}~~0 \leq x_0^2-x_1^2\leq z^2,
\ee
and meets the conformal boundary only along the light cone $x_0^2-x_1^2=0$.

\begin{figure}
    \centering
    \includegraphics[width=0.5\linewidth]{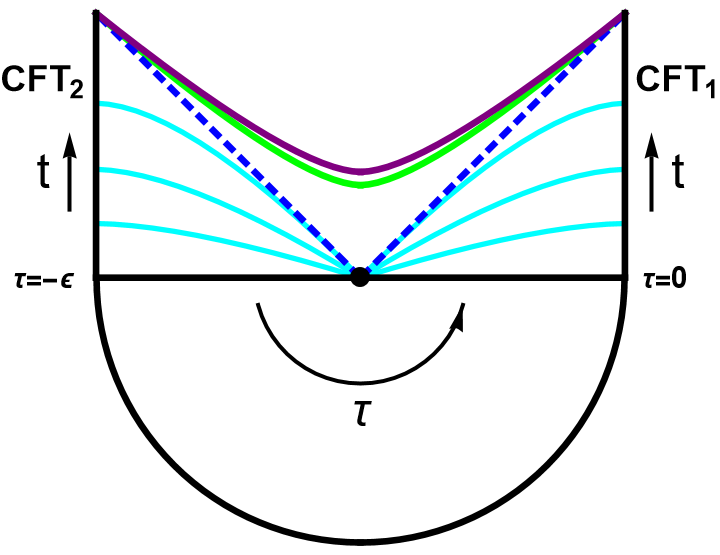}
    \caption{The gravity dual geometry of state $\ket{U_{\text{eff}}}$. $\tau$ represents the Euclidean time evolution, and $t$ is the direction of Lorentzian time evolution. The cyan, purple, blue and green curves represent the constant time slices, the black hole singularity, the event horizon, and the large $t$ behavior of the minimal surface related to the holographic entanglement entropy for sub-regions anchored on the conformal boundary \cite{Hartman:2013qma}, respectively.}
    \label{fig:Dual geometry of Ueff state}
\end{figure}

Finally, since $x_{\xi}=x_{\xi}(x)$, the metirc describing the interior and exterior of the black hole is given by
\be\label{BTZ-black hole-Exterior and Interior-metric}
    ds^2=\begin{cases}
        -dp^2+\sin^2p dT^2+\f{\cos^2p}{f(x)^2} dx^2~&\text{Interior}\\
        d\rho^2-\sinh^2\rho d\tilde{t}^2+\f{\cosh^2\rho}{f(x)^2} dx^2~&\text{Exterior}
    \end{cases},
\ee
where the envelop function is defined in (\ref{eq:envelopfunction}). The gravity dual geometry of state $\ket{U_{\text{eff}}}$ is shown in Figure \ref{fig:Dual geometry of Ueff state}.

\subsection{Asymptotic boundary metric from bulk geometry}
To reach the conformal boundary, one can take the limits $z\to 0$ for \eqref{Poincare-patch},  $r\to\infty$ for \eqref{Lorentzian-BTZ-black hole-metric} and $\rho\to\infty$ for \eqref{eq:exterior-geometry-1}, respectively. In addition, due to the fact that $z=e^{\tilde{x}_{\xi}}\f{r_0}{r}=e^{\tilde{x}_{\xi}}\text{sech}\rho$, these limits are related to each other. Let us focus on \eqref{Lorentzian-BTZ-black hole-metric} for a while. Notice that the metric has a second-order divergence at $r\to\infty$. As a result, the bulk metric does not directly yield a unique induced metric on the boundary, but gives rise to a family of boundary metrics so-called conformal structure \cite{Skenderis:2002wp}. To pick up a specific boundary metric, we need to define a ``defining function'' $D(r)$ that its inversion $D(r)^{-1}$ has a first-order divergence with respect to $r$ at $r\to\infty$ \cite{Skenderis:2002wp}. There are two constraints on $D(r)$. First, $D(r)^{-1}$ has no zeros or divergences on the boundary, $r\to\infty$. Second, $D(r)$ is positive in the bulk. Then, by choosing a proper $D(r)$ and multiplying \eqref{Lorentzian-BTZ-black hole-metric} by $D(r)^2$ and evaluating it at the conformal limit $r\to\infty$, one can uniquely determine a boundary metric from the bulk AdS metric. For our purpose, we choose $D(r)=\f{1}{r}$ such that the boundary metric is obtained as
\be\label{Effective-boundary metric}
    ds^2|_{\text{Bdy.}}=r_0^2\lim_{r\to\infty}D(r)^2ds^2=\lim_{r\to\infty}r_0^2r^{-2}ds^2=-dt^2+dx_{\xi}^2,
\ee
where $ds^2$ is given by \eqref{Lorentzian-BTZ-black hole-metric}. We take \eqref{Effective-boundary metric} as our boundary metric, which fits the quasi-particle description discussed in Section \ref{sec:interpretation}.

Furthermore, as we mentioned above, the conformal limits $z\to 0$ for Poincar\'e coordinates, $r\to\infty$ for standard BTZ coordinates, and $\rho\to\infty$ for the metric of the exterior region are related to each other. This relationship helps us define the UV regulator $\delta_{\text{UV}}$ in the Poincar\'e patch at $z\to\infty$ and IR regulator $\rho_{\text{IR}}$ in the exterior coordinates at $\rho\to\infty$, i.e.
\be\label{eq:def-regulators}
    \delta_{\text{UV}}=\text{sech}\rho_{\text{IR}}:=\lim_{r\to\infty}D(r)\to\infty,
\ee
which do not depend on the induced coordinates $(t,x_{\xi})$ at the conformal boundary.

To make the cutoff independent of the spatial location, i.e. the cutoff which we consider, we need to consider another ``defining function'', $\tilde{D}(r)=f(x)D(r)$. Recall that $x_{\xi}=\f{\xi-\bar{\xi}}{2i}$ and $dx_{\xi}^2=\f{dx^2}{f(x,\theta)^2}$, the BTZ black hole metric becomes
\be
\begin{split}
    ds^2=&\f{dr^2}{r^2-r_0^2}-(r^2-r_0^2)dt^2+\f{r^2}{f(x)^2}dx^2\\
    =&\f{dr^2}{r^2-r_0^2}+\f{r^2}{f(x)^2}\left[-(1-\f{r_0^2}{r^2})f(x)^2dt^2+dx^2\right],
\end{split}
\ee
where $x\sim x+L$. Similar to \eqref{Effective-boundary metric}, we can obtain a boundary metric by using $\tilde{D}(r)$ as
\be\label{Effective-boundary metric-2}
    ds^2|_{\text{Bdy.}}=\lim_{r\to\infty}r_0^2{\tilde{D}(r)}^2ds^2=\lim_{r\to\infty}r_0^2 f(x)^2r^{-2}ds^2=-f(x)^2dt^2+dx^2,
\ee
and define the regulators as $\delta_{\text{UV}}'=\text{sech}\rho_{\text{IR}}':=\lim_{r\to\infty}\tilde{D}(r)\to\infty$. These new expressions of regulators are related to the old ones, $\delta_{\text{UV}}=\text{sech}\rho_{\text{IR}}$, by
\be\label{eq:relations-btw-regulators}
    \delta_{\text{UV}}'=f(x)\delta_{\text{UV}}=\text{sech}\rho_{\text{IR}}'=f(x)\text{sech}\rho_{\text{IR}}.
\ee
This is significant in our later computations. In addition, we also define the cutoff
\be\label{eq:re-scaled-cutoff}
    \tilde{\delta}_{\text{UV}}=r_0\delta_{\text{UV}}=\text{sech}\tilde{\rho}_{\text{IR}}=r_0\text{sech}\rho_{\text{IR}}
\ee
in the coordinate system $(\tilde{t},\tilde{x}_{\xi})$ by multiplying a factor $r_0$, since we have re-scaled $(t,x_{\xi})$ to $\left(\tilde{t}=r_0t,\tilde{x}_{\xi}=r_0x_{\xi}\right)$.

\if[0]
\textcolor{red}{The conformal boundary can be obtained by multiplying \eqref{Poincare-patch} by a ``defining function'' squared with a first order pole on the boundary and sending $z\to 0$ \cite{Skenderis:2002wp}. For example, one can obtain the boundary metric as where the chosen defining function is $r_0r^{-1}$ due to the fact $z\approx\f{r_0}{r}$.}

We take \eqref{Effective-boundary metric} as our boundary metric, which fits the quasi-particle description discussed in Section \ref{sec:interpretation}. 
Note that originally we chose spatial locations of subsystem as $\{x\}\subset[0,L]$, and we have
\be\label{Effective x-coordinate}
    x_{\xi}(x)=\int_0^x~dx'\f{1}{f(\theta,x')}=\f{L_{\text{eff}}}{q\pi}\arctan\left(e^{2\theta}\tan\left[\f{q\pi x}{L}\right]\right).
\ee
Therefore, if we choose a subsystem with size $l_{A_j}=x_1-x_2\leq L$, we can define the effective subsystem size as
\be\label{Effective subsystem size}
\begin{aligned}
    l_{A_j}^{\text{effective}}&=\int_{x_{\xi}(x_2)}^{x_{\xi}(x_1)}dx_{\xi}=\f{L_{\text{eff}}}{q\pi}\left[\arctan\left(e^{2\theta}\tan\left[\f{q\pi x_1}{L}\right]\right)-\arctan\left(e^{2\theta}\tan\left[\f{q\pi x_2}{L}\right]\right)\right]\\
    &\overset{\theta\gg 1}{\approx}\begin{cases}
        \f{L}{2q\pi}\f{\sin\left[\f{q\pi(x_1-x_2)}{L}\right]}{\prod_{K=1}^2\sin\left[\f{q\pi X_K}{L}\right]}~&\text{no fixed point between $x_2$ and $x_1$},\\
        \f{l}{q}L_{\text{eff}}~&\text{$l$-many fixed points between $x_2$ and $x_1$}.
    \end{cases}
\end{aligned}
\ee
More precisely, we define the effective length between any two spatial locations $X,Y\in[0,L]$ as
\be\label{Effective length}
    l^{\text{effective}}(X,Y):=|x_{\xi}(X)-x_{\xi}(Y)|\leq L_{\text{eff}}.
\ee
\fi

\subsection{Lorentzian computation on the holographic entanglement entropy}
We assume that $A_{j=1.2}$ is the spatial interval between $X_2\leq x\leq X_1$ at $\f{t}{2}$ on the boundary of the first exterior, while $B_{j=1.2}$ is the one between $Y_2\leq x\leq Y_1$ on the boundary of the second exterior. Moreover, in this section, we consider the effective high temperature region, $\tau_{\text{eff}}>1$.

To compute entanglement entropy and mutual information, we need to compute the geodesic lengths of minimal surfaces. 
From \eqref{First-exterior to Poincare Patch} and \eqref{Second-exterior to Poincare Patch}, the relations between global and Poincar\'e coordinates near the boundary, $\rho=\tilde{\rho}_{\text{IR}}\to\infty$
, are given by
\be\label{Exterior to Poincare Patch Near Conformal Boundary}
\begin{aligned}
    \text{First exterior: }&z^b\approx 2e^{\tilde{x}_{\xi}-\tilde{\rho}_{\text{IR}}},~x_0^b\approx e^{\tilde{x}_{\xi}}\sinh\tilde{t},~x_1^b\approx e^{\tilde{x}_{\xi}}\cosh\tilde{t}\\
    \text{Second exterior: }&z^b\approx 2e^{\tilde{x}_{\xi}-\tilde{\rho}_{\text{IR}}},~x_0^b\approx e^{\tilde{x}_{\xi}}\sinh\tilde{t},~x_1^b\approx -e^{\tilde{x}_{\xi}}\cosh\tilde{t},
\end{aligned}
\ee
where the IR cutoff satisfies $\text{sech}\tilde{\rho}_{\text{IR}}=r_0\text{sech}\rho_{\text{IR}}$. The endpoints of $A_j$ on the boundary of the first exterior are denoted by $E_1,E_2$, and the endpoints of $B_k$ on the boundary of the second exterior are $E_3,E_4$. They are mapped to $E_{i}'$ by \eqref{Exterior to Poincare Patch Near Conformal Boundary}
\be\label{eq:Endpoints-locations-Ueff}
\begin{aligned}
    &E_1:~\left(\tilde{\rho}_{\text{IR}},\f{\tilde{t}}{2},\tilde{x}_{\xi}\left(X_2\right)\right)\to E_1':~\left(2e^{\tilde{x}_{\xi}\left(X_2\right)-\tilde{\rho}_{\text{IR}}},e^{\tilde{x}_{\xi}\left(X_2\right)}\sinh\f{\tilde{t}}{2},e^{\tilde{x}_{\xi}\left(X_2\right)}\cosh\f{\tilde{t}}{2}\right),\\
    &E_2:~\left(\tilde{\rho}_{\text{IR}},\f{\tilde{t}}{2},\tilde{x}_{\xi}\left(X_1\right)\right)\to E_2':~\left(2e^{\tilde{x}_{\xi}\left(X_1\right)-\tilde{\rho}_{\text{IR}}},e^{\tilde{x}_{\xi}\left(X_1\right)}\sinh\f{\tilde{t}}{2},e^{\tilde{x}_{\xi}\left(X_1\right)}\cosh\f{\tilde{t}}{2}\right),\\
    &E_3:~\left(\tilde{\rho}_{\text{IR}},\f{\tilde{t}}{2},\tilde{x}_{\xi}\left(Y_2\right)\right)\to E_3':~\left(2e^{\tilde{x}_{\xi}\left(Y_2\right)-\tilde{\rho}_{\text{IR}}},e^{\tilde{x}_{\xi}\left(Y_2\right)}\sinh\f{\tilde{t}}{2},-e^{\tilde{x}_{\xi}\left(Y_2\right)}\cosh\f{\tilde{t}}{2}\right),\\
    &E_4:~\left(\tilde{\rho}_{\text{IR}},\f{\tilde{t}}{2},\tilde{x}_{\xi}\left(Y_1\right)\right)\to E_4':~\left(2e^{\tilde{x}_{\xi}\left(Y_1\right)-\tilde{\rho}_{\text{IR}}},e^{\tilde{x}_{\xi}\left(Y_1\right)}\sinh\f{\tilde{t}}{2},-e^{\tilde{x}_{\xi}\left(Y_1\right)}\cosh\f{\tilde{t}}{2}\right),
\end{aligned}
\ee
where $\tilde{\rho}_{\text{IR}}\to \infty$ represents the IR-regulator for the bulk geometry. 
This can be thought of as the UV cutoff which is a parameter introduced to tame the UV divergence, 
\be \label{eq:relation-between-cutoffs}
\tilde{\delta}_{\text{UV}} \approx 2e^{-\tilde{\rho}_{\text{IR}}}\to 0,
\ee 
which is defined in (\ref{eq:re-scaled-cutoff}).

The intervals between $E_1'$ and $E_2'$ and that between $E_3'$ and $E_4'$ are intervals in the first- and second-exterior, respectively.
The static 
geodesics connecting $E_1'$ and $E_2'$ and that connecting $E_3'$ and $E_4'$ are obtained from \eqref{Poincare-patch} as
\be
\begin{aligned}
    &\text{Geodesic between $E_1'$ and $E_2'$: }z^2+\left(x_1-u\right)^2=C^2,\\
    &\text{Geodesic between $E_3'$ and $E_4'$: }z^2+\left(x_1-v\right)^2=D^2,~\text{where}\\
    &u=\left(\f{e^{\tilde{x}_{\xi}(X_1)}+e^{\tilde{x}_{\xi}(X_2)}}{2\tanh\rho}\right),~v=-\left(\f{e^{\tilde{x}_{\xi}(Y_1)}+e^{\tilde{x}_{\xi}(Y_2)}}{2\tanh\rho}\right),\\
    &C^2=\f{e^{2\tilde{x}_{\xi}(X_1)}+e^{2\tilde{x}_{\xi}(X_2)}}{2\cosh^2\rho}+\f{\left(e^{\tilde{x}_{\xi}(X_2)}-e^{\tilde{x}_{\xi}(X_1)}\right)^2}{4}\tanh^2\rho+\f{\left(e^{\tilde{x}_{\xi}(X_1)}+e^{\tilde{x}_{\xi}(X_2)}\right)^2}{4\sinh^2\rho\cosh^2\rho},\\
    &D^2=\f{e^{2\tilde{x}_{\xi}(Y_1)}+e^{2\tilde{x}_{\xi}(Y_2)}}{2\cosh^2\rho}+\f{\left(e^{\tilde{x}_{\xi}(Y_2)}-e^{\tilde{x}_{\xi}(Y_1)}\right)^2}{4}\tanh^2\rho+\f{\left(e^{\tilde{x}_{\xi}(Y_1)}+e^{\tilde{x}_{\xi}(Y_2)}\right)^2}{4\sinh^2\rho\cosh^2\rho}.
\end{aligned}
\ee
Then let $\mathcal{L}_{12}$ and $\mathcal{L}_{34}$ be the geodesic length for $A_{j=1,2}$ and $B_{j=1,2}$, respectively.
In the limit, where  $\tilde{\delta}_{\text{UV}}=2e^{-\tilde{\rho}_{\text{IR}}}\ll 1$, these geodesic lengths reduce to
\be\label{Geodesic length of disconnected pieces}
\begin{aligned}
    \mathcal{L}_{12}\approx 2\log\left(\f{2\sinh\left[\f{\tilde{x}_{\xi}(X_1)-\tilde{x}_{\xi}(X_2)}{2}\right]}{\tilde{\delta}_{\text{UV}}}\right),~\mathcal{L}_{34}\approx 2\log\left(\f{2\sinh\left[\f{\tilde{x}_{\xi}(Y_1)-\tilde{x}_{\xi}(Y_2)}{2}\right]}{\tilde{\delta}_{\text{UV}}}\right),
\end{aligned}
\ee
Then, let $\mathcal{L}_{13}$ and $ \mathcal{L}_{24}$ denote the lengths of geodesics connecting $E_1'$ and $E_3'$, and the one connecting $E_2'$ and $E_4'$, respectively.
In the limit, where  $\tilde{\delta}_{\text{UV}}=2e^{-\tilde{\rho}_{\text{IR}}}\ll 1$, these geodesic lengths reduce to
\be\label{Geodesic length of connected pieces}
\begin{aligned}
    \mathcal{L}_{13}&\approx \log\left(2\f{\cosh\left[\tilde{x}_{\xi}(Y_2)-\tilde{x}_{\xi}(X_2)\right]+\cosh\tilde{t}}{\tilde{\delta}_{\text{UV}}^2}\right),\\
    \mathcal{L}_{24}&\approx \log\left(2\f{\cosh\left[\tilde{x}_{\xi}(Y_1)-\tilde{x}_{\xi}(X_1)\right]+\cosh\tilde{t}}{\tilde{\delta}_{\text{UV}}^2}\right).
\end{aligned}
\ee

Finally, the two-interval entanglement entropy can be computed as
\be\label{Gravitational computation of two-interval EE}
    S_{A_j\cup B_k}^{i=1}=\f{\text{Min}\{\mathcal{L}_{13}+\mathcal{L}_{24},\mathcal{L}_{12}+\mathcal{L}_{34}\}}{4G_N},~G_N=\f{3}{2c},
\ee
and mutual information reads
\be\label{Gravitational computation of Mutual information}
    I_{A_j, B_k}^{i=1}=\f{\text{Max}\{\mathcal{L}_{12}+\mathcal{L}_{34}-\mathcal{L}_{13}-\mathcal{L}_{24},0\}}{4G_N}.
\ee
After mapping $(\tilde{t}, \tilde{x}_{\xi},\tilde{\delta}_{\text{UV}})$ to $(t, x_{\xi},\delta_{\text{UV}})$, and using the fact that $r_0=\f{\pi}{\epsilon}$, we can confirm that the time dependence of entanglement entropy and mutual information obtained in Lorentzian computation is consistent with that in the Euclidean computation.

For simplicity, we will call $\mathcal{L}_{13},\mathcal{L}_{24}$ and $\mathcal{L}_{12},\mathcal{L}_{34}$ as the lengths of ``connected-'' and ``disconnected-geodesics'', respectively. If there exists an exchanging time (must be non-negative) between them denoted by $t^*$, it can be solved by the following equality
\be\label{Lorentzian computation-exchanging time}
    \mathcal{L}_{13}|_{t=t^*}+\mathcal{L}_{24}|_{t=t^*}=\mathcal{L}_{12}+\mathcal{L}_{34}.
\ee
\subsubsection{Single-interval cases}
We begin by holographically computing the entanglement entropy for the single interval as the geodesic length on the background with Loretizan signature, and then check if this Lorentzian calculation is consistent with the computations done in Section \ref{Sec:Etanglement-dynamics-Ueff}.

Combining \eqref{eq:relations-btw-regulators}, \eqref{eq:re-scaled-cutoff} and \eqref{Geodesic length of disconnected pieces}, and noting that $\theta\gg 1$, the entanglement entropy for $A_1$ is given by
\be
\begin{split}
    S_{A_1}^{i=1}\approx& \f{c}{3}\log\left(\f{2\epsilon\sinh\left[\f{L}{4q\epsilon}\f{\sin\left[\f{q\pi(X_1-X_2)}{L}\right]}{\sin\left[\f{q\pi X_1}{L}\right]\sin\left[\f{q\pi X_2}{L}\right]}\right]}{\pi\delta_{\text{UV}}'\sqrt{f(X_1)f(X_2)}}\right)\approx\f{c}{3}\log\left(\f{2\epsilon}{\pi\delta_{\text{UV}}'}\right)+\f{c}{3}\log\left(2\prod_{K=1,2}\sin\left[\f{q\pi X_K}{L}\right]\right)\\
    &+\begin{cases}
        \f{cL}{12q\epsilon}\f{\sin\left[\f{q\pi(X_1-X_2)}{L}\right]}{\sin\left[\f{q\pi X_1}{L}\right]\sin\left[\f{q\pi X_2}{L}\right]}~&\text{for}~\f{L}{4q\epsilon}\f{\sin\left[\f{q\pi(X_1-X_2)}{L}\right]}{\sin\left[\f{q\pi X_1}{L}\right]\sin\left[\f{q\pi X_2}{L}\right]}\gg 1,\\
        \f{c}{3}\log\left(\f{L}{4q\epsilon}\f{\sin\left[\f{q\pi(X_1-X_2)}{L}\right]}{\sin\left[\f{q\pi X_1}{L}\right]\sin\left[\f{q\pi X_2}{L}\right]}\right)~&\text{for}~\f{L}{4q\epsilon}\f{\sin\left[\f{q\pi(X_1-X_2)}{L}\right]}{\sin\left[\f{q\pi X_1}{L}\right]\sin\left[\f{q\pi X_2}{L}\right]}\ll 1,
    \end{cases}
\end{split}
\ee
which is consistent with the entanglement entropy computed in the path-integral formalism \eqref{eq:single-interval-EE}. 
Similarly, for $A_2$ we have
\be
\begin{split}
    S_{A_2}^{i=1}\approx&\f{c}{3}\log\left(\f{2\epsilon\sinh\left[\f{l\pi}{2q\epsilon}L_{\text{eff}}\right]}{\pi\delta_{\text{UV}}'\sqrt{f(X_1)f(X_2)}}\right)\approx\f{c}{3}\log\left(\f{2\epsilon}{\pi\delta_{\text{UV}}'}\right)+\f{c}{3}\log\left(2\prod_{K=1,2}\sin\left[\f{q\pi X_K}{L}\right]\right)+\f{c\pi l\cdot L}{12q\epsilon}e^{2\theta},
\end{split}
\ee
which is also consistent with $S_{A_2}^{i=1}$ obtained in the path-integral formalism in the large $\theta$ limit, \eqref{eq:single-interval-EE}. 
\subsection{Two-interval cases:}
Now, we will calculate the holographic entanglement entropy for the two spatial intervals as the geodesic on the background with Lorentzian signature, and then check if this is consistent with the entanglement entropy obtained in the path-integral formalism.
We will calculate the holographic entanglement entropy for all configurations considered in Section \ref{Sec:Etanglement-dynamics-Ueff}.
In this section, we will report on the time dependence of the holographic entanglement entropy for the symmetric setup, and those in the other setups considered are postponed to Appendix \ref{sec:LC-on-MI}.

\subsubsection{Symmetric case}
Let us check if the Lorentzian calculation on the holographic entanglement entropy for the symmetric case is consistent with the entanglement entropy obtained in the path-integral formalism.
In the symmetric case, the locations of the edges and center of $A_{j=1,2}$ is the same as those of $B_{k=1,2}$, i.e. $Y_K=X_K$ for $K=1,2$. In this case, from \eqref{eq:re-scaled-cutoff} and \eqref{Geodesic length of connected pieces} we obtain
\be
    \mathcal{L}_{13}=\mathcal{L}_{24}\approx 2\log\left(\f{2\epsilon\cosh\left[\f{\pi t}{\epsilon}\right]}{\delta_{\text{UV}}'\pi\sqrt{f(X_1)f(X_2)}}\right),
\ee
where $\delta'_{\text{UV}}$ is given by \eqref{eq:relations-btw-regulators}. Thus, the entanglement entropy for $A_1\cup B_1$ is given by
\be
\begin{aligned}
     S_{A_1\cup B_1}^{i=1}\approx&\f{c}{6}\cdot\text{Min}\{\mathcal{L}_{13}+\mathcal{L}_{24},\mathcal{L}_{12}+\mathcal{L}_{34}\}\approx\f{2c}{3}\log\left(\f{2\epsilon}{\pi\delta_{\text{UV}}'}\right)+\f{2c}{3}\log\left(2\prod_{K=1,2}\sin\left[\f{q\pi X_K}{L}\right]\right)\\
     &+\f{2c}{3}\cdot\begin{cases}
         \log\left({\cosh\left[\f{\pi t}{2\epsilon}\right]}\right)~&t\leq t^*\\
         \log\left({\sinh\left[\f{L}{4q\epsilon}\f{\sin\left[\f{q\pi(X_1-X_2)}{L}\right]}{\sin\left[\f{q\pi X_1}{L}\right]\sin\left[\f{q\pi X_2}{L}\right]}\right]}\right)~&t> t^*,
     \end{cases}
\end{aligned}
\ee
which is same as the one given in \eqref{eq:TFD-two-interval-symmetric-EE-candidates}.
The exchanging time $t^*$ is given by
\be
\begin{aligned}
    \log\left({\cosh\left[\f{\pi t^*}{2\epsilon}\right]}\right)=\log\left({\sinh\left[\f{L}{4q\epsilon}\f{\sin\left[\f{q\pi(X_1-X_2)}{L}\right]}{\sin\left[\f{q\pi X_1}{L}\right]\sin\left[\f{q\pi X_2}{L}\right]}\right]}\right).
\end{aligned}
\ee
Correspondingly, the time dependence of the mutual information is given by
\be
\begin{aligned}
    I_{A_1\cup B_1}^{i=1}\approx\f{2c}{3}\begin{cases}
         \log\left({\sinh\left[\f{L}{4q\epsilon}\f{\sin\left[\f{q\pi(X_1-X_2)}{L}\right]}{\sin\left[\f{q\pi X_1}{L}\right]\sin\left[\f{q\pi X_2}{L}\right]}\right]}\right)-\log\left({\cosh\left[\f{\pi t}{2\epsilon}\right]}\right)~&t\leq t^*,\\
         0~&t> t^*.
     \end{cases}
\end{aligned}
\ee
Similarly, the entanglement entropy, $S_{A_2\cup B_2}^{i=1}$, results in 
\be
\begin{split}
    S_{A_2\cup B_2}^{i=1}\approx&\f{2c}{3}\log\left(\f{2\epsilon}{\pi\delta_{\text{UV}}'}\right)+\f{2c}{3}\log\left(2\prod_{K=1,2}\sin\left[\f{q\pi X_K}{L}\right]\right)\\
    &+\f{2c}{3}\begin{cases}
         \log\left({\cosh\left[\f{\pi t}{2\epsilon}\right]}\right)~&t\leq t^*,\\
         \log\left({\sinh\left[\f{\pi l\cdot L}{4q\epsilon}e^{2\theta}\right]}\right)\approx \f{\pi l\cdot L}{4q\epsilon}e^{2\theta}~&t> t^*,
     \end{cases}
\end{split}
\ee
where the exchanging time is approximately given by
\be
    \cosh\left[\f{\pi t^*}{2\epsilon}\right]\approx \sinh\left[\f{\pi l\cdot L}{4q\epsilon}e^{2\theta}\right].
\ee
The mutual information reads
\be
    I_{A_2, B_2}^{i=1}\approx\f{2c}{3}\begin{cases}
         \log\left({\sinh\left[\f{\pi l\cdot L}{4q\epsilon}e^{2\theta}\right]}\right)-\log\left({\cosh\left[\f{\pi t}{2\epsilon}\right]}\right)~&t\leq t^*,\\
         0~&t> t^*.
     \end{cases}
\ee
The above results are consistent with our previous CFT computations, i.e. \eqref{EE-with-exchange of dominance} and \eqref{MI-with-exchange of dominance}.
\subsection{Computations on Lorentzian spacetime for the boundary state}
\begin{figure}
    \centering
    \includegraphics[width=0.3\linewidth]{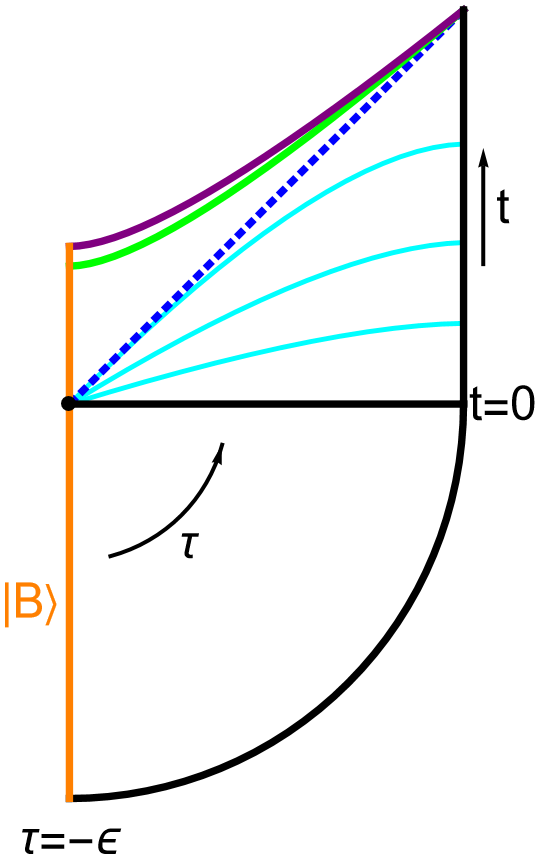}
    \caption{The holographic dual of $\ket{\Phi}$. The boundary state $\ket{B}$ is prepared for an Euclidean time $\epsilon$. The cyan, orange, purple, blue and green curves represent the constant time slices, the end-of-world-brane imposed by the boundary state, the black hole singularity, the event horizon, and the large $t$ behavior of the minimal surface  related to the holographic entanglement entropy for sub-regions anchored on the conformal boundary \cite{Hartman:2013qma}, respectively.}
    \label{fig:Dual geometry of Bdy State}
\end{figure}

\begin{figure}
    \centering
    \includegraphics[width=0.5\linewidth]{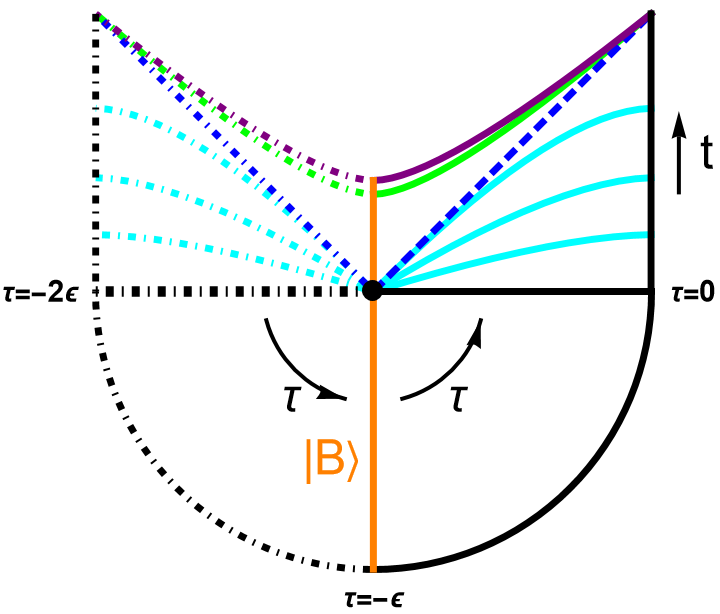}
    \caption{The holographic dual of $\ket{\Phi}$ in the method of image. The dot-dashed elements are the images of the ones in Figure \ref{fig:Dual geometry of Bdy State}. The cyan, orange, purple, blue and green curves represent the constant time slices, the end-of-world-brane imposed by the boundary state, the black hole singularity, the event horizon and the large $t$ behavior of the minimal surface, respectively.}
    \label{fig:Dual geometry of Bdy State Image}
\end{figure}

Now, we will consider the gravity dual of \eqref{eq:Bdy-state} and holographically calculate the time dependence of entanglement entropy. 
In the above, we have considered the dual bulk geometry of \eqref{eq:rm-oneside}. The state $\ket{U_{\text{eff}}}$ is the purification of \eqref{eq:rm-oneside}, which can be regarded as the thermofield double state whose Hamiltonian is given by the q-M\"obius Hamiltonian, $H_{\text{q-M\"obius}}$. It has been shown that the boundary state construction of the global quench \eqref{eq:Bdy-state} can be holographically recovered by considering the thermofield double state \cite{Hartman:2013qma,Rangamani:2016dms}. Therefore, the entanglement dynamics for \eqref{eq:Bdy-state} can be holographically calculated using the above discussions on the state $\ket{U_{\text{eff}}}$ as well.
The boundary state in \eqref{eq:Bdy-state} can be constructed by applying an identification $\mathcal{F}:\mathcal{H}_2\to\mathcal{H}_1$ to $\ket{U_{\text{eff}}}$, such that the state becomes a pure state only in $\mathcal{H}_1$ \cite{Hartman:2013qma,Rangamani:2016dms}. This is identified as state \eqref{eq:Bdy-state} with the replacements $\epsilon \to 2\epsilon$ and $t\to 2t$, which will be explained later. Correspondingly, the holographic dual of \eqref{eq:Bdy-state} is obtained by cutting off the spacetime with an end-of-world-brane, as shown by the orange curve in Figure \ref{fig:Dual geometry of Bdy State}.
Then, after applying the method of images, the geometry becomes identical to the gravity dual of $\ket{U_{\text{eff}}}$ except for the insertion of the end-of-world-brane and the replacements $\epsilon\to 2\epsilon$ and $t\to 2t$, as shown in Figure \ref{fig:Dual geometry of Bdy State Image}.
Let us explain the reasons for these replacements in detail. Through the $\mathbb{Z}_2$ projection map $\mathcal{F}$, one can prepare a state for a global quench from the time dependent thermofield double state as
\be
\begin{split}
    \mathcal{F}:~\ket{\text{TFD}(t)}=\f{1}{\sqrt{Z(\beta)}}\sum_{a}e^{-\left(\f{\beta}{2} +2it\right)H}\ket{E_a}_1\ket{E_a}_2~\mapsto~\mathcal{N}_{\beta}e^{-\left(\f{\beta}{4}+it\right)H}\ket{B},
\end{split}  
\ee
where $Z(\beta)=\Tr(e^{-\beta H})$ and $\mathcal{N}_{\beta}^{-1}=\sqrt{\bra{B}e^{-\f{\beta}{2}H}\ket{B}}$, $H$ is some Hamiltonian (in our case, $H=H_{\text{q-M\"obius}}$), $\beta$ denotes the inverse temperature (in our case, $\beta=2\epsilon$ for state $\ket{U_{\text{eff}}}$) and states $\ket{E_a}_{1,2}\in\mathcal{H}_{1,2}$ are the eigenbasis of $H$ \cite{Rangamani:2016dms}. Therefore, after applying the map $\mathcal{F}$ to \eqref{eq:dual-state-of-eff-H}, we obtain $\mathcal{N}_{\epsilon}e^{-\left(\f{\epsilon+it}{2}\right)H_{\text{q-M\"obius}}}\ket{B}$, where the normalization factor is $\mathcal{N}_{\epsilon}=1/\sqrt{\bra{B}e^{-\epsilon H_{\text{q-M\"obius}}}\ket{B}}$. Comparing this state with \eqref{eq:Bdy-state}, we can find that the replacements $\epsilon\to 2\epsilon$ and $t\to 2t$ are necessary. Hence, except for replacing $\epsilon\to 2\epsilon,~t\to 2t$, one can follow the exact same methods used for the state $\ket{U_{\text{eff}}}$ to calculate the time dependence of entanglement entropy and mutual information for $\ket{\Phi(t)}$ as well.
Here, we present the detailed computation only for the single-interval cases, i.e., $A_{j=1,2}$, because the generalization to any two-interval case is straight forward. Instead of providing detailed computations, we will briefly outline the calculations for two-interval cases in the end of this section. As a consequence of using the method of image, the second exterior of black hole is identified as the copy of the spacetime dual to \eqref{eq:Bdy-state}. The endpoints of $A_j$ at the conformal boundary are still denoted by $E_1$ and $E_2$ and their images are located on the boundary of the second exterior and denoted by $E_1^I,~E_2^I$. On a constant time slice $\tilde{t}$, the endpoints and their images are given by
\be
\begin{split}
    &E_1:~\left(\tilde{\rho}_{\text{IR}},\tilde{t},\tilde{x}_{\xi}\left(X_2\right)\right)\to E_1':~\left(2e^{\tilde{x}_{\xi}\left(X_2\right)-\tilde{\rho}_{\text{IR}}},e^{\tilde{x}_{\xi}\left(X_2\right)}\sinh\tilde{t},e^{\tilde{x}_{\xi}\left(X_2\right)}\cosh\tilde{t}\right),\\
    &E_2:~\left(\tilde{\rho}_{\text{IR}},\tilde{t},\tilde{x}_{\xi}\left(X_1\right)\right)\to E_2':~\left(2e^{\tilde{x}_{\xi}\left(X_1\right)-\tilde{\rho}_{\text{IR}}},e^{\tilde{x}_{\xi}\left(X_1\right)}\sinh\tilde{t},e^{\tilde{x}_{\xi}\left(X_1\right)}\cosh\tilde{t}\right),\\
    &E_1^I:~\left(\tilde{\rho}_{\text{IR}},\tilde{t},\tilde{x}_{\xi}\left(X_2\right)\right)\to {E_1^I}':~\left(2e^{\tilde{x}_{\xi}\left(X_2\right)-\tilde{\rho}_{\text{IR}}},e^{\tilde{x}_{\xi}\left(X_2\right)}\sinh\tilde{t},-e^{\tilde{x}_{\xi}\left(X_2\right)}\cosh\tilde{t}\right),\\
    &E_2^I:~\left(\tilde{\rho}_{\text{IR}},\tilde{t},\tilde{x}_{\xi}\left(X_1\right)\right)\to {E_2^I}':~\left(2e^{\tilde{x}_{\xi}\left(X_1\right)-\tilde{\rho}_{\text{IR}}},e^{\tilde{x}_{\xi}\left(X_1\right)}\sinh\tilde{t},-e^{\tilde{x}_{\xi}\left(X_1\right)}\cosh\tilde{t}\right),
\end{split}
\ee
respectively. The cutoff $\tilde{\rho}_{\text{IR}}$ is defined in \eqref{eq:re-scaled-cutoff}, and $\tilde{x}_{\xi}$ and $\tilde{t}$ are the same as those in \eqref{eq:Endpoints-locations-Ueff}. However, the coordinates and cutoff now are re-scaled to $\tilde{x}_{\xi}= \f{\pi x_{\xi}}{2\epsilon},~\tilde{t}= \f{\pi t}{2\epsilon}$ and $\tilde{\delta}_{\text{UV}}=\f{\pi\delta_{\text{UV}}}{2\epsilon}$ due to the replacement $\epsilon\to 2\epsilon$. Therefore, the lengths of geodesics connecting $E_1$ with $E_2$, $E_K$ with $E_K^I$ are given by 
\be
    \begin{split}
        &\mathcal{L}_{12}^{B}\approx 2\log\left(\f{4\epsilon\sinh\left[\f{\pi\left(x_{\xi}(X_1)-x_{\xi}(X_2)\right)}{4\epsilon}\right]}{\pi\delta_{\text{UV}}}\right)\approx 2\log\left(\f{4\epsilon\sinh\left[\f{\pi\left(x_{\xi}(X_1)-x_{\xi}(X_2)\right)}{4\epsilon}\right]}{\pi\delta_{\text{UV}}'\sqrt{f(X_1)f(X_2)}}\right),\\
        &\mathcal{L}_{KK}^{B}\approx2\log\left(\f{4\epsilon\cosh\left[\f{\pi t}{2\epsilon}\right]}{\pi \delta_{\text{UV}}}\right)\approx2\log\left(\f{4\epsilon\cosh\left[\f{\pi t}{2\epsilon}\right]}{\pi \delta_{\text{UV}}'f(X_K)}\right),~~~K=1,2.
    \end{split}
\ee
Then, the entanglement entropy can be evaluated by the Ryu-Takayanagi formula\footnote{Recently, the Ryu-Takayanagi formula has been investigated for $\text{AdS}_3/\text{BCFT}_2$ constructions with the emergence of the EOWs. Specifically, the authors of \cite{Geng:2021iyq, Geng:2024xpj} offered a derivation of the quantum extremal surface prescription and its geometrization in the spacetime featuring EOWs.}
\be
\begin{split}
    S_{A_j}^{i=2}\approx \f{\text{Min}\left\{\mathcal{L}_{12}^{B},\f{\mathcal{L}_{11}^{B}+\mathcal{L}_{22}^{B}}{2} \right\}}{4G_N},~~~G_N=\f{3}{2c},
\end{split}
\ee
where the extra factor $2$ in the denominator of $\f{\mathcal{L}_{11}^{B}+\mathcal{L}_{22}^{B}}{2}$ comes from the method of image. For $A_1$, if we consider the low temperature limit, $\f{L\sin\left[\f{q\pi (X_1-X_2)}{L}\right]}{8q\epsilon\prod_{K=1,2}\sin\left[\f{q\pi X_K}{L}\right]}\ll 1$, the entanglement entropy is given by
\be
    S_{A_1}^{i=2}\approx\f{c}{3}\cdot\log\left(\f{L}{q\pi \delta_{\text{UV}}'}\sin\left[\f{q\pi(X_1-X_2)}{L}\right]\right),
\ee
On the contrary, if one considers the high temperature limit, i.e. $\f{L\sin\left[\f{q\pi (X_1-X_2)}{L}\right]}{8q\epsilon\prod_{K=1,2}\sin\left[\f{q\pi X_K}{L}\right]}\gg 1$, the entanglement entropy is
\be
    \begin{split}
        S_{A_1}^{i=2}\approx&\f{c}{3}\log\left(\f{4\epsilon}{\pi\delta_{\text{UV}}'}\right)+\f{c}{6}\log\left(4\prod_{K=1,2}\sin^2\left[\f{q\pi X_K}{L}\right]\right)+\f{c}{6\epsilon}\cdot\begin{cases}
            \pi t~&\text{for}~t^*>t\gg \epsilon\\
            \f{L\sin\left[\f{q\pi (X_1-X_2)}{L}\right]}{4q\prod_{K=1,2,}\sin\left[\f{q\pi X_K}{L}\right]}~&\text{for}~t\geq t^*,
        \end{cases}
    \end{split}
\ee
where $t^*\approx \f{L\sin\left[\f{q\pi (X_1-X_2)}{L}\right]}{4q\pi\prod_{K=1,2,}\sin\left[\f{q\pi X_K}{L}\right]}$. For $A_2$, the time dependence is independent of the temperature limits, the entanglement entropy reads
\be
    S_{A_2}^{i=2}\approx\f{c}{3}\log\left(\f{4\epsilon}{\pi\delta_{\text{UV}}'}\right)+\f{c}{6}\log\left(4\prod_{K=1,2}\sin^2\left[\f{q\pi X_K}{L}\right]\right)+\f{c}{6\epsilon}\cdot\begin{cases}
            \pi t~&\text{for}~t^*>t\gg \epsilon\\
            \f{\pi l\cdot L}{4q}e^{2\theta}~&\text{for}~t\geq t^*\text{ and }0<l<\f{q}{2},\\
            \f{\pi (q-l)\cdot L}{4q}e^{2\theta}~&\text{for}~t\geq t^*\text{ and }\f{q}{2}\leq l<q,
        \end{cases}
\ee
where the exchanging time $t^*$ is given by
\be
   t^*\approx\begin{cases}
       \f{\l\cdot L}{4q}e^{2\theta}~&\text{if}~0<l<\f{q}{2}\\
       \f{(q-l)\cdot L}{4q}e^{2\theta}~&\text{if}~\f{q}{2}\leq l<q.
   \end{cases} 
\ee
These results are consistent with those in the Euclidean computations in \eqref{eq:stat-single-bs}, \eqref{eq:EE-lagth-highte} and \eqref{eq:EE-A2-BS}.

At the end of this section, instead of the explicit computations, we briefly sketch the holographic computations for two-interval cases. First, we mention that the endpoints and their images are given by
\be
\begin{split}   &E_1:~\left(\tilde{\rho}_{\text{IR}},\tilde{t},\tilde{x}_{\xi}\left(X_2\right)\right)\to E_1':~\left(2e^{\tilde{x}_{\xi}\left(X_2\right)-\tilde{\rho}_{\text{IR}}},e^{\tilde{x}_{\xi}\left(X_2\right)}\sinh\tilde{t},e^{\tilde{x}_{\xi}\left(X_2\right)}\cosh\tilde{t}\right),\\ &E_2:~\left(\tilde{\rho}_{\text{IR}},\tilde{t},\tilde{x}_{\xi}\left(X_1\right)\right)\to E_2':~\left(2e^{\tilde{x}_{\xi}\left(X_1\right)-\tilde{\rho}_{\text{IR}}},e^{\tilde{x}_{\xi}\left(X_1\right)}\sinh\tilde{t},e^{\tilde{x}_{\xi}\left(X_1\right)}\cosh\tilde{t}\right),\\  &E_3:~\left(\tilde{\rho}_{\text{IR}},\tilde{t},\tilde{x}_{\xi}\left(Y_2\right)\right)\to E_3':~\left(2e^{\tilde{x}_{\xi}\left(Y_2\right)-\tilde{\rho}_{\text{IR}}},e^{\tilde{x}_{\xi}\left(Y_2\right)}\sinh\tilde{t},e^{\tilde{x}_{\xi}\left(Y_2\right)}\cosh\tilde{t}\right),\\
&E_4:~\left(\tilde{\rho}_{\text{IR}},\tilde{t},\tilde{x}_{\xi}\left(Y_1\right)\right)\to E_4':~\left(2e^{\tilde{x}_{\xi}\left(Y_1\right)-\tilde{\rho}_{\text{IR}}},e^{\tilde{x}_{\xi}\left(Y_1\right)}\sinh\tilde{t},e^{\tilde{x}_{\xi}\left(Y_1\right)}\cosh\tilde{t}\right),\\
&E_1^I:~\left(\tilde{\rho}_{\text{IR}},\tilde{t},\tilde{x}_{\xi}\left(X_2\right)\right)\to {E_1^I}':~\left(2e^{\tilde{x}_{\xi}\left(X_2\right)-\tilde{\rho}_{\text{IR}}},e^{\tilde{x}_{\xi}\left(X_2\right)}\sinh\tilde{t},-e^{\tilde{x}_{\xi}\left(X_2\right)}\cosh\tilde{t}\right),\\
&E_2^I:~\left(\tilde{\rho}_{\text{IR}},\tilde{t},\tilde{x}_{\xi}\left(X_1\right)\right)\to {E_2^I}':~\left(2e^{\tilde{x}_{\xi}\left(X_1\right)-\tilde{\rho}_{\text{IR}}},e^{\tilde{x}_{\xi}\left(X_1\right)}\sinh\tilde{t},-e^{\tilde{x}_{\xi}\left(X_1\right)}\cosh\tilde{t}\right),\\
&E_3^I:~\left(\tilde{\rho}_{\text{IR}},\tilde{t},\tilde{x}_{\xi}\left(Y_2\right)\right)\to {E_3^I}':~\left(2e^{\tilde{x}_{\xi}\left(Y_2\right)-\tilde{\rho}_{\text{IR}}},e^{\tilde{x}_{\xi}\left(Y_2\right)}\sinh\tilde{t},-e^{\tilde{x}_{\xi}\left(Y_2\right)}\cosh\tilde{t}\right),\\
&E_4^I:~\left(\tilde{\rho}_{\text{IR}},\tilde{t},\tilde{x}_{\xi}\left(Y_1\right)\right)\to {E_4^I}':~\left(2e^{\tilde{x}_{\xi}\left(Y_1\right)-\tilde{\rho}_{\text{IR}}},e^{\tilde{x}_{\xi}\left(Y_1\right)}\sinh\tilde{t},-e^{\tilde{x}_{\xi}\left(Y_1\right)}\cosh\tilde{t}\right),
\end{split}
\ee
respectively. Second, we consider the geodesics connecting these endpoints and compute their lengths, which are denoted by
\be
\begin{split}
    &\mathcal{L}_{12}^B:~\text{the length of geodesic connecting }E_1\text{ and }E_2, \\
    &\mathcal{L}_{34}^B:~\text{the length of geodesic connecting }E_3\text{ and }E_4,\\
    &\mathcal{L}_{13}^B:~\text{the length of geodesic connecting }E_1\text{ and }E_3,\\
    &\mathcal{L}_{24}^B:~\text{the length of geodesic connecting }E_2\text{ and }E_4,\\
    &\mathcal{L}_{KK}^B,~K=1,2,3,4:~\text{the length of geodesic connecting }E_K\text{ and }E_K^I.
\end{split} 
\ee
Then, by applying the Ryu-Takayanagi formula, one can derive the explicit expression of holographic entanglement entropy by
\be\label{eq:rt-two-interval-bdy-state}
    S_{A_j\cup B}^{i=2}\approx \f{\text{Min}\left\{\mathcal{L}_{12}^B+\mathcal{L}_{34}^B,\mathcal{L}_{13}^B+\mathcal{L}_{24}^B,\f{\sum_{K=1}^4\mathcal{L}_{KK}^B}{2}\right\}}{4G_N},~~~G_N=\f{3}{2c},
\ee
where $\left(\mathcal{L}_{12}^B+\mathcal{L}_{34}^B\right)$-term and $\left(\mathcal{L}_{13}^B+\mathcal{L}_{24}^B\right)$-term correspond to the static geodesics, while $\f{\sum_{K=1}^4\mathcal{L}_{KK}^B}{2}$-term corresponds to the time dependent geodesic. One can easily verify that the results obtained using (\ref{eq:rt-two-interval-bdy-state}) are consistent with those presented in Section \ref{Sec:EE-BS-for-DI}.

\section{Line-tension picture}
In this section, we will discuss the line tension picture, an effective picture describing the entanglement dynamics in $2$d HCFT in the high temperature limit of the high effective temperature region.
The line-tension picture or membrane-tension picture was introduced as a heuristic way to interpret entanglement dynamics of operator entanglement in quantum many-body chaotic systems \cite{Jonay:2018yei,Nahum:2016muy,Mezei:2018jco}.
Here, we focus on the line-tension picture for $(1+1)$-dimensional spacetime with compact spatial direction \cite{Goto:2021gve}. In this work, 
we study the entanglement of the operator $e^{-(\epsilon+it)H_{\text{q-M\"obius}}}$ 
that is defined on the compact effective space whose coordinate is given by $(t,x_{\xi})$, since in coordinates $(t,x_{\xi})$ the spacetime remains flat and the line-tension is given by \cite{Jonay:2018yei,Goto:2021gve}
\be\label{line-tension}
    \mathcal{T}(v_{\xi})=\begin{cases}
        \log Q~&~v_{\xi}\leq 1\\
        v_{\xi}\log Q~&~v_{\xi}> 1,
    \end{cases}
\ee
where $v_{\xi}=\f{dx_{\xi}}{dt}$ denotes the velocity along the curve $C$, and 
\be\label{eq:effective-bond-dimension}
    Q\approx e^{s_{\text{eq}}}=e^{\f{c\pi}{6\epsilon}},~s_{\text{eq}}:=\f{c\pi}{6\epsilon}
\ee
are the local degrees of freedom\footnote{In some limits \cite{Nahum:2016muy}, the curve $C$ is related to the minimal cut in tensor networks, and $Q$ can be interpreted as the bond dimension.} along $C$ and thermal entropy density \cite{Jonay:2018yei}, respectively. 
Consider such a curve $C$ smoothly connecting endpoints at $(0,x_{\xi}(x))$ and $(t,x_{\xi}(y))$, where $y>x$. 
The coarse-grained 
amount of entanglement across the line $C$, which is evaluated by the line-tension picture, reads
\be\label{EE-line}
\begin{aligned}
    &S^{\text{C.G.}}(x_{\xi}(x),x_{\xi}(y),t)=\underset{C}{\text{Min}}\left[S_{\text{ini.}}(x_{\xi}(x),0)+\int_{0}^tds\mathcal{T}(v_{\xi})\right]\\
    &=S_{\text{ini.}}(x_{\xi}(x),0)+\log Q\cdot\begin{cases}
        \int_0^tds~&\text{if}~|v_{\xi}|\leq 1\\
        l^{\text{effective}}(x,y)~&\text{if}~|v_{\xi}|>1,
    \end{cases},~~\text{with}~v_{\xi}=\f{x_{\xi}(y)-x_{\xi}(x)}{t},
\end{aligned}
\ee
where $S_{\text{ini.}}(x_{\xi}(x),0)$ is from the entanglement of the initial state at the single point $x_{\xi}(x)$ when $t=0$. Note that all possible components of $C$ are straight lines, since we consider the spacetime to be flat in terms of coordinates $(t,x_{\xi})$, and $l^{\text{effective}}(x,y)$ is given by \eqref{eq:effective-length}. For multiple intervals, the curve $C$ is required to be homologous to the subsystem $\mathcal{V}$ \cite{Goto:2021sqx}. The precise definition of the operator entanglement 
in the line-tension picture is
\be\label{EE-line-tension picture}
    S_{\mathcal{V}}^{\text{C.G.}}=\underset{C\underset{\text{hom.}}{\sim} \mathcal{V}}{\text{Min}}\left[\int_{C}dt\mathcal{T}(v_{\xi})+S_{\partial C}(t=0)\right],
\ee
where $S_{\partial C}(t=0)$ is the initial entanglement entropy encoded in the endpoints of the curve $C$ at $t=0$ slice \cite{Jonay:2018yei}. In our cases, it is given by the contribution from universal pieces, i.e. $S_{\partial C}(t=0)=S_{\mathcal{V}}^{i=1;\text{U.}}$.

Before moving to the case study, we present some fundamental facts and constraints on the line-tension picture. 
The basic constraints \cite{Jonay:2018yei,Mezei:2018jco} on the line-tension are
\be
    \begin{aligned}
        &\mathcal{T}(0)=v_{E}s_{\text{eq}},~\mathcal{T}(v_{B})=v_{B}s_{\text{eq}},~\mathcal{T}(1)=\infty,~\mathcal{T}(v_{\xi})\leq |v_{\xi}|s_{\text{eq}},\\
        &\f{d\mathcal{T}(v_{\xi})}{dv_{\xi}}\bigg|_{v_{\xi}=0}=0,~\f{d\mathcal{T}(v_{\xi})}{dv_{\xi}}\bigg|_{v_{\xi}=v_B}=s_{\text{eq}},~\f{d\mathcal{T}(v_{\xi})}{dv_{\xi}}\geq 0,~\f{d^2\mathcal{T}(v_{\xi})}{dv_{\xi}^2}\geq 0,
    \end{aligned}
\ee
where in general $v_B\leq v_{\xi}^{\text{Max}}$ is the ``butterfly'' speed that is defined by out-of-order correlator, $v_E$ represents the ``entanglement speed'', which is smaller than $v_B$, and $s_{\text{eq}}$ is defined in \eqref{eq:effective-bond-dimension}. In addition, we have
\be
    \f{d\mathcal{T}(v_{\xi})}{dv_{\xi}}v_{\xi}-\mathcal{T}(v_{\xi})~\begin{cases}
        \leq 0~&\text{if}~|v_{\xi}|\leq v_B\\
        >0~&\text{if}~v_B<|v_{\xi}|\leq v_{\xi}^{\text{Max}},
    \end{cases}
\ee
where we have $v_{\xi}^{\text{Max}}=1$ as the speed of light. Specifically, in $(1+1)$-dimensional theories, $\mathcal{T}(v_{\xi})$ reduces to the form of \eqref{line-tension}, and further we have $v_E=v_B=1$ by requiring theories to be relativistic and conformal \cite{Jonay:2018yei}.
\subsection{Single-interval cases}
Here, we will present the single-interval entanglement entropy obtained in the line-tension picture.  By evaluating \eqref{EE-line-tension picture} for $A_1$ and $A_2$ defined in (\ref{eq:subsystems-considered}), we obtain the entanglement entropy as
\be\label{Single-interval-EE-line-tension}
\begin{split}
    S_{A_j}^{i=1;\text{C.G.}}=&S_{A_j}^{i=1;\text{U.}}+2s_{\text{eq}}\cdot\int_0^{\f{l_{A_j}^{\text{effective}}}{2}}dt\approx\f{c}{6}\log{\left[4\prod_{K=1,2}\sin^2{\left(\f{q\pi X_K}{L}\right)}\right]}+\f{c}{3}\log{\left(\f{2\epsilon}{\pi}\right)}\\
    &+\begin{cases}
        \f{cL}{12q\epsilon}\f{\sin\left[\f{q\pi(X_1-X_2)}{L}\right]}{\prod_{K=1}^2\sin\left[\f{q\pi X_K}{L}\right]}~&\text{for}~j=1\\
        \f{c\pi l \cdot L}{12q\epsilon}e^{2\theta}~&\text{for}~j=2,
    \end{cases}
\end{split}
\ee
where we have assumed the high temperature limit presented in (\ref{eq:high-low-temp-limits-single-interval}) and large $\theta$ limit, $\theta\gg 1$. 
This result is consistent with the one obtained in the path-integral formalism and the gravity dual with the Lorentzian signature.

\subsection{Two-interval cases}
Now, in the line-tension picture, we will calculate the entanglement entropy for the two spatial intervals for the cases considered in Section \ref{Sec:Etanglement-dynamics-Ueff}.
Here, we present the time dependence of the entanglement entropy in the simplest case, the symmetric case.
The other cases will be presented in Appendix \ref{sec:LP-on-MI}.

We begin by considering the entanglement entropy of disconnected piece defined in Section \ref{Sec:Etanglement-dynamics-Ueff}.
In the cases considered here, in the line-tension picture, it is given by 
\be\label{Disconnected-piece-line-tension}
    S_{A_j\cup B_k}^{i=1;\text{dis.}}\approx \log Q \cdot\left(l_{A_j}^{\text{effective}}+l_{B_k}^{\text{effective}}\right) = s_{\text{eq}}\cdot\left(l_{A_j}^{\text{effective}}+l_{B_k}^{\text{effective}}\right).
\ee

\subsubsection{Symmetric cases}
In the line-tension picture, we will calculate the time dependence of the entanglement entropy and mutual information in the symmetric case where 
$j=k$ and $X_K=Y_K$, $K=1,2$, and the centers of the subsystems are the same. 
Then, the effective subsystem size of $A$ is the same as that of $B$, $l_{A_j}^{\text{effective}}=l_{B_j}^{\text{effective}}$. 
Since two subsystems $A_j,B_j$ are spatially identical, for small $t$, the homologous ``line'' $C\sim\mathcal{V}$ is constructed of just two vertical straight lines connecting $\left(0,x_{\xi}(X_K)\right)$ with $\left(t,x_{\xi}(X_K)\right)$. 
Hence, the velocity along path $C$ is zero, and the line tension is determined by the expression $\mathcal{T}(0)=s_{\text{eq}}$. Then, the time dependence of the connected piece is evaluated by the line-tension picture as
\be
   S_{A_j\cup B_j}^{i=1;\text{con.}}\approx \int_{C}dt'\mathcal{T}(0) =2s_{\text{eq}}\cdot\int_0^tdt'=2s_{\text{eq}} t.
\ee
Therefore, the time dependence of the entanglement entropy and mutual information are approximately given by
\be \label{eq:EE-MI-LT-Symm}
    \begin{aligned}
        S_{A_j\cup B_j}^{i=1}\approx& S_{A_j\cup B_j}^{i=1;\text{C.G.}}\approx\f{c}{3}\log{\left[4\prod_{K=1,2}\sin^2{\left(\f{q\pi X_K}{L}\right)}\right]}+\f{2c}{3}\log{\left(\f{2\epsilon}{\pi}\right)}+ 2s_{\text{eq}}\cdot\begin{cases}
            t~&~t\leq t^*\\
            l_{A_j}^{\text{effective}}~&~t> t^*,
        \end{cases}\\
        I_{A_j,B_k}^{i=1}\approx&S_{A_j\cup B_j}^{i=1;\text{dis.}}-S_{A_j\cup B_j}^{i=1;\text{con.}} \approx2s_{\text{eq}}\cdot\begin{cases}
            l_{A_j}^{\text{effective}}-t~&~t\leq t^*\\
            0~&~t> t^*,
        \end{cases}
    \end{aligned}
\ee
where the exchange time is
\be \label{eq:ET-EE-MI-LT-Symm}
    t^*\approx l_{A_j}^{\text{effective}}\approx\begin{cases}
        \f{L}{2q\pi }\f{\sin{\left[\f{q\pi (X_1-X_2)}{L}\right]}}{\prod_{K=1,2}\sin{\left[\f{q\pi (X_K)}{L}\right]}}~&~\text{for}~j=1\\
        \f{l \cdot L}{2q}e^{2\theta}~&~\text{for}~j=2,
    \end{cases}
\ee
which is same as \eqref{eq:TFD-two-interval-exchanging-time}. Thus, in the symmetric case, the time dependence of the entanglement entropy and mutual information obtained in the line-tension picture is consistent with that in the path-integral formalism and holography.

\section{Discussions and Future direction}
In this section, we will discuss our findings in this paper and comment on the future direction.

{\bf Local quenches:} As in Section \ref{sec:interpretation}, if we start from the boundary state with the regulator determined by the Hamiltonian considered in this paper, in the low temperature regime, the state is approximately given by the product of the vacuum states for $H_{\text{SSD}}$. 
In other words, the time evolution of the system is similar to that in the local quenches.
Contrary to the local quenches considered so far, in our case, the local excitations emerge around $x=X^f_{l}$, and stay there, not propagate.
This is because the entanglement structure around $x=X^f_l$ is far from that of the vacuum state of $H_{\text{q-M\"obius}}$ inducing the time evolution, while the entanglement structure far from $x=X^f_l$ is approximately given by that of the vacuum state of $H_{\text{q-M\"obius}}$.
If we focus on the spatial subregion between $x=X^{f}_{l}$ and $x=X^{f}_{l+1}$ and think of the spatial region except for this interval as the bath and the ancilla, the time evolution considered in this paper can be regarded as the dynamics in open systems and local operation in quantum information theory. 

{\bf Linear growth of entanglement entropy:} 
As explained in Section \ref{Sec:Etanglement-dynamics-Ueff} and \ref{sec:entropy-production}, the time growth of entanglement entropy from the thermofield double and boundary states is linear in time, regardless of $q$ and $\theta$. 
In terms of gravity dual, this linear growth is induced by the wormhole growth penetrating the black hole.
This is surprising for us because as in \cite{Goto:2023wai},  if the Hamiltonian, inducing the Euclidean time evolution, is different from the one, inducing the Lorentzian time evolution, the entanglement growth is not linear.
As in \cite{2014arXiv1402.5674S,2014PhRvD..90l6007S}, the linear growth of the entanglement entropy may be caused by the growth of the computational complexity in the holographic CFTs.
Therefore, it would be interesting to explore the entanglement growth of the systems in
\be \label{eq:state-future-direction}
\begin{split}
&\ket{\Psi_{i=1,2}}=\f{1}{\sqrt{\Tr e^{-2\epsilon H_{\text{$q_1$-M\"obius}}(\theta_1)}}}\\
   &\times\begin{cases}
         e^{\f{-it}{2} \sum_{j=1}^2H^j_{\text{$q_2$-M\"obius}}(\theta_2)}\sum_{a}e^{-\epsilon H_{\text{$q_1$-M\"obius}}(\theta_1)}\ket{E^{\text{$q_1$-M\"obius}(\theta_1)}_a}_1 \otimes \ket{E^{\text{$q_1$-M\"obius}(\theta_1)*}_a}_2 & \text{for}~ i=1\\
         e^{-it H_{\text{$q_2$-M\"obius}}(\theta_2)}e^{-\epsilon H_{\text{$q_1$-M\"obius}}(\theta_1)} \ket{\text{B}},
    \end{cases}
\end{split}
\ee
More precisely, it would be interesting to explore how the time evolution of entanglement entropy and mutual information for (\ref{eq:state-future-direction}) depend on $q_{j=1,2}$ and $\theta_{j=1,2}$.
\section*{Acknowledgement}
We thank Weibo Mao, Kotaro Tamaoka, Huajia Wang, Shan-Ming Ruan, and Masataka Watanabe for useful discussions. 
M.N.~is supported by funds from the University of Chinese Academy of Sciences (UCAS), funds from the Kavli
Institute for Theoretical Sciences (KITS).

\appendix

\section{The analytic continuation\label{App:The-details-of-primary-operator}}
In this appendix, we derive the transformation rule \eqref{eq:transformation-lop} of primary operators.

For convenience, we focus on the $(z,\bar{z})$ coordinates given in \eqref{Mobius-maps}. Using the coordinates, \eqref{eq:transformation-lop} can be written as
\be \label{eq:Euc_Time_evo_in_z_coordinates}
\begin{split}
    e^{TH_{\text{q-M\"obius}}}\mathcal{O}\left(z,\overline{z}\right) e^{-TH_{\text{q-M\"obius}}}=\left|\f{d(\xi+T)}{dz}\right|^{2h_{\mathcal{O}}}\mathcal{O}(\xi+T,\overline{\xi}+T),
\end{split}
\ee
where we suppressed the subscript $X$.
After backing to $(w,\bar{w})$ by the conformal map $\left(z,\overline{z}\right)=\left(e^{\f{2\pi w}{L}},e^{\f{2\pi \overline{w}}{L}}\right)$, we get the transformation rule \eqref{eq:transformation-lop}. We show the rule by directly evaluating the left hand side of the above expression.  

First, we expand the left hand side by using the Campbell-Baker-Hausdorff formula,
 \begin{equation}
    e^{TH_{\text{q-M\"obius}} } \,  \mathcal{O}(z,\bar{z}) \, e^{-TH_{\text{q-M\"obius}}} = \sum_{n=0}^{\infty} \frac{T^{n}}{n!} \mathcal{L}_{H_{\text{q-M\"obius}}}^{n}\left[  \mathcal{O}(z,\bar{z})  \right],
\end{equation}
where $\mathcal{L}_{A}^{n}\left[  B  \right]$ denotes a $n$-th nested commutator given by
\begin{equation}
	\begin{aligned}
		\mathcal{L}_{A}\left[ B \right] &\coloneqq \left[A, B\right],\\
		 \mathcal{L}_{A}^{0}\left[ B \right] &\coloneqq B,\\
		\mathcal{L}_{A}^{n+m}\left[ B \right] &\coloneqq \mathcal{L}_{A}^{n}\left[ \mathcal{L}_{A}^{m}\left[ B \right] \right].
	\end{aligned}
\end{equation}
To evaluate the nested commutators, we use the M\"obius Hamiltonian in the $(z,\bar{z})$ coordinates,
\begin{equation}
	H_{\text{q-M\"obius}} = \f{2\pi}{L} \oint \frac{dz}{2\pi i} \,  z\left[ 1-\dfrac{\tanh (2\theta)}{2} \left( z^{q} + \dfrac{1}{z^{q}} \right) \right]T(z)+(h.c.) + (\text{const.}).
\end{equation}
From the standard OPE between the stress-energy tensor and the primary operator with the conformal dimension $\Delta_{\mathcal{O}}=2h_{\mathcal{O}}$, we can simply evaluate the first commutator,
\begin{equation}
	\begin{aligned}
		\mathcal{L}_{H_{\text{q-M\"obius}}}\left[  \mathcal{O}(z,\bar{z})  \right] &= \left[ H_{\text{q-M\"obius}} , \mathcal{O}(z,\bar{z})   \right]\\
		&= h_{\mathcal{O}} \, \partial g(z) \cdot \mathcal{O}(z,\bar{z}) + g(z) \cdot \partial \mathcal{O}(z,\bar{z}) \\
		& \qquad  + h_{\mathcal{O}} \, \bar{\partial} \bar{g}(\bar{z})  \cdot \mathcal{O}(z,\bar{z}) + \bar{g}(\bar{z})  \cdot \bar{\partial} \mathcal{O}(z,\bar{z})\\
		&= \left[   \frac{1}{ g(z)^{h_{\mathcal{O}}}} \,   g(z)  \overset{\rightarrow}{\partial}   g(z)^{h_{\mathcal{O}}} + \frac{1}{\bar{g}(\bar{z}) ^{  h_{\mathcal{O}} }} \, \bar{g}(\bar{z}) \,   \overset{\rightarrow}{\bar{\partial}} \bar{g}(\bar{z})^{ h_{\mathcal{O}} }   \right] \mathcal{O}(z,\bar{z}),
	\end{aligned}
\end{equation}
where we defined $g(z),\bar{g}(\bar{z})$ by 
\begin{equation}
	g(z)=\dfrac{2\pi}{L}z\left[ 1-\dfrac{\tanh (2\theta)}{2} \left( z^{q} + \dfrac{1}{z^{q}} \right) \right], \quad \bar{g}(\bar{z})=\dfrac{2\pi}{L}\bar{z} \left[ 1-\dfrac{\tanh (2\theta)}{2} \left( \bar{z}^{q} + \dfrac{1}{\bar{z}^{q}} \right) \right],
\end{equation}
and  $\overset{\rightarrow}{\partial}, \overset{\rightarrow}{\bar{\partial}} $  denote derivatives acting on all objects sitting on the right side, e.g., $ \overset{\rightarrow}{\partial} f_{1}(z)f_{2}(z)= \partial \left( f_{1}(z)f_{2}(z) \right)$.

By successively using the above result, we can obtain the $n$-th nested commutator,
\begin{equation}
	\begin{aligned}
		\mathcal{L}_{H_{\text{q-M\"obius}}}^{n}\left[  \mathcal{O}(z,\bar{z})  \right] &=  \left[   \frac{1}{ g(z)^{h_{\mathcal{O}}}} \,   g(z)  \overset{\rightarrow}{\partial}   g(z)^{h_{\mathcal{O}}} + \frac{1}{\bar{g}(\bar{z}) ^{  h_{\mathcal{O}} }} \, \bar{g}(\bar{z}) \,   \overset{\rightarrow}{\bar{\partial}} \bar{g}(\bar{z})^{ h_{\mathcal{O}} }   \right]^{n} \mathcal{O}(z,\bar{z}).
	\end{aligned}
\end{equation}
Then, we get 
\begin{equation}\label{eq:taylowDeriva}
	\begin{aligned}
	e^{TH_{\text{q-M\"obius}} } \,  \mathcal{O}(z,\bar{z}) \, e^{-TH_{\text{q-M\"obius}}} &= \sum_{n=0}^{\infty} \frac{T^{n}}{n!} \mathcal{L}_{H_{\text{q-M\"obius}}}^{n}\left[  \mathcal{O}(z,\bar{z})  \right]\\
	&= \exp \left[   \frac{T}{ g(z)^{h_{\mathcal{O}}}} \, g(z)  \overset{\rightarrow}{\partial}   g(z)^{h_{\mathcal{O}}} + \frac{T}{\bar{g}(\bar{z}) ^{  h_{\mathcal{O}} }} \, T \bar{g}(\bar{z}) \,   \overset{\rightarrow}{\bar{\partial}} \bar{g}(\bar{z})^{ h_{\mathcal{O}} }   \right] \mathcal{O}(z,\bar{z})\\
		&= \exp \left[   \frac{T}{ g(z)^{h_{\mathcal{O}}}} \,  g(z)  \overset{\rightarrow}{\partial}   g(z)^{h_{\mathcal{O}}} \right] \exp \left[  \frac{T}{\bar{g}(\bar{z}) ^{  h_{\mathcal{O}} } } \, \bar{g}(\bar{z}) \,   \overset{\rightarrow}{\bar{\partial}} \bar{g}(\bar{z})^{h_{\mathcal{O}} }   \right] \mathcal{O}(z,\bar{z}) \\
		& = \frac{1}{ g(z)^{h_{\mathcal{O}}} \bar{g}(\bar{z}) ^{ h_{\mathcal{O}}} } \exp \left[  T  g(z)  \overset{\rightarrow}{\partial}  \right] \exp \left[   T\bar{g}(\bar{z}) \,   \overset{\rightarrow}{\bar{\partial}}   \right]      g(z)^{h_{\mathcal{O}}} \bar{g}(\bar{z}) ^{ h_{\mathcal{O}}}  \mathcal{O}(z,\bar{z}).
	\end{aligned}
\end{equation}
In this expression, it is difficult to find the action of exponential of the derivatives with the functions, $\exp \left[  T  g(z)  \overset{\rightarrow}{\partial}  \right] \exp \left[   T\bar{g}(\bar{z}) \,   \overset{\rightarrow}{\bar{\partial}}   \right] $. We can simplify the action by moving to the coordinates $\xi ,\bar{\xi }$ such that 
\begin{equation}
	g(z)  \overset{\rightarrow}{\partial_{z}}= \partial_{\xi }, \quad \bar{g}(\bar{z}) \,   \overset{\rightarrow}{\bar{\partial}_{\bar{z}}}= \bar{\partial}_{\bar{\xi }},
\end{equation}
where we restored the sub-indices of the derivatives to distinguish $\xi ,\bar{\xi}$ from $z,\bar{z}$. Noting the chain rule of derivative, we can see that $y,\bar{y}$ must satisfy the differential equations, 
\begin{equation}\label{eq:deffexi}
	\frac{d \xi(z)}{dz} = \frac{1}{g(z)}, \quad \frac{d \bar{\xi}(\bar{z})}{d\bar{z}} = \frac{1}{\bar{g}(\bar{z})}.
\end{equation}
Thus, we can introduce the $\xi ,\bar{\xi}$-coordinates as the following solutions $G(z),\bar{G}(\bar{z})$ of the above differential equations, 
\begin{equation}\label{eq:xiz}
	\begin{aligned}
		\xi(z) &=G(z) =  \frac{L_{\text{eff}}}{2\pi q} \log\left[ \frac{  \cosh\theta\, z^{q} -\sinh\theta }{  \cosh\theta -\sinh\theta\,  z^{q} } \right]\\
	 \bar{\xi}(\bar{z}) &= \bar{G}(\bar{z}) = \frac{L_{\text{eff}}}{2\pi q} \log\left[ \frac{  \cosh\theta\, \bar{z}^{q} -\sinh\theta }{  \cosh\theta -\sinh\theta \, \bar{z}^{q} } \right].
	\end{aligned}
\end{equation}
Here, $\xi,\bar{\xi}$ are identical to those given in \eqref{Mobius-maps}.
Using these expressions, $z,\bar{z}$ can be written in terms of $\xi,\bar{\xi}$,
\begin{equation}\label{eq:zxi}
	\begin{aligned}
		z&=G^{-1}(\xi)= \left( \frac{  \cosh\theta \cdot  \exp\left( \dfrac{2\pi q}{{L_{\text{eff}}}} \xi  \right) +\sinh\theta }{  \cosh\theta +\sinh\theta \cdot   \exp\left( \dfrac{2\pi q}{{L_{\text{eff}}}} \xi \right) } \right)^{\frac{1}{q}},\\
	\bar{z}&=\bar{G}^{-1}(\bar{\xi})= \left( \frac{  \cosh\theta \cdot  \exp\left( \dfrac{2\pi q}{{L_{\text{eff}}}} \bar{\xi} \right) +\sinh\theta }{  \cosh\theta +\sinh\theta \cdot   \exp\left( \dfrac{2\pi q}{{L_{\text{eff}}}} \bar{\xi} \right) } \right)^{\frac{1}{q}}.
	\end{aligned}
\end{equation}

From the $\xi ,\bar{\xi}$-coordinates, 
the expansion \eqref{eq:taylowDeriva} reduces to the simple expression,
\begin{equation}
	\begin{aligned}
		&\frac{1}{ g(z)^{h_{\mathcal{O}}} \bar{g}(\bar{z}) ^{ h_{\mathcal{O}}} } \exp \left[  T  g(z)  \overset{\rightarrow}{\partial}_{z}  \right] \exp \left[   T\bar{g}(\bar{z}) \,   \overset{\rightarrow}{\bar{\partial}}_{\bar{z}}   \right]      g(z)^{h_{\mathcal{O}}} \bar{g}(\bar{z}) ^{ {h_{\mathcal{O}}}}  \mathcal{O}(z,\bar{z})\\
		&=\frac{1}{ g(z)^{h_{\mathcal{O}}} \bar{g}(\bar{z}) ^{ h_{\mathcal{O}}} } \exp \left[  T  \overset{\rightarrow}{\partial}_{\xi}  \right] \exp \left[   T  \overset{\rightarrow}{\bar{\partial}}_{\bar{\xi}}   \right]  g\left( G^{-1}(\xi) \right)^{h_{\mathcal{O}}} \bar{g}\left(\bar{G}^{-1}(\bar{\xi})\right) ^{ {h_{\mathcal{O}}}} \mathcal{O}(G^{-1}(\xi),\bar{G}^{-1}(\bar{\xi}))\\
		&=\frac{1}{ g(z)^{h_{\mathcal{O}}} \bar{g}(\bar{z}) ^{ h_{\mathcal{O}}} }   g\left( G^{-1}(\xi+T) \right)^{h_{\mathcal{O}}} \bar{g}\left(\bar{G}^{-1}(\bar{\xi}+T)\right) ^{ {h_{\mathcal{O}}}} \mathcal{O}(G^{-1}(\xi+T),\bar{G}^{-1}(\bar{\xi}+T))
	\end{aligned}
\end{equation}
where, in the final line, we used the relation
\begin{equation}
	\exp\left[ a \overset{\rightarrow}{\frac{d}{dx}} \right] f(x)=f(x+a).
\end{equation}

Thus, we obtain a kind of the transformation law,
\begin{equation}\label{eq:transRatio}
	\begin{aligned}
		&e^{-TH_{\text{q-M\"obius}} } \,  \mathcal{O}(z,\bar{z}) \, e^{TH_{\text{q-M\"obius}}}\\
		 &= \left( \f{ g\left( G^{-1}(\xi+T) \right) }{g\left(z\right)} \right)^{ {h_{\mathcal{O}}} } \left( \f{ \bar{g}\left( \bar{G}^{-1}(\bar{\xi}+T) \right) }{\bar{g}\left( \bar{z}\right)} \right)^{ {h_{\mathcal{O}}} }  \mathcal{O}\left(G^{-1}(\xi+T),\bar{G}^{-1}(\bar{\xi}+T)\right)\\
		 &=\left( \frac{d  G^{-1} (\xi+T) }{d z} \right)^{ {h_{\mathcal{O}}} } \left(  \frac{d \bar{G}^{-1}(\bar{\xi}+T)}{d\bar{z}} \right)^{ {h_{\mathcal{O}}} }  \mathcal{O}\left(G^{-1}(\xi+T),\bar{G}^{-1}(\bar{\xi}+T)\right),
	\end{aligned}
\end{equation}
where, in the final line, we used \eqref{eq:deffexi} and the fact that, from the expressions \eqref{eq:deffexi}, \eqref{eq:zxi} and the inverse function theorem\footnote{That is, $\dfrac{dG^{-1}(\xi)}{d\xi} = \left( \dfrac{d G(z)}{dz} \right)^{-1}$.}, the functions $g(z),\bar{g}(\bar{z})$ can be written as
\begin{equation}
	g(z)=  \frac{d G^{-1}(\xi)}{d\xi}, \quad \bar{g}(\bar{z})=  \frac{d \bar{G}^{-1}(\bar{\xi})}{d\bar{\xi}},
\end{equation}
leading to 
\begin{equation}
	\begin{aligned}
		g\left( G^{-1}\left(\xi +T\right) \right)&= \frac{d  G^{-1} (\xi+T) }{d\xi},\\
		\bar{g}\left(\bar{G}^{-1} (\bar{\xi}+T)\right)&=  \frac{d \bar{G}^{-1}(\bar{\xi}+T)}{d\bar{\xi}}.
	\end{aligned}
\end{equation}

Finally, we rewrite the overall factor such that the transformation rule has the standard form \eqref{eq:Euc_Time_evo_in_z_coordinates}. 
To this end, we consider the following conformal map in the right hand side of \eqref{eq:transRatio},
\begin{equation}
	\begin{aligned}
		&\left( \frac{d  G^{-1} (\xi+T) }{d z} \right)^{ {h_{\mathcal{O}}} } \left(  \frac{d \bar{G}^{-1}(\bar{\xi}-T)}{d\bar{z}} \right)^{ {h_{\mathcal{O}}} }  \mathcal{O}\left(G^{-1}(\xi+T),\bar{G}^{-1}(\bar{\xi}+T)\right)\\
		&=\left( \frac{d  G^{-1} (\xi+T) }{d z} \right)^{ {h_{\mathcal{O}}} } \left(  \frac{d \bar{G}^{-1}(\bar{\xi}+T)}{d\bar{z}} \right)^{ {h_{\mathcal{O}}} }   \left( \frac{d(\xi+T)}{d G^{-1}(\xi+T)}\right)^{ {h_{\mathcal{O}}} } \left( \frac{d(\bar{\xi}+T)}{d \bar{G}^{-1}(\bar{\xi}+T)}\right)^{ {h_{\mathcal{O}}} }\mathcal{O}\left(\xi+T,\bar{\xi}+T\right)\\
		&=  \left( \frac{d(\xi+T)}{d z}\right)^{ {h_{\mathcal{O}}} } \left( \frac{d(\bar{\xi}+T)}{d \bar{z}}\right)^{ {h_{\mathcal{O}}} }\mathcal{O}\left(\xi+T,\bar{\xi}+T\right).
	\end{aligned}
\end{equation}
Thus, we obtain \eqref{eq:Euc_Time_evo_in_z_coordinates}.


\section{Lorentzian computation on the mutual information \label{sec:LC-on-MI}}
In this section, we will report the time dependence of the mutual information in the asymmetric and disjoint cases.
\subsection{Disjoint cases}
In this section, we will calculate the holographic entanglement entropy for the disjoint intervals considered in Section \ref{sec:disjoint-case}.
The geodesic lengths corresponding to the connected pieces are given by
\be
\begin{aligned}
    \mathcal{L}_{13}&\approx \log\left(2\epsilon^2\f{\cosh\left[\f{\pi(x_{\xi}(Y_2)-x_{\xi}(X_2))}{\epsilon}\right]+\cosh\left[\f{\pi t}{\epsilon}\right]}{{\delta_{\text{UV}}'}^2\pi^2 f(X_2)f(Y_2)}\right),\\
    \mathcal{L}_{24}&\approx \log\left(2\epsilon^2\f{\cosh\left[\f{\pi(x_{\xi}(Y_1)-x_{\xi}(X_1))}{\epsilon}\right]+\cosh\left[\f{\pi t}{\epsilon}\right]}{{\delta_{\text{UV}}'}^2\pi^2 f(X_1)f(Y_1)}\right),
\end{aligned}
\ee
where we have used \eqref{eq:re-scaled-cutoff}. Note that both $\mathcal{L}_{13},\mathcal{L}_{24}$ are monotonically-increasing functions with time $t$, such that they take their minimal values at $t=0$ as 
\be
\begin{aligned}
    \mathcal{L}_{13}^{\text{Min}}=\mathcal{L}_{13}(t=0)&\approx \log\left(2\epsilon^2\f{\cosh\left[\f{\pi}{\epsilon}(x_{\xi}(Y_2)-x_{\xi}(X_2))\right]}{{\delta_{\text{UV}}'}^2\pi^2 f(X_2)f(Y_2)}\right),\\
    \mathcal{L}_{24}^{\text{Min}}=\mathcal{L}_{24}(t=0)&\approx \log\left(2\epsilon^2\f{\cosh\left[\f{\pi}{\epsilon}(x_{\xi}(Y_1)-x_{\xi}(X_1))\right]}{{\delta_{\text{UV}}'}^2\pi^2f(X_1)f(Y_1)}\right).
\end{aligned}
\ee
Then, let us consider the geodesic lengths of disconnected pieces, and they are read from \eqref{Geodesic length of disconnected pieces}. 
To determine the extreme surface, we need to compare $\mathcal{L}_{13}^{\text{Min}}+\mathcal{L}_{24}^{\text{Min}}$ with the value of $\mathcal{L}_{12}+\mathcal{L}_{34}$. 
Since $x_{\xi}(x)$ is a monotonically-increasing function with respect to $x$, and suppose that $X_2<X_1<Y_2<Y_1$.
Then we obtain an inequality about $x_{\xi}(X_{j=1,2})$ and $x_{\xi}(Y_{j=1,2})$ 
\be
\begin{aligned}
    \sum_{K=1,2}l^{\text{effective}}(Y_K,X_K)>l^{\text{effective}}_{A_j}+l^{\text{effective}}_{B_k},
\end{aligned}
\ee
where we have used $\delta_{\text{UV}} \ll 1$, and \eqref{eq:effective-subsystem-size}-\eqref{eq:effective-length}. When we consider spatially-disjoint subsystems, the $S_{A_{j}\cup B_k}^{i=1}$ is determined by 
\be
    S_{A_j\cup B_k}^{i=1}\approx\f{c}{6}\cdot\left(\mathcal{L}_{12}+\mathcal{L}_{34}\right),
\ee
where $\mathcal{L}_{12},\mathcal{L}_{34}$ are given by \eqref{Geodesic length of disconnected pieces}. Furthermore, by substituting \eqref{eq:effective-subsystem-size} and \eqref{eq:effective-length} into $\mathcal{L}_{12},\mathcal{L}_{34}$, $S_{A_j\cup B_k}^{i=1}$ is consistent with the CFT computation.

\subsection{Asymmetric cases}
The asymmetric settings refer to the subsystems $A_j\cup B_k$ with spatial overlaps (or one subsystem is included by the other one). The subsystems can be chosen as $A_{j=1,2}\cup B_{k=j}$ and  $A_{j=1,2}\cup B_{k\neq j}$. Let us discuss them case by case.
\begin{itemize}
    \item $A_{j=1,2}\cup B_{k=j}$: in this case, let us consider $X_2<Y_2<X_1<Y_1$. Two subsystems are partly overlapped with each other. Similar to the disjoint cases, we first need to compare $\mathcal{L}_{13}^{\text{Min}}+\mathcal{L}_{24}^{\text{Min}}$ with $\mathcal{L}_{12}+\mathcal{L}_{34}$, because the minimal values of $\mathcal{L}_{13}^{\text{Min}},\mathcal{L}_{24}^{\text{Min}}$ are their initial values, and they monotonically grow with time.
    If
    \be
    \sum_{K=1}^2l^{\text{effective}}(Y_K,X_K)\geq l^{\text{effective}}_{A_j}+l^{\text{effective}}_{B_j},
    \ee
    then the static geodesics $\mathcal{L}_{12}+\mathcal{L}_{34}$ dominate during the evolution, and mutual information is zero. In contrast, if
    \be
    \sum_{K=1}^2l^{\text{effective}}(Y_K,X_K)<l^{\text{effective}}_{A_j}+l^{\text{effective}}_{B_j},
    \ee
    the entanglement entropy is
    \be\label{eq:two-interval-EE-gravity-computation-asymmetric-case-overlap}
    \begin{aligned}
        S_{A_j\cup B_j}^{i=1}\approx\f{c}{6}\begin{cases}
            \mathcal{L}_{13}+\mathcal{L}_{24}~&\text{for}~t\leq t^*,\\
            \mathcal{L}_{12}+\mathcal{L}_{34}~&\text{for}~t> t^*,
        \end{cases}
    \end{aligned}
    \ee
    where $t^*$ is determined by the equality $\mathcal{L}_{13}+\mathcal{L}_{24}=\mathcal{L}_{12}+\mathcal{L}_{34}$. The time dependence of the mutual information is given by
    \be\label{eq:two-interval-MI-gravity-computation-asymmetric-case-overlap}
    \begin{aligned}
        I_{A_j, B_j}^{i=1}\approx\f{c}{6}\begin{cases}
            \mathcal{L}_{12}+\mathcal{L}_{34}-\mathcal{L}_{13}-\mathcal{L}_{24}~&\text{for}~t\leq t^*,\\
            0~&\text{for}~t> t^*,
        \end{cases}
    \end{aligned}
    \ee
     where $\mathcal{L}_{12},~\mathcal{L}_{34},~\mathcal{L}_{13}$ and $\mathcal{L}_{24}$ can be explicit computed by using \eqref{Geodesic length of disconnected pieces}, \eqref{Geodesic length of connected pieces} and \eqref{eq:re-scaled-cutoff}. After explicit calculations, one can easily find that the time dependence of $S_{A_j\cup B_j}^{i=1}$ and $I_{A_j,B_j}^{i=1}$ is consistent with \eqref{EE-with-exchange of dominance} and \eqref{MI-with-exchange of dominance}. Since all calculations can be carried out in the same manner as in the symmetric or disjoint cases, we do not provide the detailed calculation here.

     Similarly, one can also consider the spatial locations as $Y_2<X_2<X_1<Y_1$, which means one subsystem includes another. In this case, since the inequality,
     \be
        \sum_{K=1}^2l^{\text{effective}}(Y_K,X_K)< l^{\text{effective}}_{A_j}+l^{\text{effective}}_{B_j},
    \ee
    holds, the entanglement entropy and mutual information are the same as those in \eqref{eq:two-interval-MI-gravity-computation-asymmetric-case-overlap} and \eqref{eq:two-interval-EE-gravity-computation-asymmetric-case-overlap}.
    
    \item $A_{j=1,2}\cup B_{k\neq j}$: now, without loss of generality, in the large $\theta$ limit we consider $A_1\cup B_2$ that satisfies the relations $X_2<Y_2<X_1<Y_1$ or $Y_2<X_2<X_1<Y_1$. For the former, $A_j$ is partly overlapped with $B_k$, and the candidates of minimal geodesic length are given by
    \be
    \begin{aligned}
        \mathcal{L}_{12}+\mathcal{L}_{34}\approx& 2\log\left(\f{2\epsilon\sinh\left[\f{L}{4q\epsilon}\f{\sin\left[\f{q\pi(X_1-X_2)}{L}\right]}{\sin\left[\f{q\pi X_1}{L}\right]\sin\left[\f{q\pi X_2}{L}\right]}\right]}{\delta_{\text{UV}}'\pi \sqrt{f(X_1)f(X_2)}}\right)+2\log\left(\f{2\epsilon\sinh\left[\f{l\pi}{2q\epsilon}L_{\text{eff}}\right]}{\delta_{\text{UV}}'\pi \sqrt{f(Y_1)f(Y_2)})}\right),\\
        \mathcal{L}_{13}+\mathcal{L}_{24}\approx& \log\left(\f{2\epsilon^2\left(\cosh\left[\f{L}{2q\epsilon}\f{\sin\left[\f{q\pi(Y_2-X_2)}{L}\right]}{\sin\left[\f{q\pi Y_2}{L}\right]\sin\left[\f{q\pi X_2}{L}\right]}\right]+\cosh\left[\f{\pi t}{\epsilon}\right]\right)}{{\delta_{\text{UV}}'}^2\pi^2 f(X_2)f(Y_2)}\right)\\
        &+\log\left(\f{2\epsilon^2\left(\cosh\left[\f{l\pi}{q\epsilon}L_{\text{eff}}\right]+\cosh\left[\f{\pi t}{\epsilon}\right]\right)}{{\delta_{\text{UV}}'}^2\pi^2 f(X_1)f(Y_1)}\right),
    \end{aligned}
    \ee
    where we have applied \eqref{eq:re-scaled-cutoff}. Recall that $\mathcal{L}_{13}+\mathcal{L}_{24}$ takes its minimal value at $t=0$ since $\mathcal{L}_{13}$ and $\mathcal{L}_{24}$ are monotonically-growing functions with $t$. Again, we define $\mathcal{L}_{24}^{\text{Min}}:=\mathcal{L}_{24}(t=0)$ and $\mathcal{L}_{13}^{\text{Min}}:=\mathcal{L}_{13}(t=0)$, then compare $\mathcal{L}_{13}^{\text{Min}}+\mathcal{L}_{24}^{\text{Min}}$ with $\mathcal{L}_{12}+\mathcal{L}_{34}$.
    As the value of the latter exceeds that of the former, the entanglement entropy can be determined by
    \be
        \begin{aligned}
        &S_{A_1\cup B_1}^{i=1}\approx\f{c}{6}\begin{cases}
            \mathcal{L}_{13}+\mathcal{L}_{24}~&\text{for}~t\leq t^*,\\
            \mathcal{L}_{12}+\mathcal{L}_{34}~&\text{for}~t> t^*,
        \end{cases}\\
        &I_{A_1, B_1}^{i=1}\approx\f{c}{6}\begin{cases}
            \mathcal{L}_{12}+\mathcal{L}_{34}-\mathcal{L}_{13}-\mathcal{L}_{24}~&\text{for}~t\leq t^*,\\
            0~&\text{for}~t> t^*.
            \end{cases}
    \end{aligned}
    \ee

    For the case where $B_k$ completely contains $A_j$ with the relation $Y_2<X_2<X_1<Y_1$, since $l^{\text{effective}}_{A_j}>l^{\text{effective}}(X_2,Y_2)$, the time dependence of the entanglement entropy and mutual information are determined by
    \be
        \begin{aligned}
        &S_{A_1\cup B_1}^{i=1}=\f{c}{6}\begin{cases}
            \mathcal{L}_{13}+\mathcal{L}_{24}~&\text{for}~t\leq t^*,\\
            \mathcal{L}_{12}+\mathcal{L}_{34}~&\text{for}~t> t^*,
        \end{cases}\\
        &I_{A_1, B_1}^{i=1}=\f{c}{6}\begin{cases}
            \mathcal{L}_{12}+\mathcal{L}_{34}-\mathcal{L}_{13}-\mathcal{L}_{24}~&\text{for}~t\leq t^*,\\
            0~&\text{for}~t> t^*,
            \end{cases}
    \end{aligned}
    \ee
    where $\mathcal{L}_{12},~\mathcal{L}_{34},~\mathcal{L}_{13}$ and $\mathcal{L}_{24}$ can be explicit computed by using \eqref{Geodesic length of disconnected pieces}, \eqref{Geodesic length of connected pieces} and \eqref{eq:re-scaled-cutoff}. These results have the same time dependence as \eqref{EE-with-exchange of dominance} and \eqref{MI-with-exchange of dominance}. As all calculations can be followed in the same detailed steps as in the symmetric or disjoint cases, we do not provide the similar calculation for $j\neq k$ cases as well.
\end{itemize}

\section{Line-tension picture for the mutual information \label{sec:LP-on-MI}}
Here, we will report the detailed calculation, in the line tension picture, of the mutual information for the asymmetric and disjoint cases.

\subsection{Disjoint cases}

Here, in the line-tension picture, we will calculate the time dependence of the entanglement entropy and mutual information in the disjoint interval case considered in Sections \ref{Sec:Etanglement-dynamics-Ueff} and  \ref{sec:gravity-dual}.
In this case $X_2<X_1<Y_2<Y_1$, due to the holomogous condition for $C$ and the flatness for coordinates $(t,x_{\xi})$, $C$ comprises of two straight lines, denoted by $C_K$, connecting $\left(0,x_{\xi}(X_K)\right)$ with $\left(t,x_{\xi}(Y_K)\right)$. The speed along $C_K$ is obtained by
\be
    v_{\xi}^{(K)}=\f{x_{\xi}(Y_K)-x_{\xi}(X_K)}{t},
\ee
which is initially larger than the speed of light. Additionally, the connected piece monotonically increases with $t$. Therefore, $S_{A_j\cup B_k}^{i=1;\text{dis.}}$ dominates during the time evolution, so that mutual information is simply zero.
We can see from the behavior of the entanglement entropy and mutual information that the results in the line-tension picture are consistent with those in the path-integral formalism and Lorentzian holography.

\subsection{Asymmetric cases}
We close this section by calculating the time dependence of the entanglement entropy and mutual information in the asymmetric cases considered in the previous sections.
Then, we will check if the results in the line-tension picture are consistent with those obtained in the previous sections. 


In the cases that one subsystem is included into the other, for simplicity, we take the endpoints of the subsystems to be $X_2=Y_2<X_1<Y_1$. In this case, $C$ comprises of two straight lines denoted by $C_{K=1,2}$. 
One of the straight lines, $C_2$, is a vertical line connecting points $\left(0,x_{\xi}(X_2)\right)$ and $\left(t,x_{\xi}(X_2)\right)$ with slope $v_{\xi}^{(2)}=0$, and the other, $C_1$,  is a line connecting $\left(0,x_{\xi}(X_1)\right)$ and $\left(t,x_{\xi}(Y_1)\right)$ with slope 
\be
    v_{\xi}^{(1)}=\f{x_{\xi}(Y_1)-x_{\xi}(X_1)}{t}.
\ee
Note that the disconnected pieces are given by \eqref{Disconnected-piece-line-tension}. 
For connected pieces, the effective time evolution operator is initially independent of time, so that $S_{A_j\cup B_k}^{i=1;\text{con.}}(t=0)\approx l^{\text{effective}}(X_1,Y_1)$, which $v_{\xi}^{(2)}=0,v_{\xi}^{(1)}>1$. When $0<t\leq l^{\text{effective}}(X_1,Y_1)$, $v_{\xi}^{(2)}=0,v_{\xi}^{(1)}\geq 1$ remain unchanged, and entanglement entropy is linearly increasing with slope $s_{\text{eq}}$ due to the contribution from $C_2$. 
Finally, when $t>l^{\text{effective}}$, we have $v_{\xi}^{(2)}=0,0\leq v_{\xi}^{(1)}<1$.
Since both contributions from $C_1,C_2$ depend on time, the connected piece grows linearly at twice the previous rate, $S_{A_j\cup B_k}^{i=1;\text{con.}}\approx 2s_{\text{eq}}t$. After minimization over connected- and disconnected pieces, by line-tension picture we obtain the entanglement entropy and mutual information as
\be
    \begin{aligned}
        S_{A_j\cup B_k}^{i=1}\approx&\f{c}{6}\sum_{Z=X,Y}\log{\left[4\prod_{K=1,2}\sin^2{\left(\f{q\pi Z_K}{L}\right)}\right]}+\f{2c}{3}\log{\left(\f{2\epsilon}{\pi}\right)}\\
        &+s_{\text{eq}}\cdot\begin{cases}
            t+l^{\text{effective}}(X_1,Y_1)~&~0< t\leq l^{\text{effective}}(X_1,Y_1)\\
            2t~&~l^{\text{effective}}(X_1,Y_1)< t\leq \f{l_{A_j}^{\text{effective}}+l_{B_k}^{\text{effective}}}{2}\\
            l_{A_j}^{\text{effective}}+l_{B_k}^{\text{effective}}~&~ t> \f{l_{A_j}^{\text{effective}}+l_{B_k}^{\text{effective}}}{2},
        \end{cases}\\
        I_{A_j, B_k}^{i=1}\approx &s_{\text{eq}}\cdot\begin{cases}
            l_{A_j}^{\text{effective}}+l_{B_k}^{\text{effective}}-l^{\text{effective}}(X_1,Y_1)-t~&~0< t\leq l^{\text{effective}}(X_1,Y_1)\\
            l_{A_j}^{\text{effective}}+l_{B_k}^{\text{effective}}-2t~&~l^{\text{effective}}(X_1,Y_1)< t\leq \f{l_{A_j}^{\text{effective}}+l_{B_k}^{\text{effective}}}{2}\\
            0~&~ t> \f{l_{A_j}^{\text{effective}}+l_{B_k}^{\text{effective}}}{2},
        \end{cases}
    \end{aligned}
\ee
where we have assumed that
\be
    \f{l_{A_j}^{\text{effective}}+l_{B_k}^{\text{effective}}}{2}>l^{\text{effective}}(X_1,Y_1).
\ee
Similarly, if we assume that
\be
    \f{l_{A_j}^{\text{effective}}+l_{B_k}^{\text{effective}}}{2}<l^{\text{effective}}(X_1,Y_1),
\ee
the entanglement entropy and mutual information read
\be
    \begin{aligned}
        S_{A_j\cup B_k}^{i=1}\approx&\f{c}{6}\sum_{Z=X,Y}\log{\left[4\prod_{K=1,2}\sin^2{\left(\f{q\pi Z_K}{L}\right)}\right]}+\f{2c}{3}\log{\left(\f{2\epsilon}{\pi}\right)}\\ &+s_{\text{eq}}\cdot\begin{cases}
            t+l^{\text{effective}}(X_1,Y_1)~&~0<t\leq l_{A_j}^{\text{effective}}+l_{B_k}^{\text{effective}}-l^{\text{effective}}(X_1,Y_1)\\
            l_{A_j}^{\text{effective}}+l_{B_k}^{\text{effective}}~&~ t> l_{A_j}^{\text{effective}}+l_{B_k}^{\text{effective}}-l^{\text{effective}}(X_1,Y_1),
        \end{cases}\\
        I_{A_j, B_k}^{i=1}\approx &s_{\text{eq}}\cdot\begin{cases}
            l_{A_j}^{\text{effective}}+l_{B_k}^{\text{effective}}-l^{\text{effective}}(X_1,Y_1)-t~&~0<t\leq l_{A_j}^{\text{effective}}+l_{B_k}^{\text{effective}}-l^{\text{effective}}(X_1,Y_1)\\
            0~&~ t> l_{A_j}^{\text{effective}}+l_{B_k}^{\text{effective}}-l^{\text{effective}}(X_1,Y_1).
        \end{cases}
    \end{aligned}
\ee

We can also assume $X_2<Y_2<X_1<Y_1$ such that $A_j$ and $B_k$ overlap with each other. Due to the homologous condition, the curve $C$ comprises of two lines, $C=C_1\cup C_2$. 
The velocities associated with $C_{K=1,2}$ are given by
\be
    v_{\xi}^{(K)}=\f{x_{\xi}(Y_K)-x_{\xi}(X_K)}{t}.
\ee
Then, the connected piece is evaluated as
\be
\begin{aligned}
    S_{A_j\cup B_k}^{i=1; \text{ con.}}\approx s_{\text{eq}}\cdot\begin{cases}
        \sum_{K=1}^2l^{\text{effective}}(X_K,Y_K)~&~0< t\leq t^*_1\\
        t+t^*_1~&~t^*_1< t\leq t^*_2\\
        2t~&~t> t^*_2,
    \end{cases}
\end{aligned}
\ee
where the exchanging time is given by
\be
\begin{aligned}
    t^*_1&=\text{Min}\{l^{\text{effective}}(X_1,Y_1),l^{\text{effective}}(X_2,Y_2)\},\\
    t^*_2&=\text{Max}\{l^{\text{effective}}(X_1,Y_1),l^{\text{effective}}(X_2,Y_2)\}.
\end{aligned}
\ee
Therefore, if we assume that
\be
    \f{l_{A_j}^{\text{effective}}+l_{B_k}^{\text{effective}}}{2}>t^*_2,
\ee
the entanglement entropy and mutual information in the line-tension picture are approximately given by
\be
\begin{aligned}
    S_{A_j\cup B_k}^{i=1}\approx&\f{c}{6}\sum_{Z=X,Y}\log{\left[4\prod_{K=1,2}\sin^2{\left(\f{q\pi Z_K}{L}\right)}\right]}+\f{2c}{3}\log{\left(\f{2\epsilon}{\pi}\right)}\\ &+s_{\text{eq}}\cdot\begin{cases}
       \sum_{K=1}^2l^{\text{effective}}(X_K,Y_K)~&~0< t\leq t^*_1\\
        t+t^*_1~&~t^*_1< t\leq t^*_2\\
        2t~&~ t^*_2<t<\f{l_{A_j}^{\text{effective}}+l_{B_k}^{\text{effective}}}{2}\\
        l_{A_j}^{\text{effective}}+l_{B_k}^{\text{effective}}~&~ t>\f{l_{A_j}^{\text{effective}}+l_{B_k}^{\text{effective}}}{2},
    \end{cases}\\
    I_{A_j, B_k}^{i=1}\approx&s_{\text{eq}}\cdot\begin{cases}
       l_{A_j}^{\text{effective}}+l_{B_k}^{\text{effective}}-\sum_{K=1}^2l^{\text{effective}}(X_K,Y_K)~&~0< t\leq t^*_1\\
        l_{A_j}^{\text{effective}}+l_{B_k}^{\text{effective}}-t^*_1-t~&~t^*_1< t\leq t^*_2\\
        l_{A_j}^{\text{effective}}+l_{B_k}^{\text{effective}}-2t~&~ t^*_2<t<\f{l_{A_j}^{\text{effective}}+l_{B_k}^{\text{effective}}}{2}\\
        0~&~ t>\f{l_{A_j}^{\text{effective}}+l_{B_k}^{\text{effective}}}{2}.
    \end{cases}
\end{aligned}
\ee
We note that $\f{l_{A_j}^{\text{effective}}+l_{B_k}^{\text{effective}}}{2}>t^*_1$ always hold, and then assume that 
\be
    t^*_1<\f{l_{A_j}^{\text{effective}}+l_{B_k}^{\text{effective}}}{2}<t^*_2.
\ee
In this case, the time dependence of the entanglement entropy and mutual information is approximately given by 
\be
\begin{aligned}
    S_{A_j\cup B_k}^{i=1}\approx&\f{c}{6}\sum_{Z=X,Y}\log{\left[4\prod_{K=1,2}\sin^2{\left(\f{q\pi Z_K}{L}\right)}\right]}+\f{2c}{3}\log{\left(\f{2\epsilon}{\pi}\right)}\\ &+s_{\text{eq}}\cdot\begin{cases}
       \sum_{K=1}^2l^{\text{effective}}(X_K,Y_K)~&~0< t\leq t^*_1\\
        t+t^*_1~&~t^*_1< t\leq l_{A_j}^{\text{effective}}+l_{B_k}^{\text{effective}}-t^*_1\\
        l_{A_j}^{\text{effective}}+l_{B_k}^{\text{effective}}~&~ t>l_{A_j}^{\text{effective}}+l_{B_k}^{\text{effective}}-t^*_1,
    \end{cases}\\
    I_{A_j, B_k}^{i=1}\approx&s_{\text{eq}}\cdot\begin{cases}
       l_{A_j}^{\text{effective}}+l_{B_k}^{\text{effective}}-\sum_{K=1}^2l^{\text{effective}}(X_K,Y_K)~&~0< t\leq t^*_1\\
        l_{A_j}^{\text{effective}}+l_{B_k}^{\text{effective}}-t^*_1-t~&~t^*_1< t\leq l_{A_j}^{\text{effective}}+l_{B_k}^{\text{effective}}-t^*_1\\
        0~&~ t>l_{A_j}^{\text{effective}}+l_{B_k}^{\text{effective}}-t^*_1.
    \end{cases}
\end{aligned}
\ee


\bibliographystyle{ieeetr}
\bibliography{reference.bib}
\end{document}